\documentclass[AMA,STIX1COL]{WileyNJD-v2}

\articletype{Research Article}

\received{<day> <Month>, <year>}
\revised{<day> <Month>, <year>}
\accepted{<day> <Month>, <year>}

\usepackage{bm}
\usepackage{subcaption}
\usepackage{hyperref}
\usepackage{color}
\usepackage{tikz}
\usepackage{pgfplots}
\usetikzlibrary{spy}
\usepackage{esdiff}
\usepackage{amsmath}
\usepackage{multirow}
\usetikzlibrary{calc}

\newcommand{\adaptation}[1]{\emph{#1}-adaptation}
\newcommand{\nekmesh}{\emph{NekMesh}}
\newcommand{\nektar}{\emph{Nektar++}}

\begin{document}

\title{\adaptation{rp} for compressible flows} 

\author[1]{Julian Marcon}
\author[1]{Giacomo Castiglioni}
\author[2]{David Moxey}
\author[1]{Spencer J. Sherwin}
\author[1]{Joaquim Peir\'{o}*}

\authormark{JULIAN MARCON \textsc{et al}}

\address[1]{
  \orgdiv{Aeronautics},
  \orgname{Imperial College London},
  \orgaddress{
    \state{London},
    \country{United Kingdom}
  }
}
\address[2]{
  \orgdiv{Engineering},
  \orgname{University of Exeter},
  \orgaddress{
    \state{Exeter},
    \country{United Kingdom}
  }
}

\corres{
* Joaquim Peir\'{o},
South Kensington Campus,
London SW7 2AZ, United Kingdom.
\email{j.peiro@imperial.ac.uk}
}

\presentaddress{
  South Kensington Campus,
  London SW7 2AZ, United Kingdom
}

\abstract[Abstract]{
  We present an \adaptation{rp} strategy for high-fidelity simulation of compressible inviscid flows with shocks.
  The mesh resolution in regions of flow discontinuities is increased by using a variational optimiser to \(r\)-adapt the mesh and cluster degrees of freedom there.
  In regions of smooth flow, we locally  increase or decrease the local resolution through increasing or decreasing the polynomial order of the elements, respectively.
  This dual approach allows us to take advantage of the strengths of both methods for best computational performance, thereby reducing the overall cost of the simulation.
  The adaptation workflow uses a sensor for both discontinuities and smooth regions that is cheap to calculate, but the framework is general and could be used in conjunction with other feature-based sensors or error estimators.
  We demonstrate this proof-of-concept using two geometries at transonic and supersonic flow regimes.
  The method has been implemented in the open-source spectral/\emph{hp} element framework \nektar{}, and its dedicated high-order mesh generation tool \nekmesh{}.
  The results show that the proposed \adaptation{rp} methodology is a reasonably cost-effective way of improving accuracy.
}


\jnlcitation{\cname{
\author{J. Marcon},
\author{G. Castiglioni},
\author{D. Moxey},
\author{S.J. Sherwin}, and
\author{J. Peir\'{o}}}
(\cyear{2019}),
\ctitle{\adaptation{rp} for compressible flows},
\cjournal{International Journal for Numerical Methods in Engineering} <year> <vol> Page <xxx>-<xxx>}

\footnotetext{\textbf{Abbreviations:}
  BC:\@ boundary conditions;
  CAD:\@ computer-aided design;
  CFD:\@ computational fluid dynamics;
  CPU:\@ central processing unit;
  DG:\@ discontinuous Galerkin;
  DOF:\@ degrees of freedom;
  HLLC:\@ Harten-Lax-van Leer-Contact;
  HPC:\@ high-performance computing;
  IC:\@ initial conditions;
  I/O:\@ input-output;
  LDG:\@ local discontinuous Galerkin;
  NACA:\@ National Advisory Committee for Aeronautics;
  PDE:\@ partial differential equations
}

\maketitle

\section{Introduction}\label{sec:intro}

The accurate and high-fidelity simulation of high-speed compressible flows is, at present, a problem of
significant interest to the aeronautics community, particularly in relation to aviation in which such conditions
are routinely encountered. The complex and interdependent fluid phenomena found in this regime pose a
difficult challenge for numerical modelling, with a stark contrast between regions of smooth flow, boundary
layers near solid walls where large velocity gradients are present, and the interaction with shock waves or
shear layers where the fluid properties change sharply in a discontinuous manner.

The use of high-order spectral/\emph{hp} element methods in the simulation of compressible fluid dynamics
is now becoming increasingly common for high-fidelity large-eddy simulations and direct numerical simulations
of realistic aeronautical configurations~\cite{degrazia-2016}.
As in traditional low-order methods, the domain of interest is partitioned into finite elements; however, these
elements are also equipped with high-order polynomial expansions, as opposed to traditional linear shape functions.
This yields several advantages in terms of computational performance, as well as enhanced numerical resolution as \(p\) is increased.
However, in the presence of shocks and discontinuities, the latter advantage will not be realised, and can lead to significant issues in terms of stability and accuracy in the resolution of shocks.

A common approach used in the resolution of discontinuous features is to refine these regions in an adaptive manner, so that the mesh resolution around the features is increased.
In broad terms, the error of a computed solution which is sufficiently smooth can be roughly expressed as \(e \approx k h^p\), where \(k\) is a constant related to the measure of the solution regularity, \(h\) is the mesh size, and \(p\) is the polynomial order.
Mesh adaptation is concerned with achieving increased resolution by either locally reducing the mesh size, \(h\), or locally increasing the polynomial order, \(p\).
Due to its higher convergence rates, \adaptation{p} is typically preferred over \adaptation{h} for smooth flow regions~\cite{Burbeau2005,Li2010,Ekelschot2016}, whereas the opposite is true where flow discontinuities exist.
The reason for the latter --- \adaptation{h} being preferred for flow discontinuities --- is that the representation of shocks by high-order discretisations leads to numerical oscillations that must be smoothed out by the addition of high-order dissipation terms.
This effectively means that the high-order degrees of freedom (DOF) are wasted in the vicinity of shocks.

To address these issues, we present a proof-of-concept strategy based on \adaptation{rp} to best take advantage of \emph{h}-type, through \adaptation{r}, and \emph{p}-type local resolution modifications.
For the \adaptation{r} procedure, a variational optimiser is used to deform the mesh~\cite{Marcon2017}.
By targeting a small element size in regions of shocks, the optimiser deforms the mesh and clusters nodes in said regions.
By effectively redistributing DOF, \emph{h}-type refinement is obtained at flow discontinuities.
We then apply \adaptation{p}~\cite{Ekelschot2016} to this adapted mesh to better resolve regions of smooth flow.
Throughout the work, we focus on the simulaton of inviscid flows and focus on the challenge of efficiently modelling smooth flows with embedded discontinuities.

The success of an adaptation procedure depends on the use of reliable error indicators.
Different types of error indicators have been studied over the years, each with their own pros and cons.
In a first category, we identify indicators based on flow or solution features, such as boundary layer, multiphase interfaces and vortices~\cite{Mitran2001}. These can be costly to evaluate and are often not robust.
Another type of indicator looks at the discretisation error, namely the difference between the exact and the discrete solution.
The exact solution is typically unknown though and a practical workaround is to compare the solution at two different levels of accuracy, e.g. the solution and its projection onto a coarser mesh~\cite{Oden1995} or onto a lower polynomial space~\cite{Persson2006}.
These indicators are typically cheaper to compute but only highlight regions of high local error, even convected error.
Finally, goal oriented indicators provide sensitivities of a target quantity of interest to local mesh changes~\cite{Fidkowski2011}.
These are often based on an adjoint solution~\cite{Yano2012}, which is expensive to obtain, but they give
an accurate indicator of solution error since they incorporate the physics of the problem through the computed adjoint sensitivities.
However, for the problems we consider here, where shocks are a dominant feature of the flow physics, an adjoint-based error indicator may not yield the desired increase in accuracy, as shown by Ekelschot~\emph{et~al.}~\cite{Ekelschot2016}.
For a more complete review of existing error indicators for high-order computational fluid dynamics (CFD) solutions, we refer the reader to the work of Naddei~\emph{et~al.}~\cite{Naddei2019}.

In this article, we use a discontinuity sensor~\cite{Persson2006} that is easily computed as it essentially looks at the energy of the higher modes to determine the level of resolution of the solution.

Though originally intended as a \emph{shock sensor}, the sensor is in fact a resolution indicator based on the decay of modal energy.
This characteristic makes it applicable to regions of smooth flows as well as regions containing shocks.
This kind of indicator has been successfully used in various contexts, including incompressible, \emph{h}-adaptive spectral element solvers~\cite{offermans2020adaptive}, smooth compressible finite difference solvers~\cite{jacobs2018error}, and incompressible~\cite{Moxey2017} and smooth compressible~\cite{Moxey2017,Naddei2019}, \emph{p}-adaptive spectral element solvers.

Because of its versatility, the purposes of this sensor in the current work are threefold:
firstly it adds artificial viscosity to the governing equations, based on values of the sensor, to stabilise the solution in the presence of shocks;
secondly it identifies regions of flow discontinuities based on values of the sensor, as used for the artificial viscosity, to drive \adaptation{r};
and thirdly it locally increases or decreases the local polynomial approximation based on the values of the sensor.

We present the proposed proof-of-concept methodology as follows.
Sect.~\ref{sec:gov} introduces the governing equations in continuous and discrete forms.
Sect.~\ref{sec:discr} describes the spectral/\emph{hp} discontinuous Galerkin (DG) discretisation used in \nektar{}, with Sect.~\ref{sec:visc} covering the formulation of the discontinuity sensor and the artificial viscosity.
Sect.~\ref{sec:r-adapt} recalls previous work on variational \adaptation{r}~\cite{Marcon2017}.
Sect.~\ref{sec:p-adapt} summarises the \adaptation{p} strategy~\cite{Ekelschot2016}.
Sect.~\ref{sec:workflow} describes the novel dual \adaptation{rp} workflow.
Finally, we present two numerical examples in Sect.~\ref{sec:num-ex}: a transonic flow past a NACA 0012 profile at a free-stream Mach number of 0.8 an incidence of 1.25 degrees, and a supersonic flow at a free-stream Mach number of 3 past an engine intake that exhibits a complex shock pattern in its diffuser.

\section{Governing equations}\label{sec:gov}

The Euler equations of inviscid compressible flow are written, in a two-dimensional Cartesian frame of reference with coordinates \mbox{\(\bm{x}=(x_1,x_2)\)} within a domain \(\Omega \) with boundary \(\Gamma \),  as
\begin{equation}
  \label{eq:EU}
  \frac{\partial {\bf u}}{\partial t} + \nabla \cdot {\bf F} =
  \frac{\partial {\bf u}}{\partial t} + \nabla \cdot \left [ \, {\bf F}_c({\bf u}) + {\bf F}_d({\bf u}, \nabla {\bf u}) \, \right ] = {\bf 0}
\end{equation}
Here \mbox{\({\bf u} =  {[ \rho, \rho v_1, \rho v_2, \rho E ]}^t\)} is the vector of conserved variables, where
\(\rho \) is the density, the Cartesian components of the velocity are \(\bm{v}=(v_1,v_2)\), and \(E\) is the total energy.
The terms \({\bf F}_c\) and \({\bf F}_d\) denote the convective and dissipative fluxes, respectively.
A dissipative flux is required to stabilise the solution in the presence of shocks which is chosen to be of the form
\begin{equation}
  \label{eq:artdis}
  {\bf F}_d = -  \mu_a({\bf u})  \nabla {\bf u}
\end{equation}
where \(\mu_a\) is an artificial viscosity coefficient that will be discussed in detail in Sect.~\ref{sec:visc}.
The components of the convective flux, \({\bf F}_c = ( {\bf f}_1, {\bf f}_2 )\), are given by
\begin{equation}
  \label{eq:convflux}
  {{\bf f}}_1=\left \{ \begin{array}{c}
    \rho v_1 \\ P+\rho v_1^2 \\ \rho v_1 v_2 \\ \rho v_1 H
  \end{array}
  \right \},\quad
  {{\bf f}}_2=\left \{ \begin{array}{c}
    \rho v_2 \\ \rho v_1v_2 \\ P + \rho v_2^2  \\ \rho v_2 H
  \end{array}
  \right \}
\end{equation}
where \(H\) is the total enthalpy and \(P\) is the pressure.
The total enthalpy is defined as
\begin{equation}
  H = E + \frac{P}{\rho}
\end{equation}
and, to close the system, the pressure for a perfect gas is given by
\begin{equation}
  P  = (\gamma - 1) \rho \left(E-\frac{v_1^2+v_2^2}{2}\right)
\end{equation}
where \(\gamma \) is the ratio of specific heats and its value for air is \(\gamma = 1.4\).

The setting of the problem is completed through a suitable choice of initial and boundary conditions (IC/BC).
Given that only steady-state problems are of interest here, we start the simulation with a uniform flow at the given free-stream Mach number and flow incidence.
Solid walls are modeled through the no-flow BC, \(\bm{v} \cdot \bm{n} = 0\), where \(\bm{n}\) denotes the wall outer normal.
Far-field boundaries are weakly imposed through the normal boundary fluxes by specifying free-stream BC, \({\bf u} = {\bf u}_\infty \), outside the boundary and evaluating the normal fluxes through a Riemann solver that accounts for the propagation of information across the boundary.

\section{Discontinuous Galerkin discretisation}\label{sec:discr}

To obtain a discrete solution of~\eqref{eq:EU} via a high-order spectral/\(hp\) DG discretisation, we assume that the computational domain, \(\Omega \), is subdivided into \(N_{el}\) non-overlapping elements, so that
\(\Omega = \bigcup_{e=1}^{N_{el}}\Omega^e\) and \(\Omega^{e_1}\bigcap\Omega^{e_2}=\emptyset \) for \(e_1,e_2=1,\ldots,N_{el}\) and \(e_1\neq e_2\).
We adopt a mixed formulation similar to~\cite{Bassi1996} and write~\eqref{eq:EU} as
\begin{eqnarray}
  {\bf g} - \nabla {\bf u}&= & \bf {\bf 0}\label{eq:EUmix1}\\
  \frac{\partial {\bf u}}{\partial t} + \nabla \cdot  \left [ \, {\bf F}_c({\bf u}) + {\bf F}_d({\bf u}, {\bf g} ) \, \right ] &= &{\bf 0}\label{eq:EUmix2}
\end{eqnarray}

We seek a discrete approximation within an element, \(\Omega^e\), of the form
\begin{equation}
  \label{eq:discrete}
  \mathbf{u}(\bm{x},t) \approx \mathbf{u}_h^e(\bm{x},t)= \sum_{i=1}^{N_{el}}  \mathbf{u}_i^e(t) w_i^e(\bm{x}); \quad \bm{x} \in \Omega^e
\end{equation}
where \(w_i^e(\bm{x}); i=1,\ldots, N_{el}\) represent the elemental expansion functions for the high-order spectral/{\(hp\)} DG method available in \nektar{}~\cite{Cantwell2015,moxey2020}.
Both the solution and test functions are discontinuous at the interface between elements.

Following the standard Galerkin procedure, a weak form of the mixed formulation~\eqref{eq:EUmix1}--\eqref{eq:EUmix2} is obtained as follows.
The discrete version of equation~\eqref{eq:EUmix1} reads
\begin{equation}
  \label{eq:eq1coup}
  \sum_{e=1}^{N_{el}} \int_{\Omega_e} w_i^e \left( {\bf g}_h^e-\nabla{\bf u}_h^e \right) d\Omega^e = \mathbf{0} \, ; \quad i=1,\ldots, N_{el}
\end{equation}
Using an approximation of the form~\eqref{eq:discrete} for both \({\bf u}_h^e\) and \({\bf g}_h^e\), and applying Gauss' theorem this equation becomes
\begin{equation}
  \label{eq:eq1coupled}
  \sum_{e=1}^{N_{el}}
  \int_{\Omega^e}   \left ( w_i^e  \sum_{j=1}^{N_{el}} {\bf g}_j^e  w_j^e +
  \nabla w_i^e \sum_{j=1}^{N_{el}} {\bf u}_j^e w_j^e \right ) d\Omega^e
  - \sum_{e=1}^{N_{el}}
  \int_{\Gamma^e} {w_i^e} \left ( \sum_{j=1}^{N_{el}} {\bf u}_j^e w_j^e \right ) \bm{n}\, d\Gamma^e
  = \mathbf{0} \, ; \quad i=1,\ldots, N_{el}
\end{equation}
where \(\Gamma^e\) denotes the boundary faces of the element \(\Omega^e\).
The solution of this equation give us the discrete values of the first-order derivatives \({\bf g}_h^e\).
To evaluate the integral expressions, we use an auxiliary mapping \(\phi_M: \bm{x} \mapsto \bm{\xi}\) to transform the local element coordinates \(\bm{x}=(x_1, x_2)\) to reference element coordinates \(\bm{\xi}=(\xi_1,\xi_2)\) such that \(-1\leq \xi_1,\xi_2\leq 1\) and all required operations take place in the reference element \(\Omega_{\textrm{st}}\), see Fig.~\ref{fig:mapping}.

The weak form of equation~\eqref{eq:EUmix2} is obtained in a similar fashion to give
\begin{equation}
  \label{eq:discEuler}
  \sum_{e=1}^{N_{el}} \int_{\Omega^e} w_i^e \sum_{j=1}^{N_{el}} \frac{d {\bf u}_j^e }{dt}  w_j^e \ d{\Omega}
  - \sum_{e=1}^{N_{el}} \int_{\Omega^e} \nabla w_i^e \sum_{j=1}^{N_{el}} {\bf F}_j^e w_j^e \ d{\Omega}
  + \sum_{e=1}^{N_{el}} \int_{\Gamma_e} w_i^e  \sum_{j=1}^{N_{el}} ( {\bf F}_j^e \cdot \bm{n}) w_j^e \ d\Gamma  = 0
  \, ; \quad i=1,\ldots, N_{el}
\end{equation}
The solution is discontinuous at the interface between the elements and the integrand in the boundary integral of~\eqref{eq:discEuler} is substituted by a numerical flux function.
The convective normal flux at an interface is approximated by a numerical flux calculated via a Riemann solver
\begin{equation}
  \label{eq:RiemannGov}
  {\left [ {\left ( {\bf F}_c \right )}_i^e \cdot \bm{n} \right ]}_{\Gamma_e} \approx \mathcal{H}^c({\bf u}_e,{\bf u}_{e^+};\bm{n})
\end{equation}
where \({\bf u}_{e^+}\) and \({\bf u}_e\) are the values of the conservative variables on the external and internal sides of the interface with respect to the \(e\)-th element.
This mechanism allows information to pass from one element to the other.
The evaluation of the diffusive normal flux at the interface follows the local discontinuous Galerkin (LDG) formulation~\cite{CockburnShu1997}, where it is approximated by
\begin{equation}
  \label{eq:viscflux}
  {\left [ {\left ( {\bf F}_d \right )}_i^e \right ]}_{\Gamma_e}
  = \{ \! \{ {\bf F}_d \} \! \} + {\bf C}_{12} [\![{\bf F}_d]\!] + C_{11} [\![\bf{u}]\!],
\end{equation}
and similarly
\begin{equation}
  \label{eq:viscflux_aux}
  {\left [ {\left ( {\bf u} \right )}_i^e  \right ]}_{\Gamma_e}
  = \{ \! \{ {\bf u} \} \! \} - {\bf C}_{12} [\![{ \bf u}]\!],
\end{equation}
where \({\bf C}_{12} = \frac{1}{2} {\bf n}\), and \(C_{11}\) is an order 1 constant.
The average and jump operators are defined as
\begin{align}
  \left \{ \! \left \{ u \right \} \! \right \}         & = \frac{1}{2} \left(  u^+ +u^- \right) ,
                                                        & \left [\! \left [ u \right ]\! \right ] =  \left(  u^+\bm{n}^+ +u^-\bm{n}^- \right),                                         \\
  \left \{ \! \left \{\mathbf{u} \right \} \! \right \} & = \frac{1}{2} \left(  \mathbf{u}^+ +\mathbf{u}^- \right) ,
                                                        & \left [\! \left [ \mathbf{u} \right  ]\! \right ] =  \left(  \mathbf{u}^+ \cdot\bm{n}^+ +\mathbf{u}^- \cdot\bm{n}^- \right).
\end{align}

\subsection{Shock capturing via a discontinuity sensor}\label{sec:visc}

Our DG discretisation of the Euler equations requires the addition of the diffusion flux, \({\bf F}_d\) to stabilize the solutions in the presence of shock waves.
The term \(\mu_a\) in~\eqref{eq:artdis} is an artificial viscosity coefficient that allows dissipation to be selectively applied to shocks.
For consistency \(\mu_a \sim h/p\), we use~\cite{barter2010shock}
\begin{equation}
  \mu_a \sim \frac{h}{p} \lambda_{\max},
\end{equation}
where \(\lambda_{\max} = |u|+c\) is the local maximum wave speed of the system.
The characteristic cell length \(h\) is chosen as the minimum edge length of an element.
Finally, for the artificial viscosity to vanish outside shocks it needs to be proportional to a shock sensor, \(S\), such as
\begin{equation}
  \mu_a = \mu_0 \frac{h}{p} \lambda_{\max} S,
\end{equation}
where \(\mu_0=O(1)\) is a constant.
To build the shock sensor, we adopt a modal resolution-based indicator~\cite{Persson2006} which is element-wise constant and defined via an intermediary term
\begin{equation}\label{eq:sensor}
  s_e = \log_{10}\left( \frac{\langle q - \tilde{q}, q - \tilde{q}  \rangle}{\langle q, q \rangle} \right) ,
\end{equation}
where \(\langle \cdot, \cdot \rangle \) represents a \(L^2\) inner product, \(q\) and \(\tilde{q}\) are the full and truncated expansions of a state variable
\begin{equation}
  q(x) = \sum_{i=1}^{N(P)} \hat{q}_i \phi_i , \quad \tilde{q}(x) = \sum_{i=1}^{N(P-1)} \hat{q}_i \phi_i ,
\end{equation}
where \(\phi_i\) are the basis functions, \(\hat{q}_i\) the associated coefficients, and \(N(P)\) the size of the expansion of order \(P\).
In our case the test variable \(q\) is chosen to be the density \(\rho \).
To spatially smooth out the variation of the values of the sensor, the constant element-wise sensor is computed as follows
\begin{equation}
  S =
  \left \{ \begin{array}{lr}
    0  ,                                                                                  \\
    \frac{1}{2} \left( 1 + \sin \frac{ \pi \left( s_e - s_0 \right) }{2 \kappa} \right) , \\
    1 ,
  \end{array}
  \right.
  \begin{array}{lr}
    s_e \leq s_0 - \kappa ,      \\
    | s_e - s_0 | \leq  \kappa , \\
    s_e \geq s_0 + \kappa ,      \\
  \end{array}
\end{equation}
with \(s_0 \sim \log_{10}(p^4)\) from an analogy to Fourier coefficients decaying as \(1/p^2\), and \(\kappa \) needs to be sufficiently large to obtain a smooth shock profile.
We select
\begin{equation}
  \quad s_0 = - s_\kappa - 4.25 \, \log_{10} \, p   ,
\end{equation}
where \(s_\kappa \) and \(\kappa \) can be adjusted for a specific problem.
We describe how to choose these parameters in Sect.~\ref{sec:workflow}.

\section{\emph{r}-adaptation}\label{sec:r-adapt}

In \emph{r}-adaptation we are aiming at increasing resolution by locally reducing the mesh size, \(h\), whilst keeping the number of DOF in the mesh constant.
This effectively requires us to cluster mesh nodes in the vicinity of those regions where additional resolution is required, e.g.\ shocks.
We propose to accomplish this by adapting a variational framework for the optimisation of high-order meshes~\cite{Turner2018}.

\subsection{Variational mesh optimisation}\label{sec:var-opti}

The objective of this variational framework~\cite{Turner2018} is to improve the quality of high-order \textit{curvilinear} elements using a node-based optimisation approach using a formulation based on the energy of deformation.
An important aspect of such energy-based formulation is that a suitable choice of the energy functional, namely polyconvex, would guarantee the existence of a minimum and therefore of a solution to the minimisation problem.

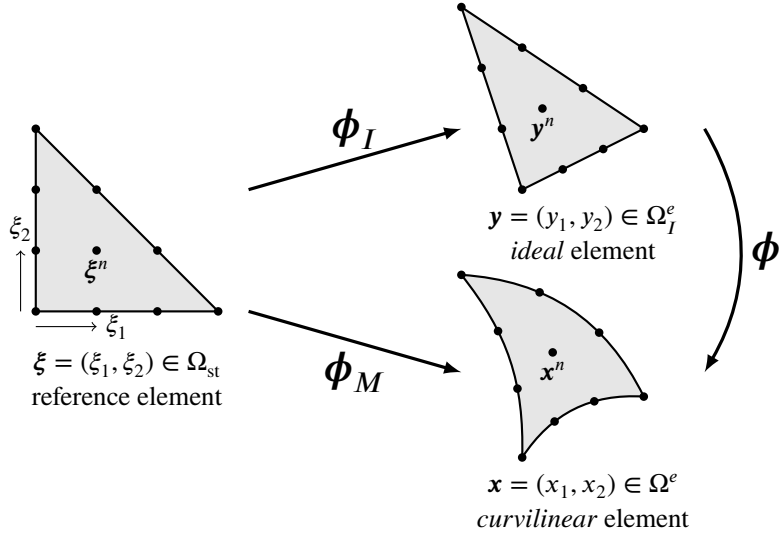
\begin{figure}[htbp!]
  \centering
  \begin{tikzpicture}[scale=0.8]
    \begin{scope}
      \node[align=center,below] at (1.5,-0.5) {$\bm{\xi} = (\xi_1,\xi_2) \in \Omega_{\text{st}}$\\reference element};
      \draw[->] (0,-0.25) -- (1,-0.25) node[right] {$\xi_1$} ;
      \draw[->] (-0.25,0) -- (-0.25,1) node[above] {$\xi_2$} ;
      \fill[black!10] (0,0) -- (3,0) -- (0,3) -- cycle;
      \draw[thick] (0,0) -- (3,0) -- (0,3) -- cycle;

      \fill[black] (0,0) circle (2pt);
      \fill[black] (1,0) circle (2pt);
      \fill[black] (2,0) circle (2pt);
      \fill[black] (3,0) circle (2pt);
      \fill[black] (0,1) circle (2pt);
      \fill[black] (1,1) circle (2pt) node[below] {$\bm{\xi}^n$};
      \fill[black] (2,1) circle (2pt);
      \fill[black] (0,2) circle (2pt);
      \fill[black] (1,2) circle (2pt);
      \fill[black] (0,3) circle (2pt);
    \end{scope}

    \begin{scope}[xshift=8cm, yshift=2cm]
      \node[align=center,below] at (1,-0.1) {$\bm{y} = (y_1,y_2) \in \Omega_{I}^e$\\\textit{ideal} element};
      \fill[black!10] (0,0) -- (2,1) -- (-1,3) -- cycle;
      \draw[thick] (0,0) -- (2,1) -- (-1,3) -- cycle;

      \fill[black] (0,0) circle (2pt);
      \fill[black] (0.667,0.333) circle (2pt);
      \fill[black] (1.333,0.667) circle (2pt);
      \fill[black] (2,1) circle (2pt);
      \fill[black] (-0.333,1) circle (2pt);
      \fill[black] (0.333,1.333) circle (2pt) node[below] {$\bm{y}^n$};
      \fill[black] (1,1.667) circle (2pt);
      \fill[black] (-0.667,2) circle (2pt);
      \fill[black] (0,2.333) circle (2pt);
      \fill[black] (-1,3) circle (2pt);
    \end{scope}

    \begin{scope}[xshift=8cm, yshift=-2.4cm]
      \node[align=center,below] at (1,-0.1) {$\bm{x} = (x_1,x_2) \in \Omega^e$\\\textit{curvilinear} element};
      \filldraw[fill=black!10,draw=black,thick]
	      (0,0) 
		  to [bend left] coordinate[pos=0.33] (A1) coordinate[pos=0.66] (A2) (2,1)
		  to [bend right=30] coordinate[pos=0.33] (B1) coordinate[pos=0.66] (B2) (-1,3)
		  to [bend left=20] coordinate[pos=0.33] (C1) coordinate[pos=0.66] (C2) cycle;
	  \path (B1) to [bend right=20] coordinate[pos=0.5] (D1) (C2);

      \fill[black] (0,0) circle (2pt);
      \fill[black] (A1) circle (2pt);
      \fill[black] (A2) circle (2pt);
      \fill[black] (2,1) circle (2pt);
      \fill[black] (B1) circle (2pt);
      \fill[black] (B2) circle (2pt);
      \fill[black] (C1) circle (2pt);
      \fill[black] (C2) circle (2pt);
      \fill[black] (D1) circle (2pt) node[below] {$\bm{x}^n$};
      \fill[black] (-1,3) circle (2pt);
    \end{scope}
	
    \draw[very thick,-latex] (3.5, 2) -- node[above] {\Large $\bm{\phi}_{I}$} (7,3);
    \draw[very thick,-latex] (3.5, 0) -- node[below] {\Large $\bm{\phi}_M$} (7,-1);
    \draw[very thick,-latex] (11, 3) to[bend left] node[right] {\Large $\bm{\phi}$} (11,-1);
  \end{tikzpicture}
  \caption{Existing isoparametric mappings between the reference, the \textit{ideal} and the \textit{curvilinear} elements. The \textit{ideal} and \textit{curvilinear} elements become respectively the \textbf{target} and \textbf{adapted} elements in the framework of \adaptation{r}.}\label{fig:mapping}
\end{figure}

The deformation of the mesh and the displacements of its nodes are represented via \textit{isoparametric} mappings as follows.
Fig.~\ref{fig:mapping} shows that a mapping \( \bm{\phi}_M \) exists from a reference element \( \bm{\Omega}_{st} \) to a \textit{curvilinear} high-order element \( \bm{\Omega}^e \).
We can further decompose the mapping \( \bm{\phi}_M \) into two distinct  mappings:
a mapping \( \bm{\phi}_I \) from reference to \textit{ideal} elements and a mapping \( \bm{\phi} \) from the \textit{ideal} to the \textit{curvilinear} elements.
We define the \textit{ideal} element as the high-order linear element, which after minimisation will be the element that the optimiser seeks to achieve.

From this \textit{ideal} element, we calculate the deformation energy.
The mesh is deformed to obtain a new set of nodal coordinates that minimise an energy functional \( \mathcal{E} (\nabla \bm{\phi}) \):
\begin{equation}
  \mathrm{find} \, \min_{\bm{\phi}} \mathcal{E} (\nabla \bm{\phi}) = \sum_e
  \int_{\bm{\Omega}^e} W(\nabla \bm{\phi})\, d\bm{y}
\end{equation}
where \( W (\nabla \bm{\phi}) \) is a formulation of the deformation energy.
Several formulations were tested and it was found that the best results were obtained when using a hyperelastic model~\cite{Turner2018}.
For this model, the strain energy takes the form
\begin{equation}
  W = \frac{\mu}{2} (I_1^{\bm{C}} - 3) - \mu \ln~J + \frac{\lambda}{2} {(\ln~J)}^2
\end{equation}
where \( \lambda \) and \( \mu \) are material constants, \( \bm{C} \) is the right Cauchy-Green tensor, \( I_1^{\bm{C}} \) is its trace and \( J \) is the determinant of the Jacobian matrix \( \bm{J} = \nabla \bm{\phi} \). We use this formulation in the work that follows.

\subsection{Improving the shock resolution via element scaling}\label{sec:shock-capt}

\begin{figure}[htbp!]
  \centering
  \begin{tikzpicture}[scale=0.8]
    \begin{scope}
      \node[align=center,below] at (1.5,-0.5) {$\bm{\xi} = (\xi_1,\xi_2) \in \Omega_{\text{st}}$\\reference element};
      \draw[->] (0,-0.25) -- (1,-0.25) node[right] {$\xi_1$} ;
      \draw[->] (-0.25,0) -- (-0.25,1) node[above] {$\xi_2$} ;
      \fill[black!10] (0,0) -- (3,0) -- (0,3) -- cycle;
      \draw[thick] (0,0) -- (3,0) -- (0,3) -- cycle;

      \fill[black] (0,0) circle (2pt);
      \fill[black] (3,0) circle (2pt);
      \fill[black] (0,3) circle (2pt);
    \end{scope}

    \begin{scope}[xshift=8cm, yshift=2cm]
      \node[align=center,below] at (1,0.5) {$\bm{y} = (y_1,y_2) \in \Omega_{T}^e$\\\textbf{target} element};
      \fill[black!10] (0.5,0.67) -- (1.5,1.17) -- (0.0,2.17) -- cycle;
      \draw[thick] (0.5,0.67) -- (1.5,1.17) -- (0.0,2.17) -- cycle;

      \fill[black] (0.5,0.67) circle (2pt);
      \fill[black] (1.5,1.17) circle (2pt);
      \fill[black] (0.0,2.17) circle (2pt);
    \end{scope}

    \begin{scope}[xshift=8cm, yshift=-2.4cm]
      \node[align=center,below] at (1,-0.1) {$\bm{x} = (x_1,x_2) \in \Omega^e$\\\textbf{adapted} element};
      \fill[black!10] (0.5,0) -- (1.5,1) -- (0.0,3) -- cycle;
      \draw[thick] (0.5,0) -- (1.5,1) -- (0.0,3) -- cycle;

      \fill[black] (0.5,0) circle (2pt);
      \fill[black] (1.5,1) circle (2pt);
      \fill[black] (0.0,3) circle (2pt);
    \end{scope}
	
    \draw[very thick,-latex] (3.5, 2) -- node[above] {\Large $\bm{\phi}_{T}$} (7,3);
    \draw[very thick,-latex] (3.5, 0) -- node[below] {\Large $\bm{\phi}_M$} (7,-1);
    \draw[very thick,-latex] (11, 3) to[bend left] node[right] {\Large $\bm{\phi}$} (11,-1);
  \end{tikzpicture}
  \caption{Existing mappings between the reference, the \textbf{target} and the \textbf{adapted} elements for \adaptation{r}.}\label{fig:mapping-r}
\end{figure}
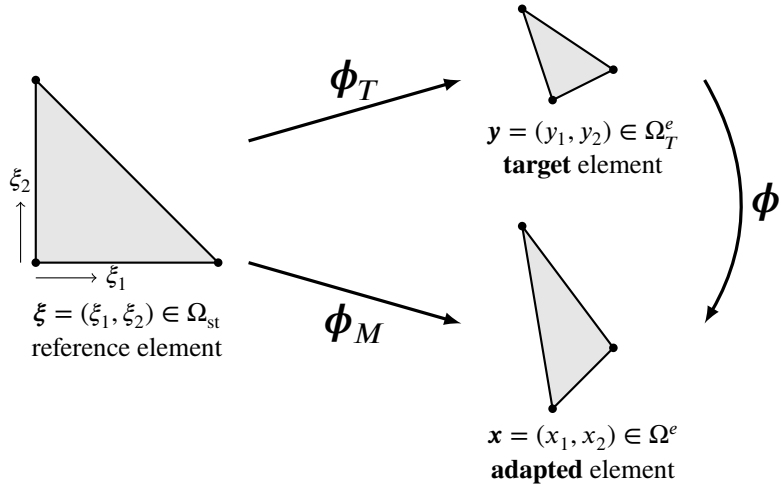

To achieve \adaptation{r}, we change this \textit{ideal} element and make it an arbitrary \textbf{target} element~\cite{Marcon2017}.
The optimiser now aims at \textbf{adapting} element \( \bm{\Omega}^e \) towards a size and a shape similar to the \textbf{target} element \( \bm{\Omega}_T^e \); see Fig.~\ref{fig:mapping-r}.
Previous work~\cite{Marcon2017} has shown that performing \adaptation{r} on the high-order mesh yielded spurious deformations.
For this reason, \adaptation{r}, in this work, is performed on the linear mesh before it is enriched to high-order.
This allows us to obtain high-order meshes of good quality, even after \adaptation{r}.

Although in principle the \textbf{target} element \( \bm{\Omega}_T^e \) can take any shape and size, we have adopted a practical approach in this work that aims at avoiding too large deformations.
The rationale for this is that the definition of a \textbf{target} element \( \bm{\Omega}_T^e \) that is very different from the \textit{ideal} element \( \bm{\Omega}_I^e \) --- i.e.\ the initial linear element before \adaptation{r} --- introduces extra energy in the system that the optimiser has to minimise and thus slows down the process.
For this reason, we define a \textbf{target} element \( \bm{\Omega}_T^e \) with respect to the \textit{ideal} element \( \bm{\Omega}_I^e \).
The \textit{ideal} element \( \bm{\Omega}_I^e \) can be manipulated anisotropically~\cite{Marcon2017} by applying a metric tensor \( \bm{M} \) to the Jacobian of the mapping, \( \bm{J} = \nabla \bm{\phi} \).
We transform the \textit{ideal} element \( \bm{\Omega}_I^e \) into the \textbf{target} element \( \bm{\Omega}_T^e \) through
\begin{equation}
  \bm{J}_T = \bm{M} \bm{J}_I
\end{equation}

We do not consider directionality in this work and only \emph{shrink} elements where additional resolution is required, i.e.\ in the shock regions.
In this case, the Jacobian is simply scaled by a linear shrinking factor \( r_{\text{scale}} \).
The metric tensor \( M \) is simplified to \( r_{\text{scale}} \bm{I} \) to obtain
\begin{equation}
  \bm{J}_T = (r_{\text{scale}} \bm{I}) \bm{J}_I
\end{equation}

This framework was implemented in \nekmesh{}, an open-source software solution for the generation of geometry-accurate high-order meshes, part of the \nektar{} platform~\cite{Cantwell2015,moxey2020}.

\section{\emph{p}-adaptation}\label{sec:p-adapt}

To enhance resolution in regions of smooth flow through local \adaptation{p}, we use the following procedure~\cite{Ekelschot2016,Moxey2017}.
The strategy is fairly straightforward.
We increase the local resolution by increasing the polynomial order within the elements where the local error is estimated to be high and we decrease the local resolution or, equivalently, the elemental polynomial order within those elements where the local error is estimated to be low.

We summarise this procedure in Algorithm~\ref{alg:p-adapt} where \(e\) denotes an individual element, \(s_e\) and \(p_e\) are its associated error indicator and polynomial order, \(\varepsilon_u\) and \(\varepsilon_l\) are the upper and lower error thresholds, and \(p_{\max}\) and \(p_{\min}\) are the maximum and minimum polynomial orders allowed.

At every iteration (see outer loop), the polynomial order of individual elements is increased or decreased by 1 single order at most.
It may also remain the same if neither the upper nor the lower threshold is reached.
Following this change in element polynomial orders, we project the steady state solution of the previous iteration onto the new polynomial space and the process continues.
In this work, we use the formulation of sensor \(s_e\) in equation~\eqref{eq:sensor} to drive the local change in polynomial order.

\begin{algorithm}
  \caption{The \textit{p}-adaptive procedure.}\label{alg:p-adapt}
  \begin{algorithmic}
    \Repeat{}
    \State{calculate the steady-state solution}
    \ForAll{\(e\)}
    \State{calculate \(s_e\)}
    \If{\(s_e>\varepsilon_u \text{ and } p_e<{p_{\max}}\)}
    \State{increment \(p_e\)}
    \ElsIf{\(s_e<\varepsilon_l \text{ and } p_e>{p_{\min}}\)}
    \State{decrement \(p_e\)}
    \Else{}
    \State{maintain \(p_e\)}
    \EndIf{}
    \EndFor{}
    \Until{no \(p_e\) is modified}
  \end{algorithmic}
\end{algorithm}

\section{Workflow for \emph{rp}-adaptation}\label{sec:workflow}

Finally we attempt to combine the best properties of the two previous strategies in the same simulation to maximize their effect in increasing the resolution of compressible flow simulations.
More specifically, \adaptation{r} will be responsible for the resolution of shocks whereas  \adaptation{p} will resolve smooth flow regions.
In the proposed \emph{rp}-adaptation workflow these adaptative techniques will be alternatively applied in a sequence of steps that is described in the following.

As noted in Sect.~\ref{sec:intro}, the sensor of Eq.~\ref{eq:sensor} is used, at the same time, to add artificial viscosity, to move mesh DOF towards shocks and to drive local \adaptation{p}.
We emphasise that although the original intention of this sensor in the work of Persson \& Peraire~\cite{Persson2006} was to identify regions of discontinuity, in this work we will also use the sensor to drive a $p$-adaptation process.
This is a valid strategy since the sensor is ultimately a resolution indicator that is based on the decay of modal energy, which therefore makes it applicable for usage in general regimes, and not only those that contain shocks.
This is directly validated by the work of Naddei~\emph{et~al.}~\cite{Naddei2019}.
In addition, indicators based on this concept have been successfully leveraged in both spectral element solvers of smooth flows, where the efficacy is demonstrated from the context of incompressible $h$-adaptivity~\cite{offermans2020adaptive}, as well as a broader numerical context of e.g. high-order finite difference simulations of smooth compressible flows~\cite{jacobs2018error}.

We first generate an initial high-order mesh for the domain.
We anticipate the requirements of \adaptation{r} and the need for DOF to be moved around when deforming the mesh.
For this reason, we generate a relatively coarse mesh, but with enough resolution to allow for nodes movement.
We then proceed to run the solver on this initial mesh and obtain a flow solution which represents our base solution.
During this step, we have to determine appropriate parameter values for the artificial viscosity.

The artificial viscosity term (\ref{eq:artdis}) depends on three parameters: \(\mu_0\), \(s_\kappa \) and \(\kappa \).
As is common practice in codes based on artificial viscosity, we start with \nektar{} default values, and then tune the parameters for our specific problem.
The default values are determined via one-dimensional analysis of the scaling and smoothing of the solution by its modal decay, as described for instance by Kl\"ockner~\emph{et~al.}\cite{Kloeckner2011}.
The level of artificial viscosity (\(\mu_0\)) is chosen empirically to be sufficiently large so as to obtain a sharp but smooth shock profile and it is tuned so that the shock is stable but not overly dissipative.
In the sensor equation (\ref{eq:sensor}), the parameter \(\kappa \) is the width of the activation window of shock sensor values and the parameter \(s_\kappa \), together with \(\kappa \), sets the threshold above which the shock is detected.
The artificial viscosity parameters (\(s_\kappa \) and \(\kappa \)) are adjusted to ensure that artificial viscosity is only triggered in the direct vicinity of shock waves, by excluding values of large gradients elsewhere, e.g. near stagnation points or trailing wakes.
The final values of the tuned parameters are given in Sect.~\ref{sec:num-ex} for the two test cases considered in this work.

From this base solution, we apply \adaptation{r} to the mesh.
We first extract the list of elements where artificial viscosity was added during the initial simulation.
If the run was set up properly, these elements only represent the regions where a shock is present.
From these elements, we extract their barycentres and assign an isotropic shrinking factor \(r_{\text{scale}}\) to them (see Sect.~\ref{sec:shock-capt}).
This shrinking factor \(r_{\text{scale}}\) is currently chosen empirically in such a way that as many nodes as possible are pulled inside the shock area, without compromising the quality of the mesh.
From experience, values as low as \(r_{\text{scale}}=0.1\) can be used in open geometries while more moderates values must be used in closed geometries.
Both situations are illustrated below.
For all the other elements, we also extract the barycentres and assign a factor \(r_{\text{scale}}=1\).
In practical terms, we force elements in the shock regions to shrink and pull mesh nodes from all other parts of the mesh.
This field of \(r_{\text{scale}}\) factors is then supplied to the variational \adaptation{r} code which is then run on the linear mesh.
The variational framework optimises the mesh so that each element is as close as possible to its target size, effectively moving nodes from areas of \(r_{\text{scale}}=1\) to areas of \(r_{\text{scale}}<1\).
We also note that \adaptation{r} is run on the linear mesh before making the adapted mesh high-order again.
This significantly speeds up the optimisation procedure and improves the validity of the final mesh.
We then run the solver on the adapted high-order mesh and obtain a new solution with enhanced shock resolution.
This procedure can optionally be repeated based on the new solution.

From this solution on the adapted mesh, we can run \adaptation{p} as described in Sect~\ref{sec:p-adapt}.
At the end of each cycle, a sensor value is computed for each element and the local polynomial order of that element is decreased, kept the same or increased based on the value of the sensor.
In this work, we use \nektar{} default values for the adaptation parameters: \(\varepsilon_u = -6\), \(p_{\max} = 6\), \(\varepsilon_l = -8\) and \(p_{\min} = 2\).
The simulation then proceeds onto a new cycle and the process is repeated until a steady solution is obtained and the local polynomial orders do not vary.

We also allow the user a choice to restrict the polynomial order of elements within the shock regions.
These are the zones that have been previously identified in the \adaptation{r} procedure.
The local polynomial order of the elements in these regions is then set to a user-defined value, which should be typically low (\(p<3\)).
The proposed \adaptation{rp} workflow is summarised in Algorithm~\ref{alg:rp-adapt}.

\begin{algorithm}
  \caption{The \textit{rp}-adaptive workflow.}\label{alg:rp-adapt}
  \begin{algorithmic}
    \State{generate an initial high-order mesh}
    \State{calculate the steady-state solution}
    \Repeat{}
    \State{extract shock areas based on sensor values}
    \State{apply \adaptation{r} in shock areas to linear mesh and re-project to high-order}
    \State{calculate the steady-state solution}
    \Until{shocks are well captured}
    \State{apply \adaptation{p} as described in Algorithm~\ref{alg:p-adapt}}
    \State{calculate the final solution}
  \end{algorithmic}
\end{algorithm}

\section{Numerical examples}\label{sec:num-ex}

In this section, we use two different test cases to demonstrate the \adaptation{rp} workflow:
a NACA 0012 aerofoil in transonic regime and a supersonic intake.
Different difficulties arise for each of these test cases as we will discuss below.
Most importantly, we take slightly different approaches when it comes to \adaptation{r} and \adaptation{p}.

The transonic flow in the first test develops two simple shocks: a strong shock on the upper surface of the aerofoil and a weak shock on the lower surface.
The disparity of strengths between the two shocks permits us to verify the shock capturing ability of the artificial viscosity. From the point of view of the adaptative procedures, the purpose of the test case is twofold. In the first instance, we chose a fine mesh for the flow simulation and run a single round of \adaptation{r} to
show that the variational framework can easily pull mesh nodes from the smooth regions towards the shock and that, in our implementation, the movement of the nodes is compatible with the computer-aided design (CAD) definition of the boundaries of the computational domain. Further, our choice of a fine mesh allows us to use this flow simulation as an accuracy benchmark for subsequent simulations. In the second instance, we use this test case to analyse the effectiveness of our order restriction strategy in \adaptation{p}. We study three different approaches to order restriction.
In the first approach, no restriction is applied and the local polynomial order of elements in shock areas is left free to increase.
In the second one, we keep the local polynomial order of these elements constant, i.e.\ the order of the initial simulation.
In the third and final one, we immediately decrease the local polynomial order of these elements to the minimum allowed in the run.

In the second test, the supersonic inflow ($M_\infty=3$) at the entrance of an intake generates a complex diamond-like pattern
of oblique shocks inside the intake due to reflections of the shocks at its internal walls and their interactions. In this example,
the number of nodes available for deforming the mesh is limited, which places additional stress on the variational optimisation
process.  We therefore decide to take a two-step approach to \adaptation{r} where each step is run with a milder shrinking
factor \(r_{\text{scale}}\) in order to retain good mesh quality.

\subsection{NACA 0012}\label{sec:naca}

We first demonstrate the new technology on a canonical aeronautical test case: a transonic (\( M=0.8 \))
inviscid flow past a NACA 0012 aerofoil at \( 1.25^\circ \) angle of attack.
This configuration produces two shocks~\cite{Vassberg2010}: a strong shock on the suction side and
a weak shock on the pressure side at approximately 60\% and 35\% of the chord respectively.
This provides a relatively easy test case to showcase the technology where the shocks are quasi-vertical.
The main difficulty lies in the relative weakness of the shock on the pressure side and in capturing it appropriately.

The domain used has external boundaries at a distance of \(40c\) from the aerofoil, where \(c\) denotes the chord length.
We discretise the domain uniformly along the chord with an element of size \( \sim 0.5c \) on the aerofoil boundary and create a smooth progression towards an element of size \( \sim 10c \) on the outer boundary.
The automatic sizing of elements in the field is determined through an octree system~\cite{Turner2016a}.
The mesh is curvilinear of order \( p=4 \) and it is optimised using the variational framework described in Sect.~\ref{sec:var-opti}.
Fig.~\ref{fig:naca-r-mesh} (left) shows what the mesh looks like in the near field.
The starting mesh is relatively coarse but it is run through the solver at uniform \( p=4 \) order.

It is important to note the importance of having sufficient resolution (either through \textit{h} or \textit{p}) in the initial mesh in order to distinguish shocks, i.e.\ actual discontinuities, from smooth high-gradient regions when looking at high discontinuity sensor values.
The necessity of having a sufficient number of DOF in the mesh arises from the nature of the numerical representation of shocks.
More specifically, shocks are captured over a fixed number of elements in the mesh that depends on the resolution capabilities of the particular CFD algorithm employed.
The requirement of enough resolution in the initial mesh is general and applies to any type of adaptivity strategy used.
Although one may sensibly consider including other forms of adaptivity (e.g.\ \emph{h}- or \emph{p}-adaptation) before improving the resolution of shocks using \adaptation{r} only, these considerations are beyond the scope of this paper, which looks at the feasibility of the novel proposed adaptation strategy.

We first run the solver on the initial mesh to obtain a base solution.
We impose slip wall BC on the surface of the profile and far-field BC at the external boundaries of the domain.
We use the Harten-Lax-van Leer-Contact (HLLC) Riemann solver~\cite{Toro2009}.
For the artificial viscosity, we tuned the solver parameters to \( s_\kappa = -1.2 \), \( \kappa = 0.7 \) and \( \mu_0 = 1.0 \).
Fig.~\ref{fig:naca-r-sensor}~(left) shows that large values of the sensor are obtained in both shock areas but also near the leading and trailing edges.
However, Fig.~\ref{fig:naca-r-visc}~(left) shows that artificial viscosity only triggers in the vicinity of the two shocks, proving adequate tuning of the artificial viscosity parameters.
The flow solution in Fig.~\ref{fig:naca-r-mach}~(left) displays very thick shocks as expected on this relatively coarse mesh.
We can also observe oscillations in the field past the strong shock caused by the under-resolution of the shock and the generation of entropy.

\subsubsection{\emph{r}-adaptation}\label{sec:naca-r}

\begin{figure}[htbp!]
  \begin{center}

    \begin{subfigure}[]{\textwidth}
      \begin{center}
        \includegraphics[width=0.49\textwidth]{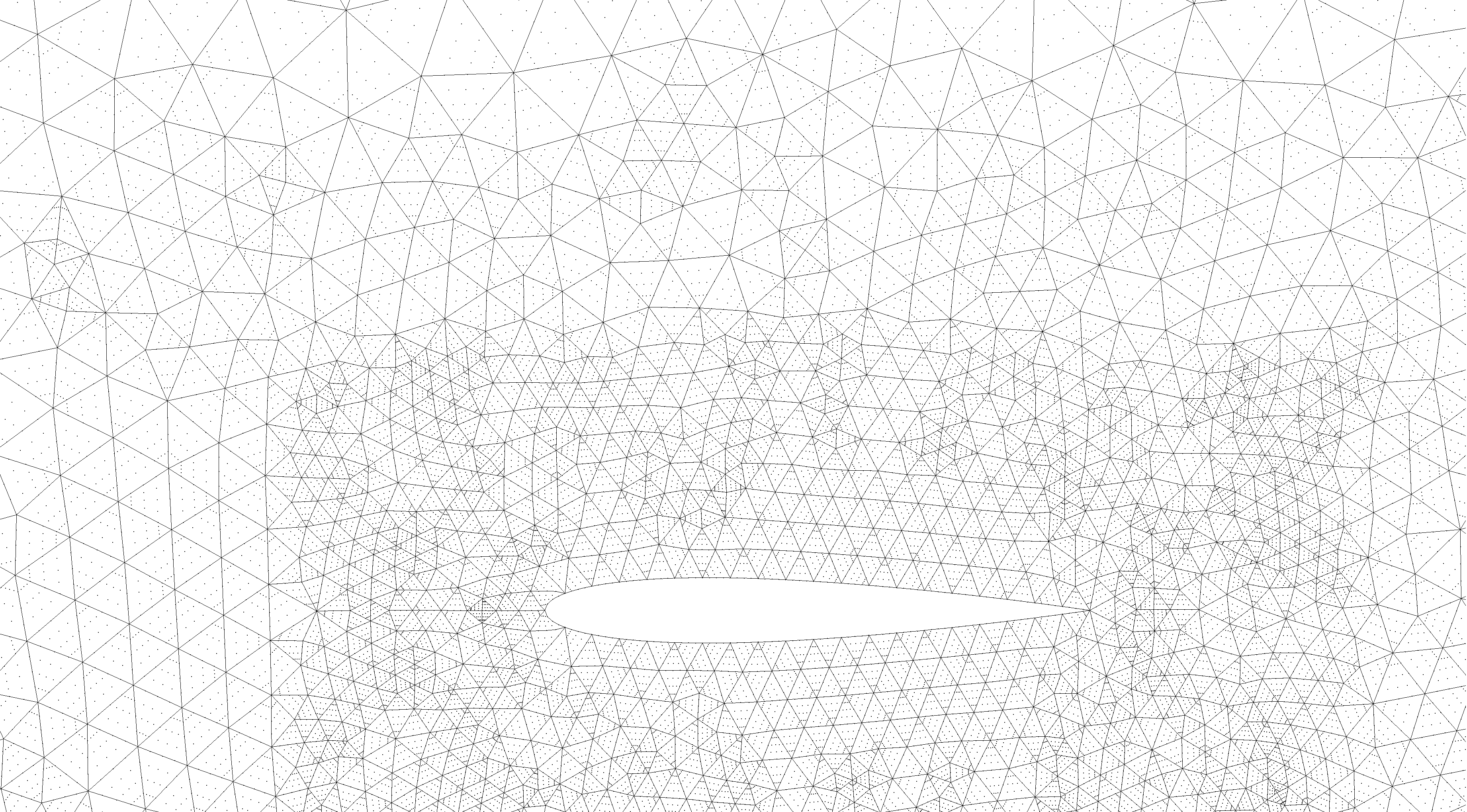}
        \includegraphics[width=0.49\textwidth]{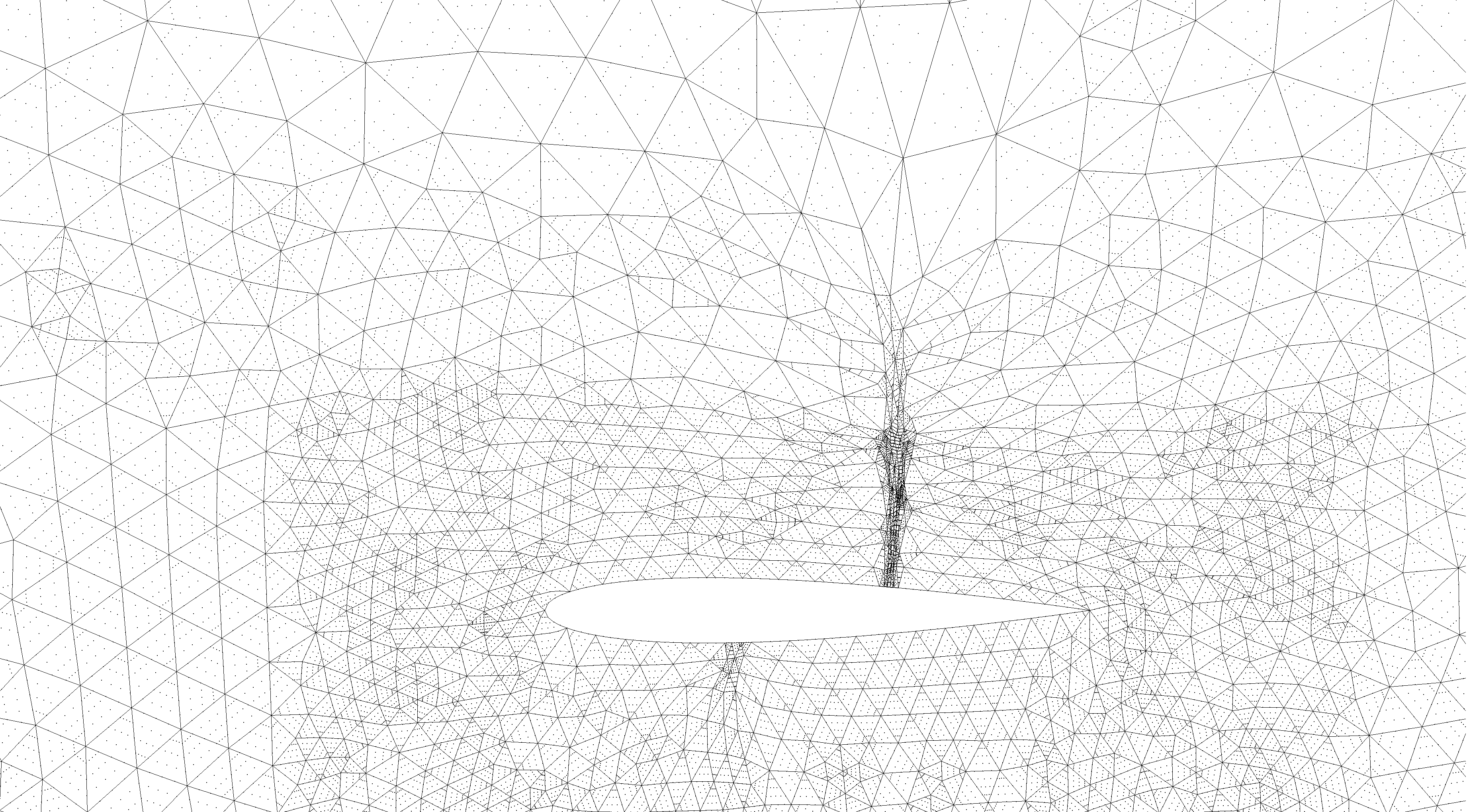}
        \caption{Mesh.}\label{fig:naca-r-mesh}
      \end{center}
    \end{subfigure}

    \begin{subfigure}[]{\textwidth}
      \begin{center}
        \includegraphics[width=0.49\textwidth]{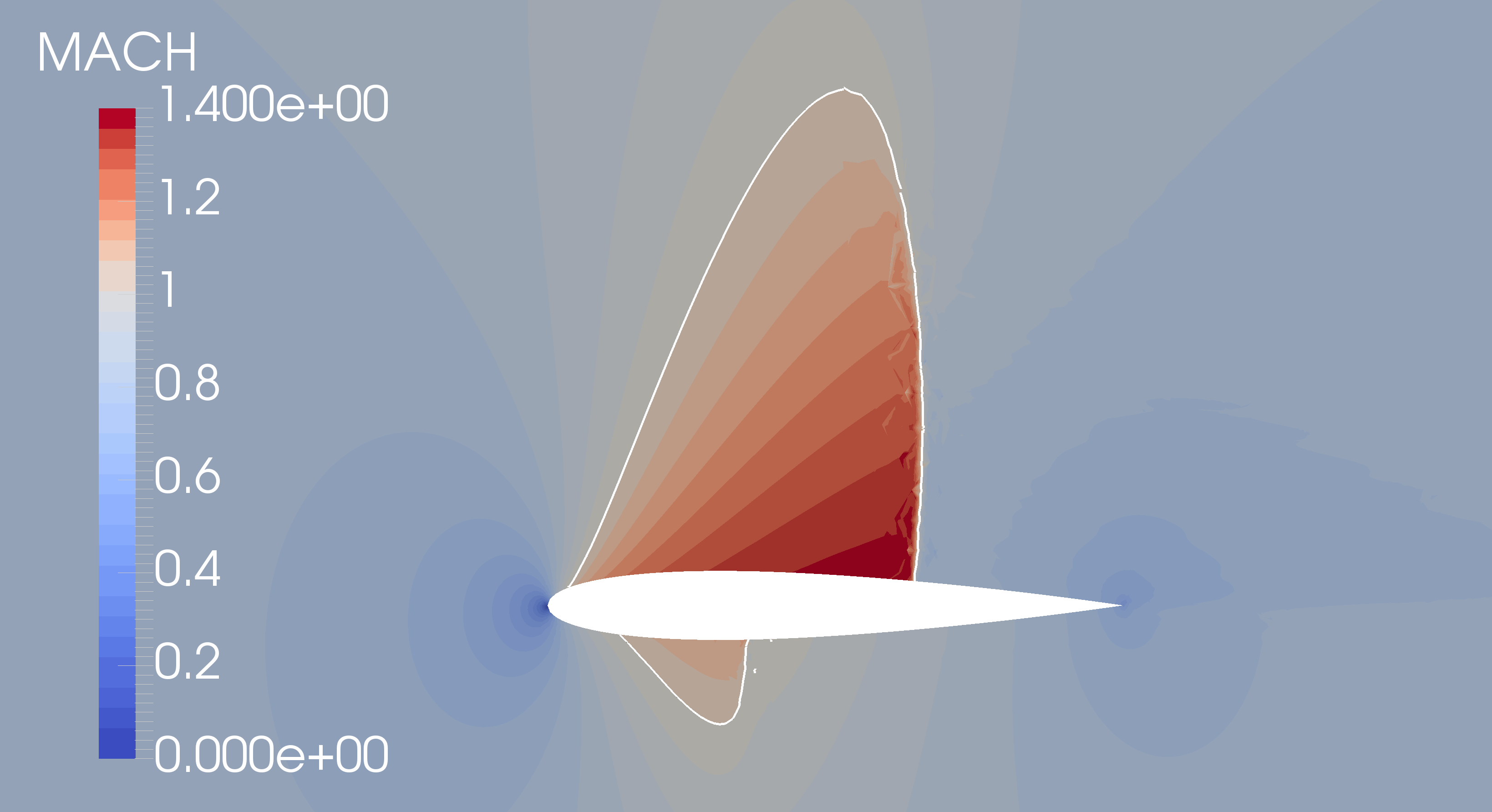}
        \includegraphics[width=0.49\textwidth]{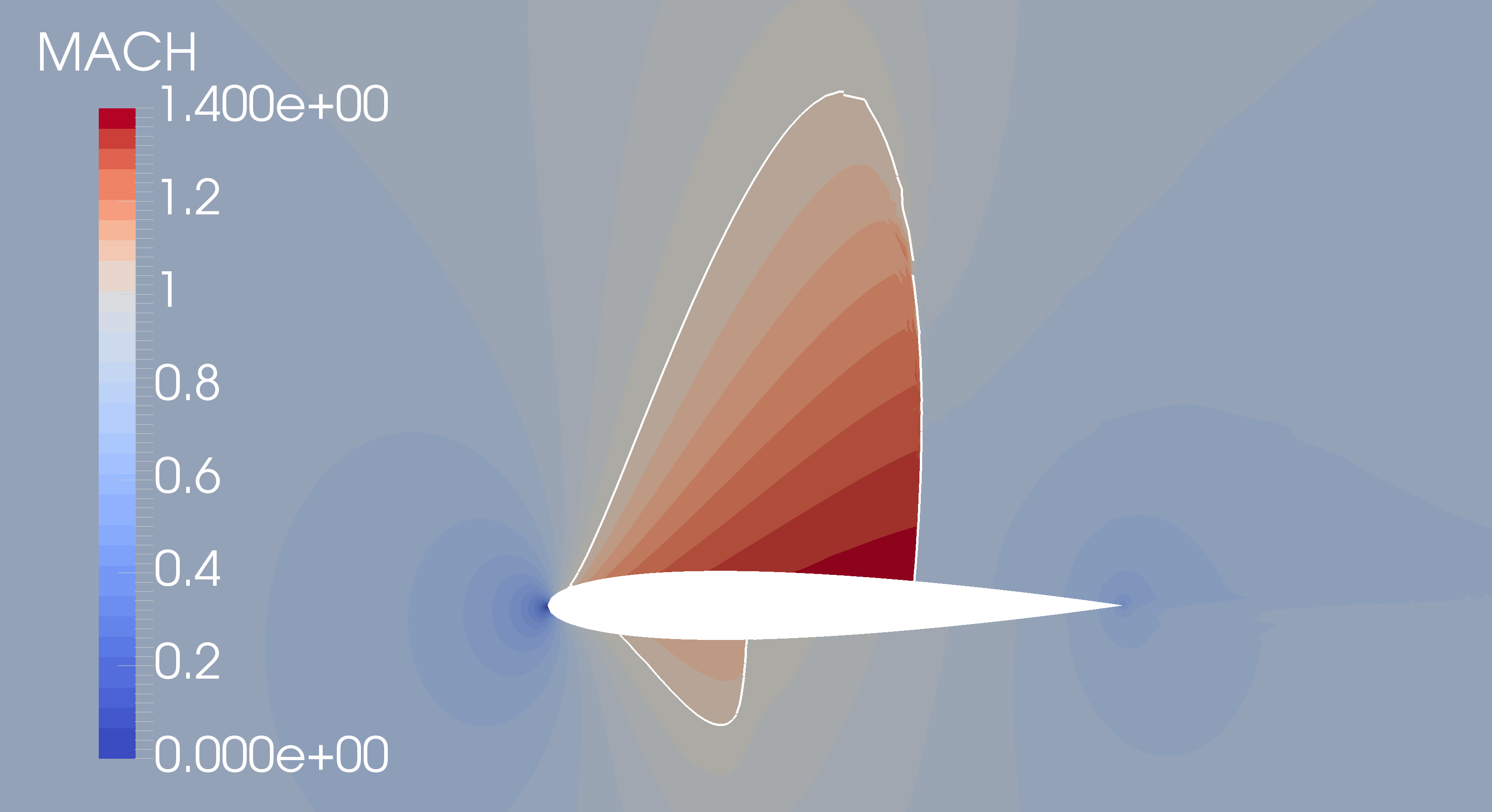}
        \caption{Mach number field.}\label{fig:naca-r-mach}
      \end{center}
    \end{subfigure}

    \begin{subfigure}[]{\textwidth}
      \begin{center}
        \includegraphics[width=0.49\textwidth]{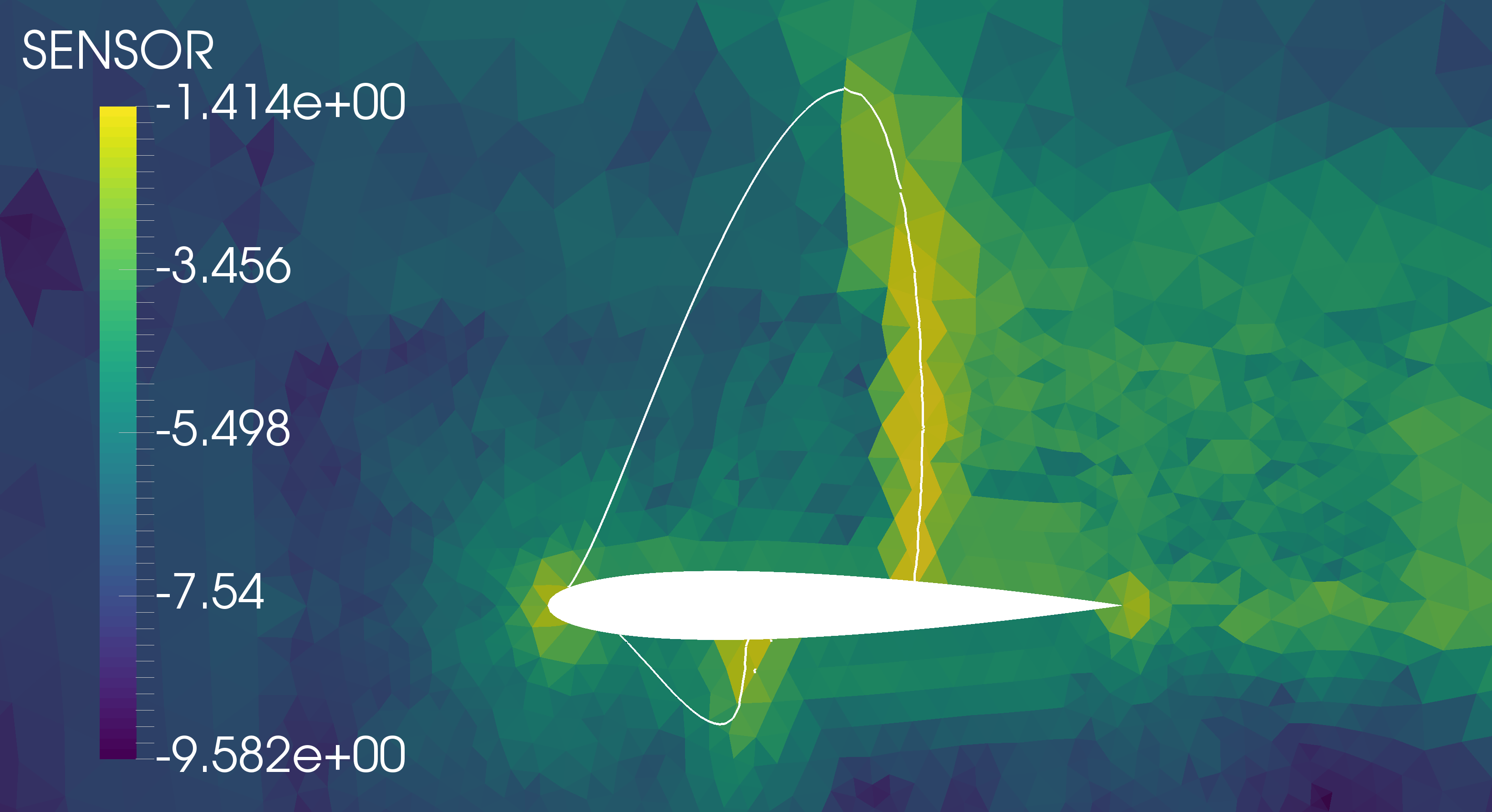}
        \includegraphics[width=0.49\textwidth]{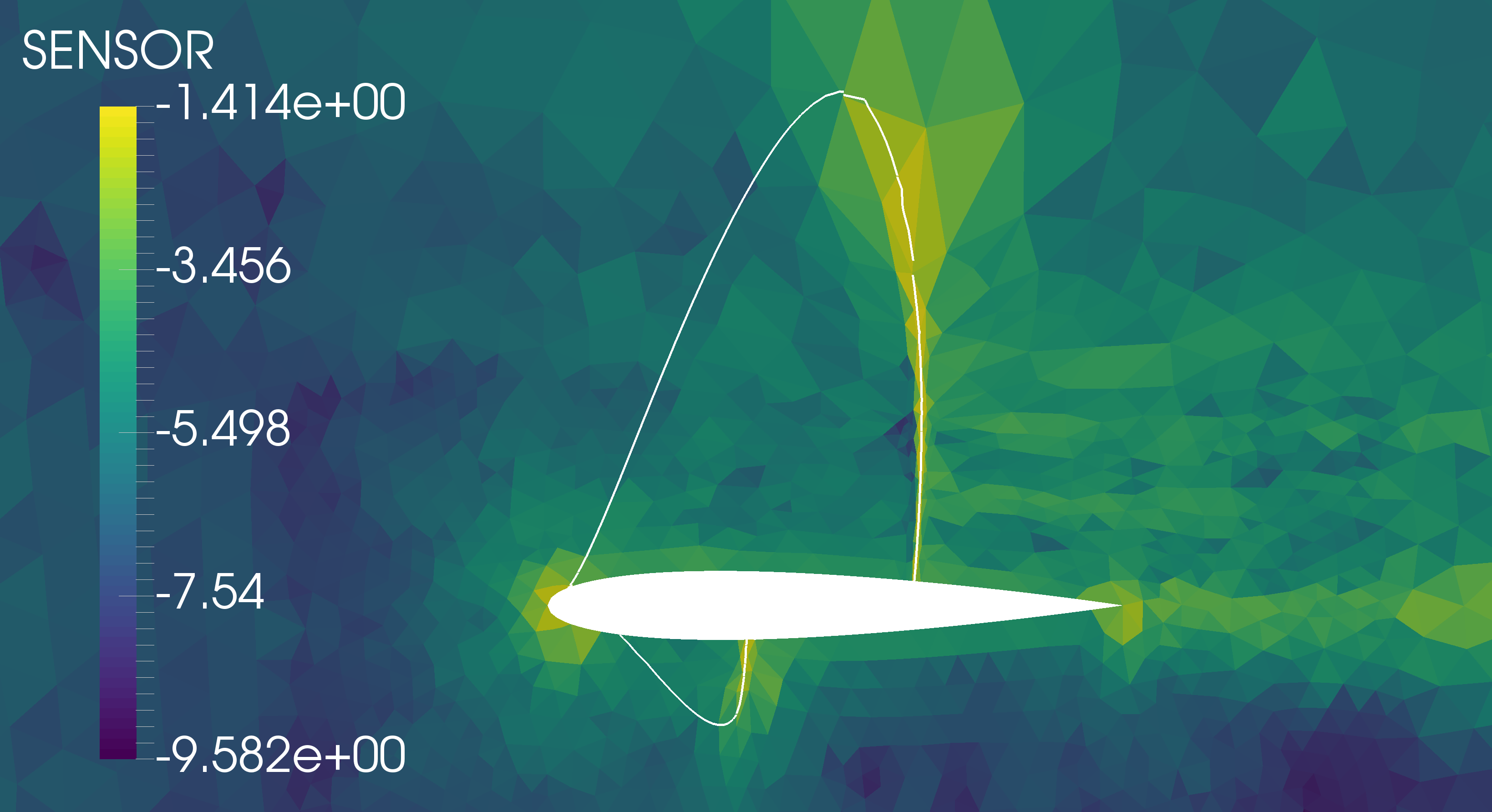}
        \caption{Sensor field.}\label{fig:naca-r-sensor}
      \end{center}
    \end{subfigure}

    \begin{subfigure}[]{\textwidth}
      \begin{center}
        \includegraphics[width=0.49\textwidth]{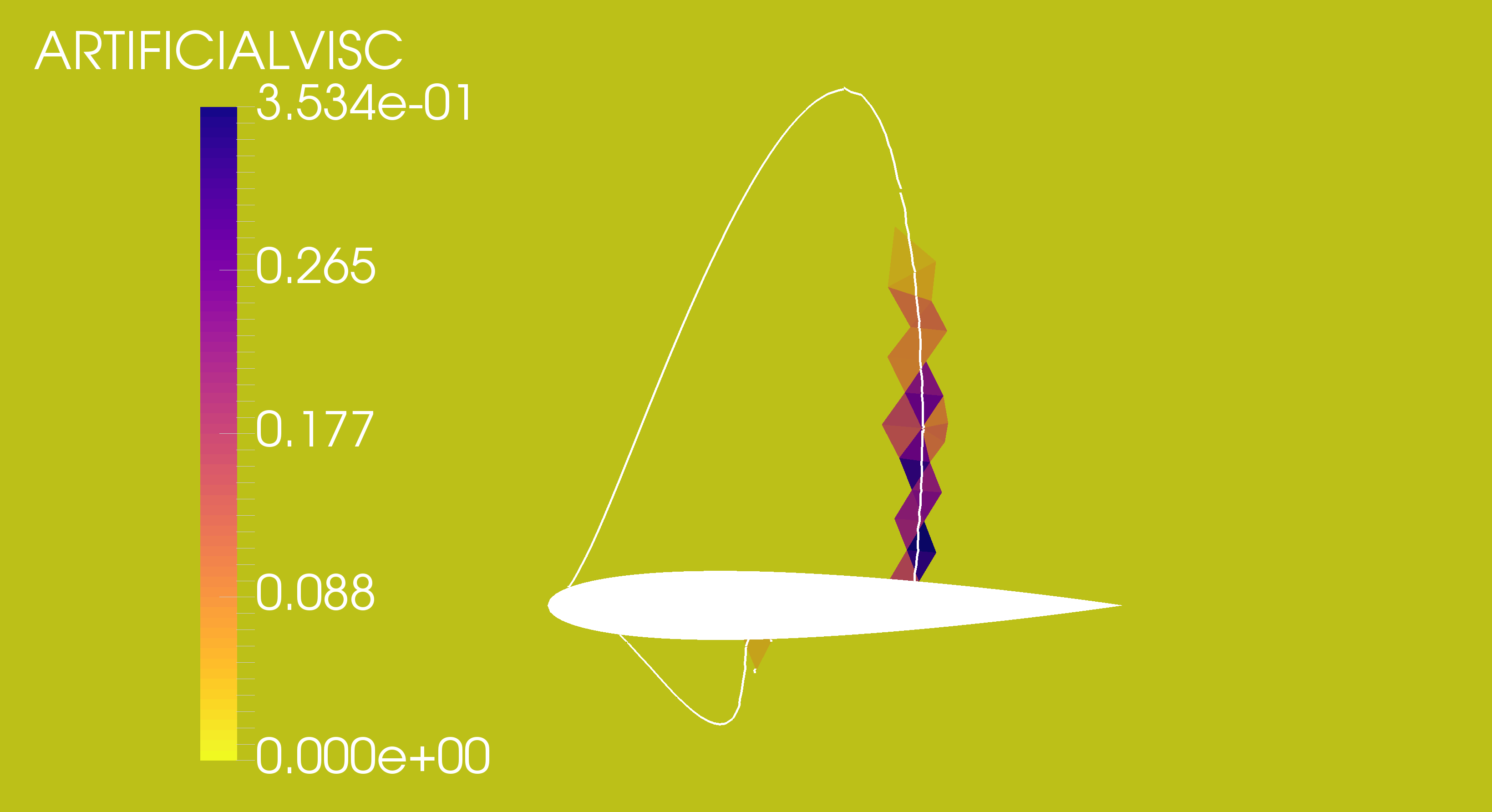}
        \includegraphics[width=0.49\textwidth]{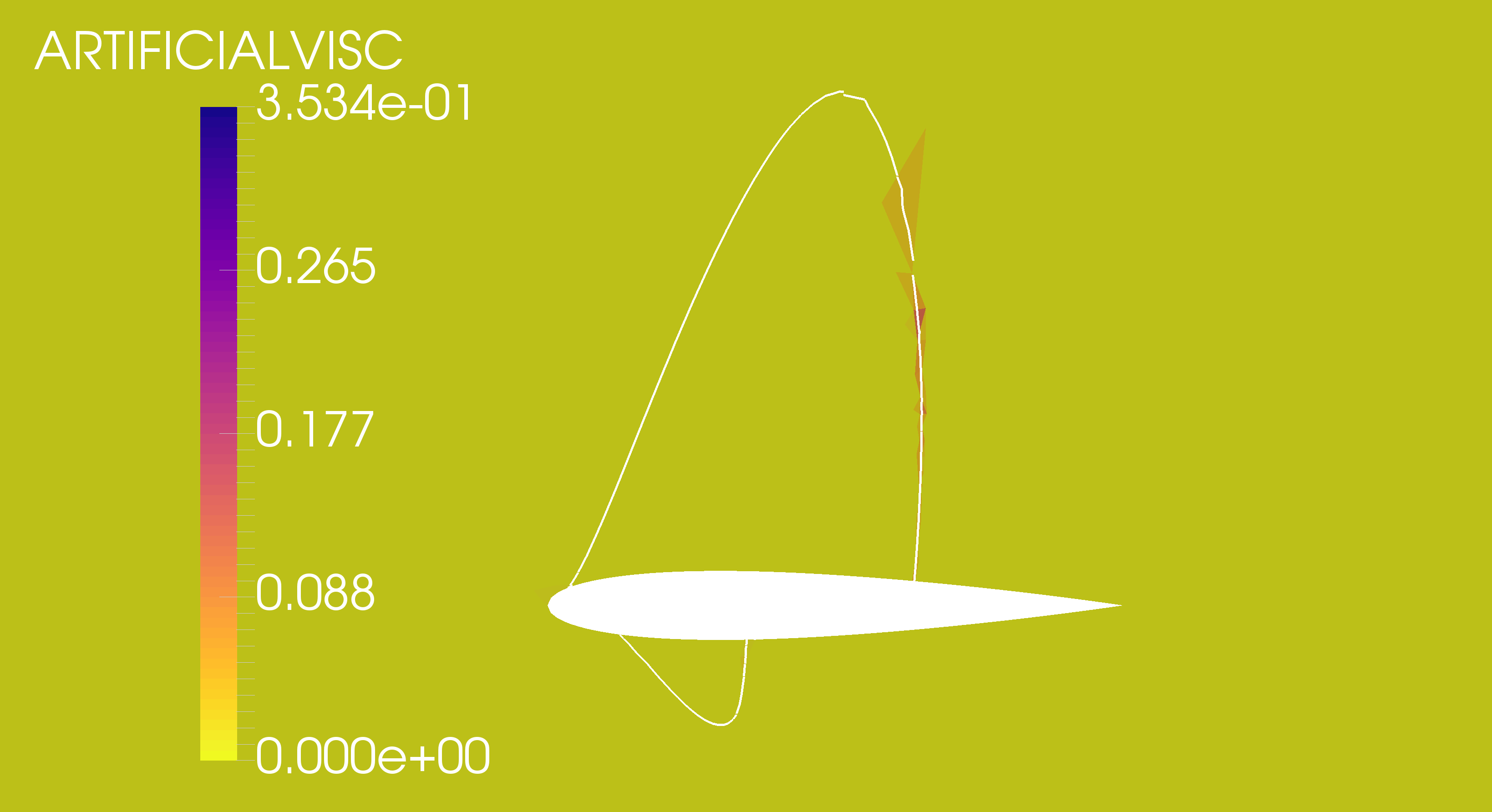}
        \caption{Artificial viscosity field.}\label{fig:naca-r-visc}
      \end{center}
    \end{subfigure}

    \caption{Comparison of the mesh and fields for the NACA 0012 profile before (left) and after (right) \adaptation{r}. A white line denotes the \( M=1 \) line in Figs.~\ref{fig:naca-r-mach}--\ref{fig:naca-r-visc}.}\label{fig:naca-r-adapt}
  \end{center}
\end{figure}

From the base solution, we follow the workflow explained in Sect.~\ref{sec:workflow}.
We first extract the shock regions: these correspond to the elements of non-null artificial viscosity in Fig.~\ref{fig:naca-r-visc}~(left).
To these regions, we assign a shrinking factor \(r_{\text{scale}}=0.1\) and run the \adaptation{r} procedure.
We obtain a new mesh which, for quality considerations, we re-optimise before simulation.
The new mesh shown in Fig.~\ref{fig:naca-r-mesh}~(right) shows refinement in the shock areas and consequently a slight coarsening outside of those zones.
Shrinking is also observed, to a smaller extent, in the vertical direction due to the isotropy of the \adaptation{r} approach.
However, the resulting mesh is clearly anisotropic and aligned to the presence of the shock.

We now interpolate the old solution onto the adapted mesh and run the simulation again.
In order to avoid any instability of the solver due to the interpolation of the under-resolved shock onto the new mesh, we first run the solver over a few hundred time steps with a decreased step size.
We then run the simulation, using the exact same artificial viscosity parameters, until steadiness is achieved.
The flow solution in Fig.~\ref{fig:naca-r-mach}~(right) shows better resolution of both shocks as seen by the sharpness of the shocks.
We also observe reduced oscillations in the wake of the strong shock.
Figs.~\ref{fig:naca-r-sensor}~\&~\ref{fig:naca-r-visc}~(right) finally show that discontinuity, as per the sensor, is now observed in a narrower area and that the artificial viscosity reaches lower values.

\begin{figure}[htbp!]
  \begin{center}
    \begin{tikzpicture}[spy using outlines={rectangle,magnification=2.7,connect spies}]
      \begin{axis}[
          xmin=0, xmax=1, ymin=0, ymax=1.6,
          width=0.99\textwidth,
          xlabel=\(x/c\),
          ylabel=\(M\),
          legend pos=north west
        ]

        \addplot[color=olive,no marks]
        table[x=x,y=Mach,col sep=comma] {plots/naca_suction_ref.csv};
        \addlegendentry{Reference}

        \addplot[color=red,no marks]
        table[x=x,y=Mach,col sep=comma] {plots/naca_suction_init.csv};
        \addlegendentry{Initial mesh}

        \addplot[color=blue,no marks]
        table[x=x,y=Mach,col sep=comma] {plots/naca_suction_uni.csv};
        \addlegendentry{\emph{r}-adapted mesh}

        \addplot[color=olive,no marks]
        table[x=x,y=Mach,col sep=comma] {plots/naca_pressure_ref.csv};
        \addplot[color=red,no marks]
        table[x=x,y=Mach,col sep=comma] {plots/naca_pressure_init.csv};
        \addplot[color=blue,no marks]
        table[x=x,y=Mach,col sep=comma] {plots/naca_pressure_uni.csv};

        \coordinate (spypoint1) at (axis cs:0.345,1.04);
        \coordinate (spyviewer1) at (axis cs:0.345,0.44);
        \spy[width=0.4\textwidth,height=0.4\textwidth] on (spypoint1) in node [fill=white] at (spyviewer1);

        \coordinate (spypoint2) at (axis cs:0.615,1.375);
        \coordinate (spyviewer2) at (axis cs:0.85,1.25);
        \spy[width=0.15\textwidth,height=0.3\textwidth] on (spypoint2) in node [fill=white] at (spyviewer2);

        \coordinate (spypoint3) at (axis cs:0.66,0.755);
        \coordinate (spyviewer3) at (axis cs:0.85,0.35);
        \spy[width=0.15\textwidth,height=0.3\textwidth] on (spypoint3) in node [fill=white] at (spyviewer3);

      \end{axis}
    \end{tikzpicture}
    \caption{Plot of the Mach number \(M\) before and after \adaptation{r}.}\label{fig:naca-plot}
  \end{center}
\end{figure}
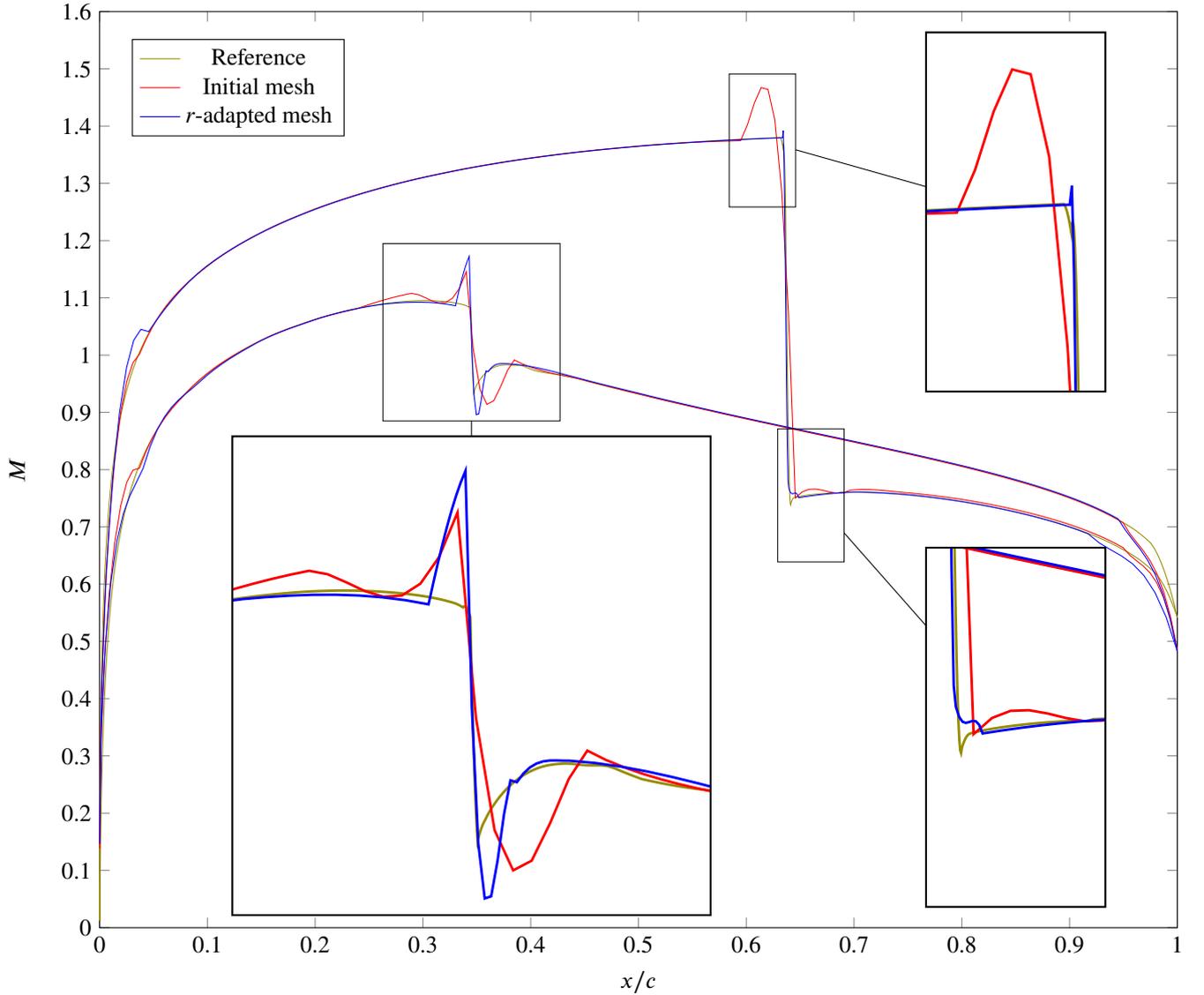

The improvement of the resolution of the shock can be better seen by visualisation of quantities along the wing surface.
Fig~\ref{fig:naca-plot} shows the Mach number on the surface of the profile.
Because elements are so large in the initial mesh, the solution shows strong oscillatory behaviour, known as the Gibbs phenomenon described in Sect.~\ref{sec:intro}, inherent to the low-dissipation discretisation used in this work.
The artificial viscosity~\ref{eq:artdis} introduced to stabilise the solution reduces this phenomenon but does not totally eliminate it.
At this point, we do not seek to further increase the artificial viscosity, which would cause more dissipation.
Instead, our goal is to simply stabilise the solution.

After \adaptation{r}, elements in the region of the shock are much smaller and, although oscillations are still present, they have both a smaller amplitude and a narrower range, thanks to increased resolution in the region.
This confirms the qualitative observation of the increased sharpness of the shock seen in Fig.~\ref{fig:naca-r-mach}.
Because there is less mesh movement at the weak shock, the reduction in the Gibbs phenomenon is also smaller.

\begin{figure}[htbp!]
  \begin{center}
    \includegraphics[trim={10cm 0 10cm 0},clip,width=0.49\textwidth]{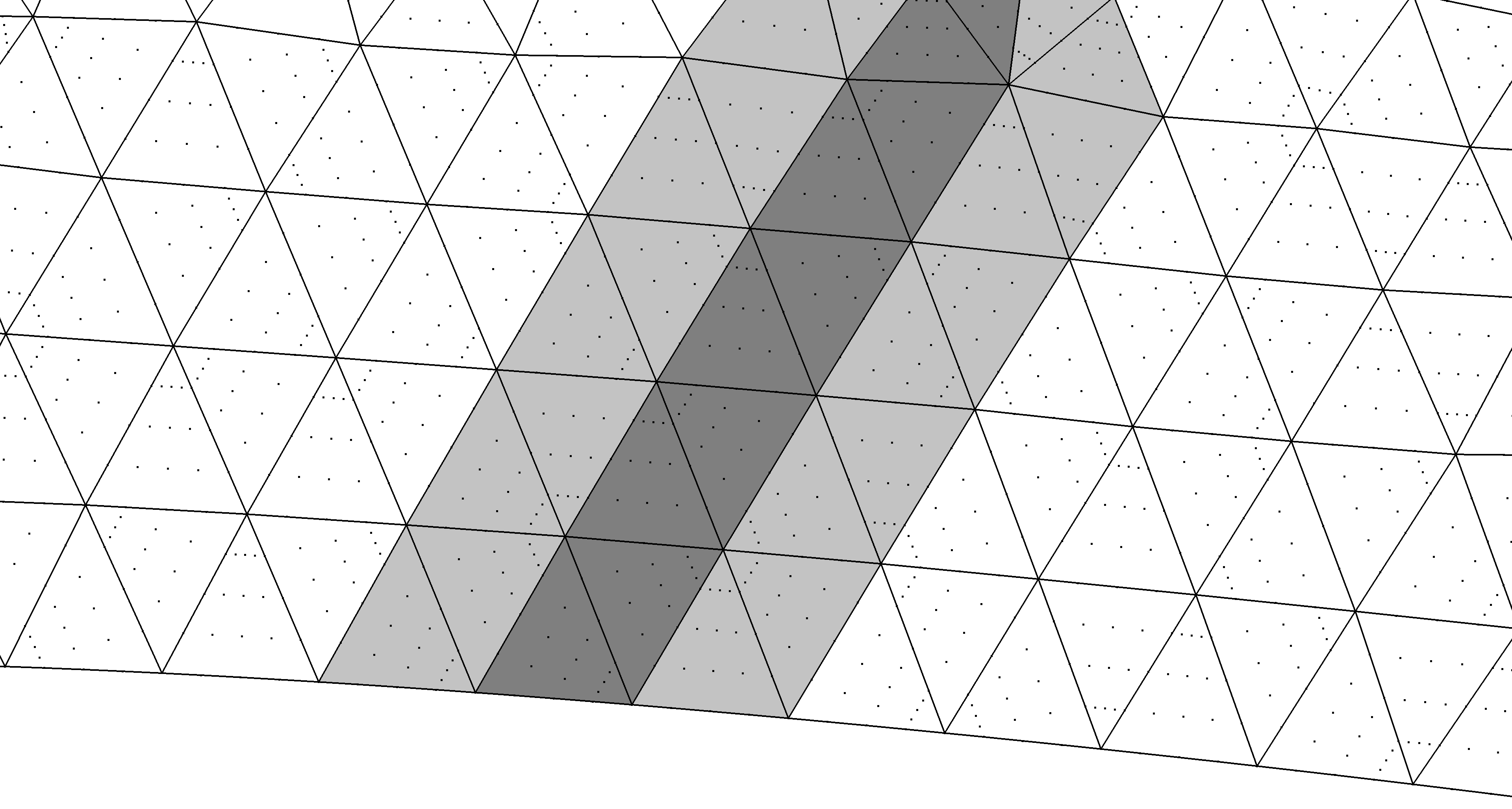}
    \includegraphics[trim={10cm 0 10cm 0},clip,width=0.49\textwidth]{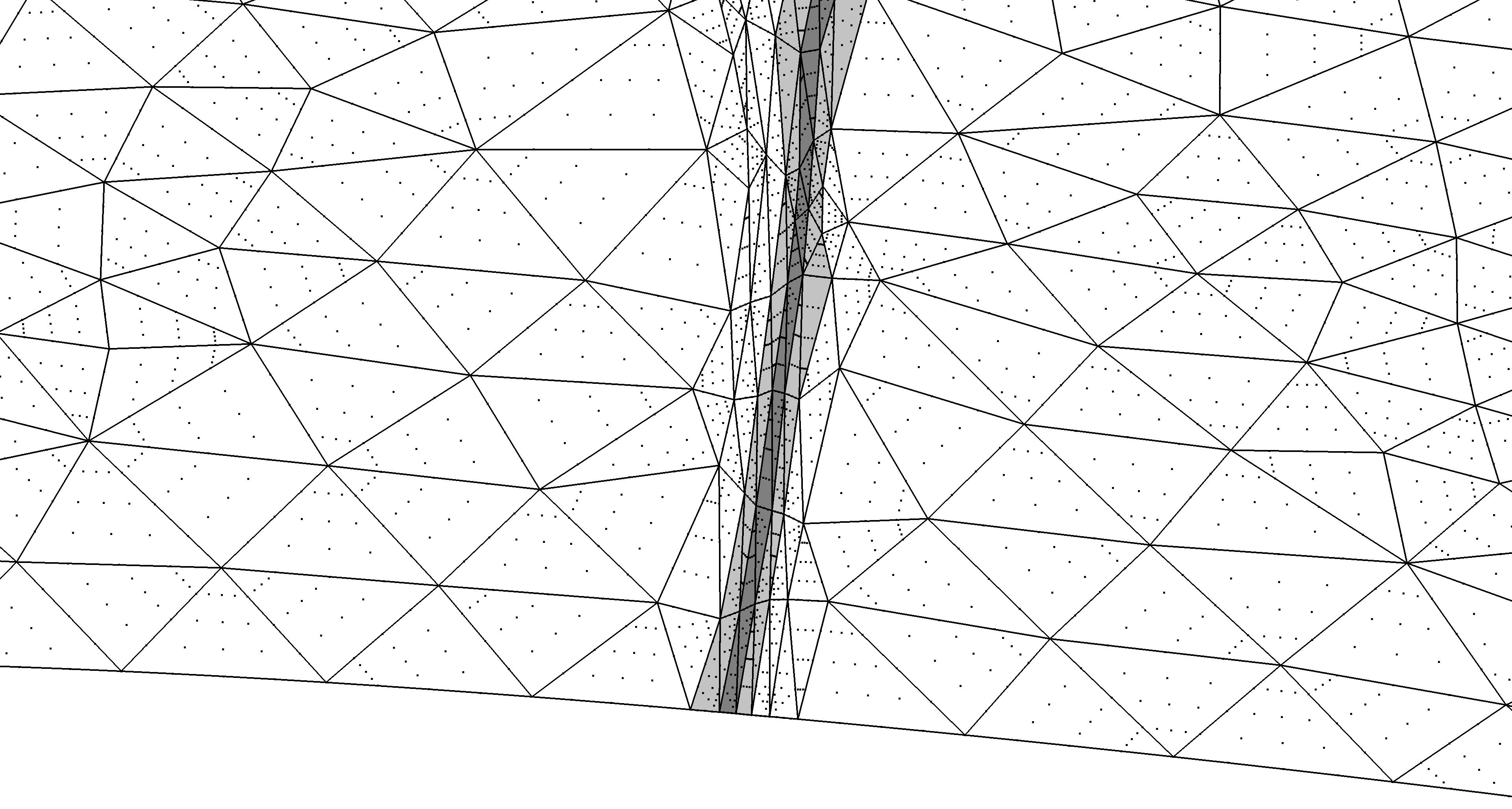}
    \caption{Magnified view of boundary elements with \textit{CAD sliding} enabled nodes for the NACA 0012 test case: original (left) and adapted (right) meshes. Three rows of elements have been coloured to highlight the large movement of nodes.}\label{fig:cad-sliding}
  \end{center}
\end{figure}

We also draw the reader's attention to the importance of the presence of CAD data for the refinement process, as in order to retain accurate boundary representation, it is important that the \adaptation{r} code has access to a CAD system.
In this instance, \nekmesh{}~\cite{Turner2016a} and the variational optimiser have been implemented to use the OpenCASCADE~\cite{OpenCascadeSAS2018} framework as its CAD engine.
This allows \nekmesh{} to query the geometry and ensure that all nodes remain on the boundaries at all time.
The capability to slide nodes along the surface (\textit{CAD sliding}) is shown in Fig.~\ref{fig:cad-sliding} where nodes remain on the aerofoil surface throughout the \adaptation{r} process.
Fig.~\ref{fig:cad-sliding} also shows that the optimiser is able to move nodes across large distances, as seen through the row of coloured elements before (left) and after (right) \adaptation{r}.

\subsubsection{\emph{p}-adaptation}\label{sec:naca-p}

\begin{figure}[htbp!]
  \begin{center}

    \begin{subfigure}[]{\textwidth}
      \begin{center}
        \includegraphics[width=0.49\textwidth]{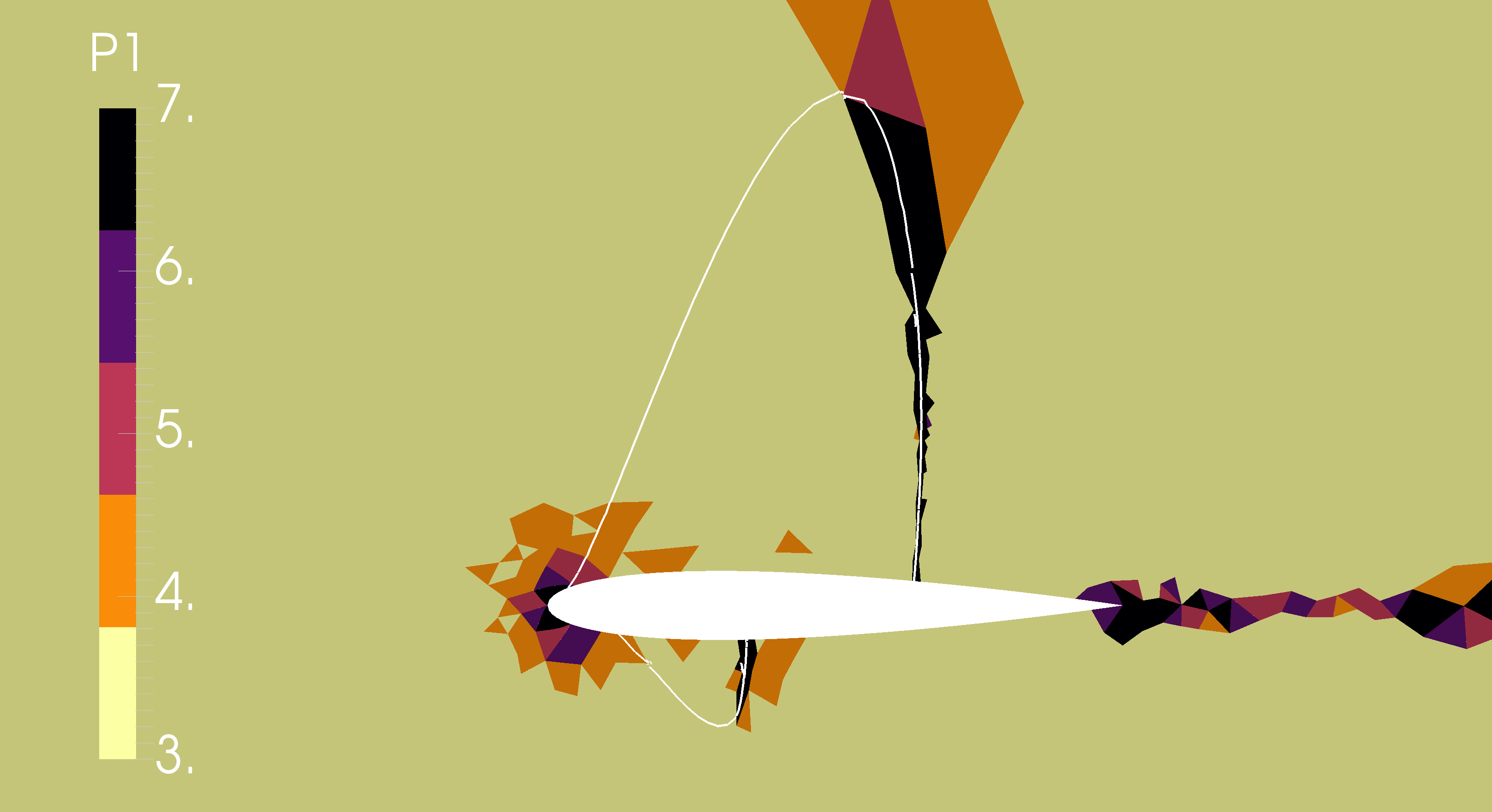}
        \includegraphics[width=0.49\textwidth]{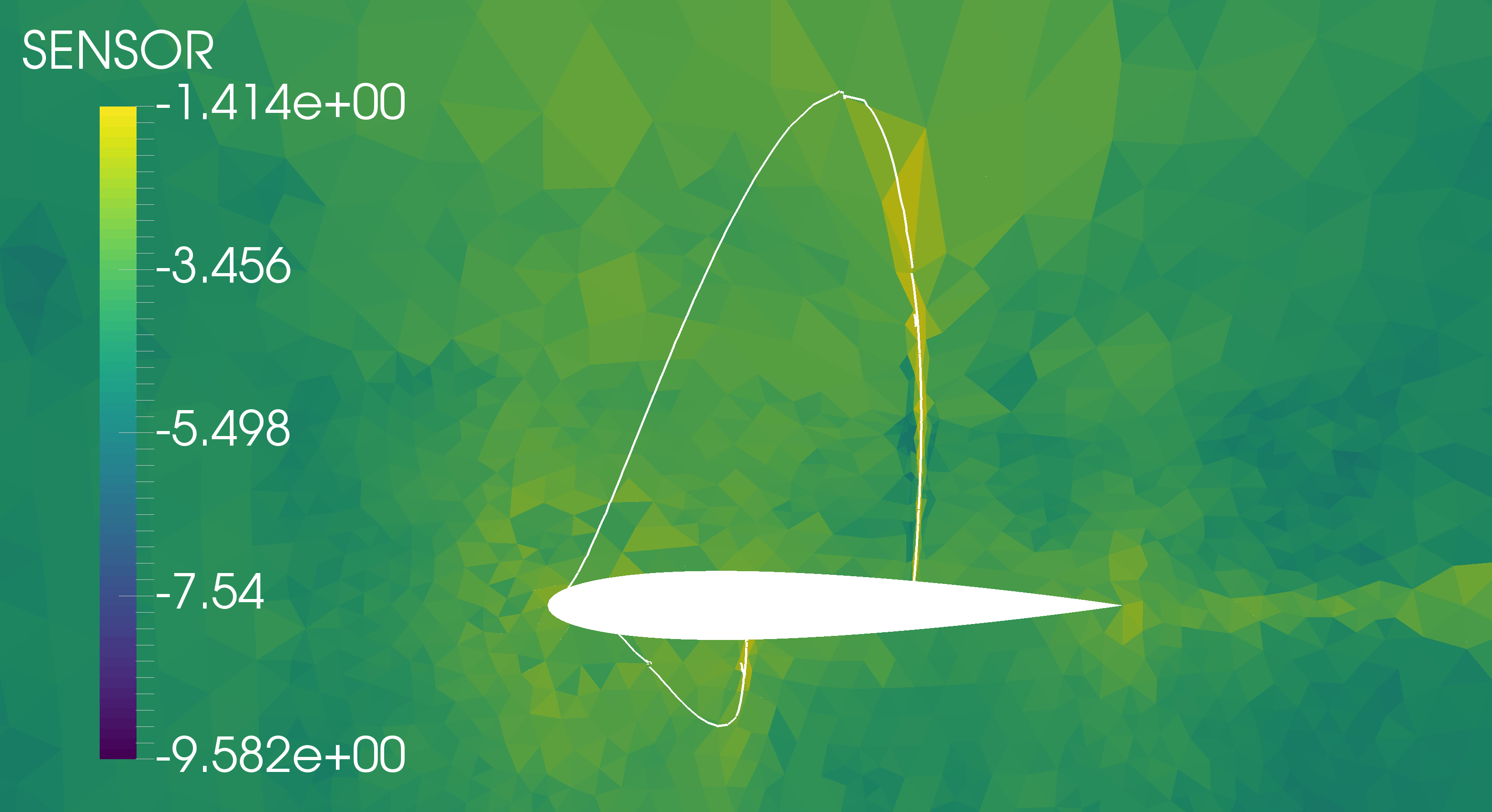}
        \caption{Full \adaptation{p}.}\label{fig:naca-p-modes-2}
      \end{center}
    \end{subfigure}

    \begin{subfigure}[]{\textwidth}
      \begin{center}
        \includegraphics[width=0.49\textwidth]{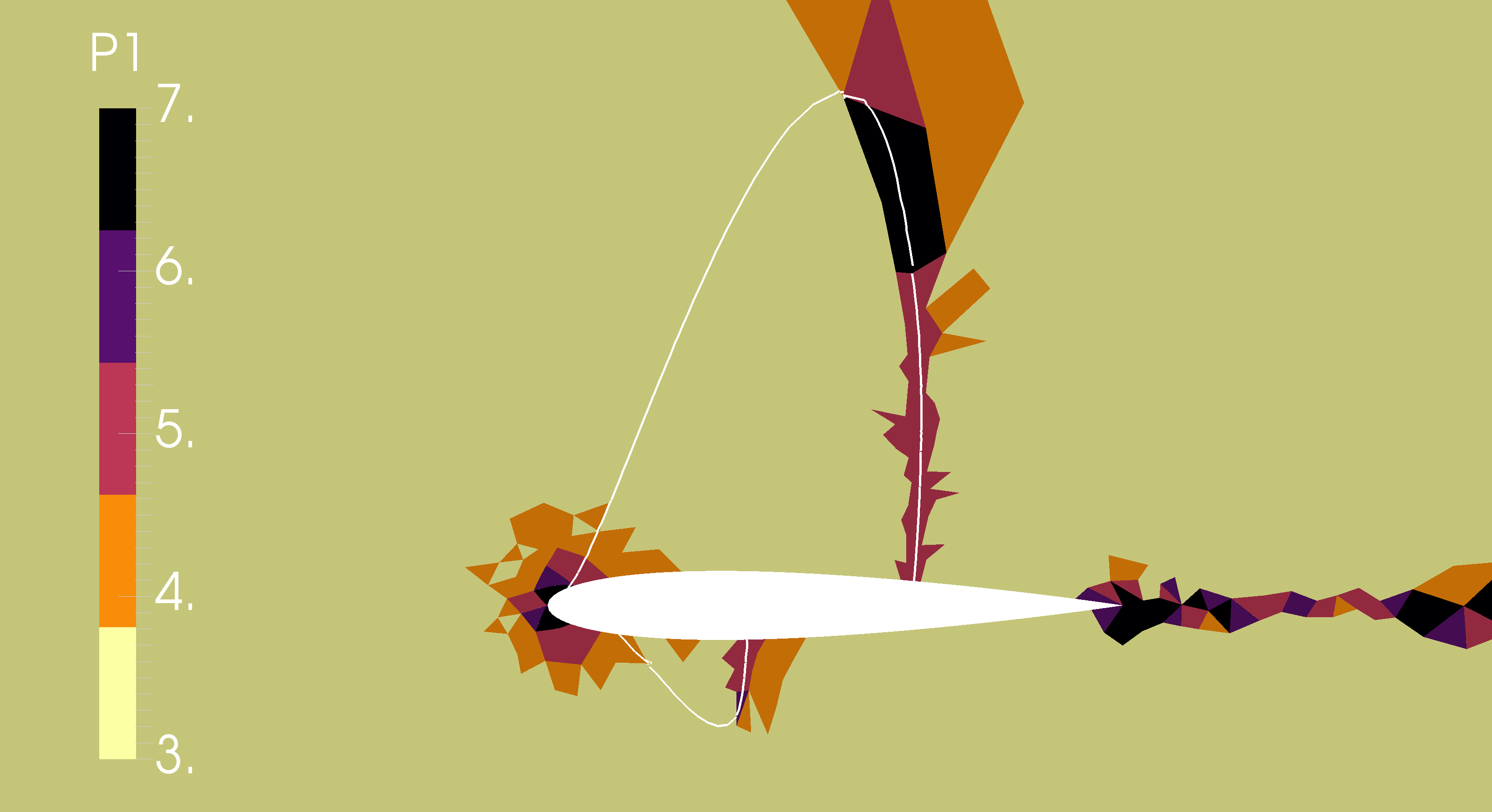}
        \includegraphics[width=0.49\textwidth]{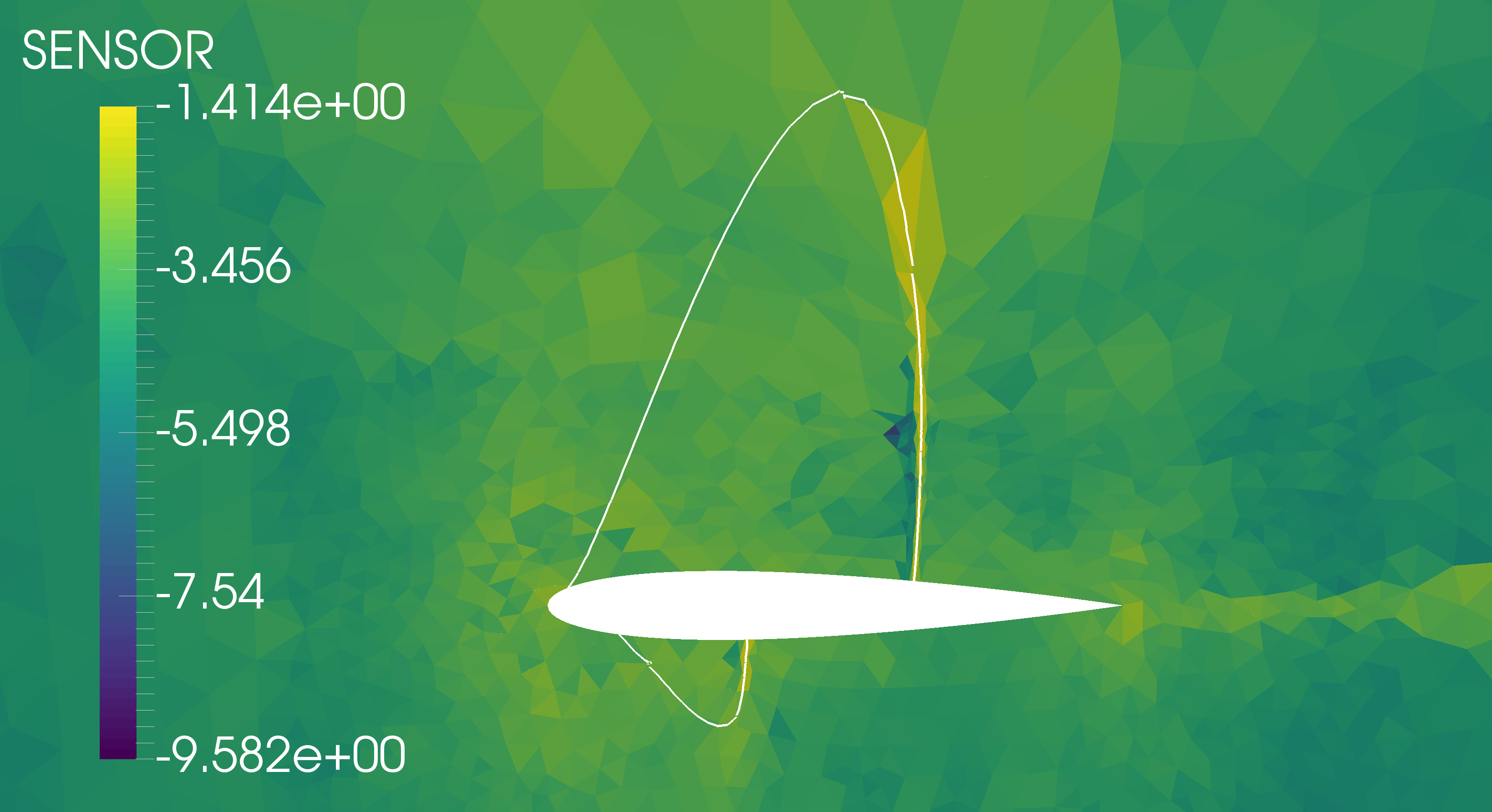}
        \caption{\adaptation{p} with original order restriction.}\label{fig:naca-p-modes-3}
      \end{center}
    \end{subfigure}

    \begin{subfigure}[]{\textwidth}
      \begin{center}
        \includegraphics[width=0.49\textwidth]{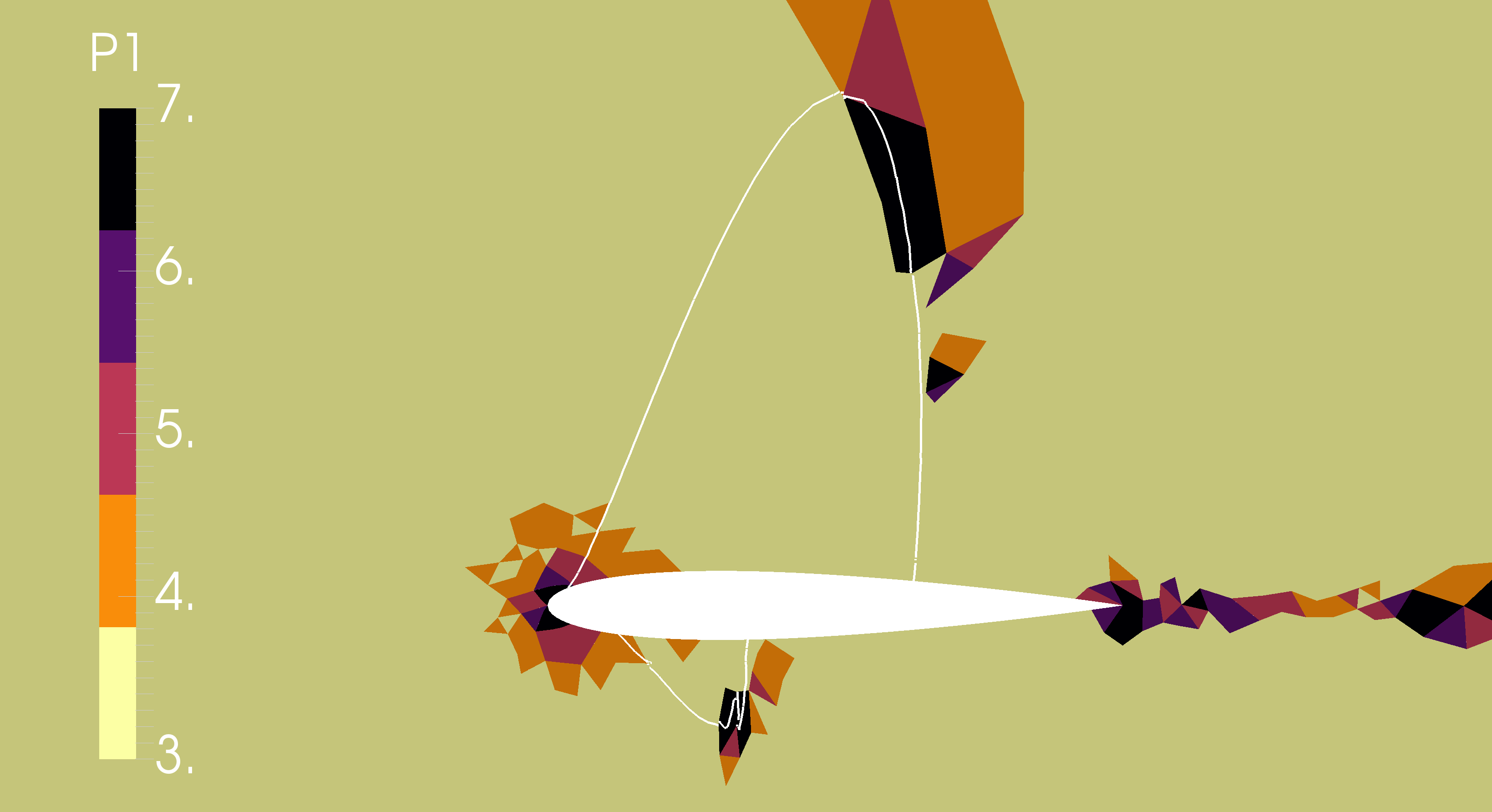}
        \includegraphics[width=0.49\textwidth]{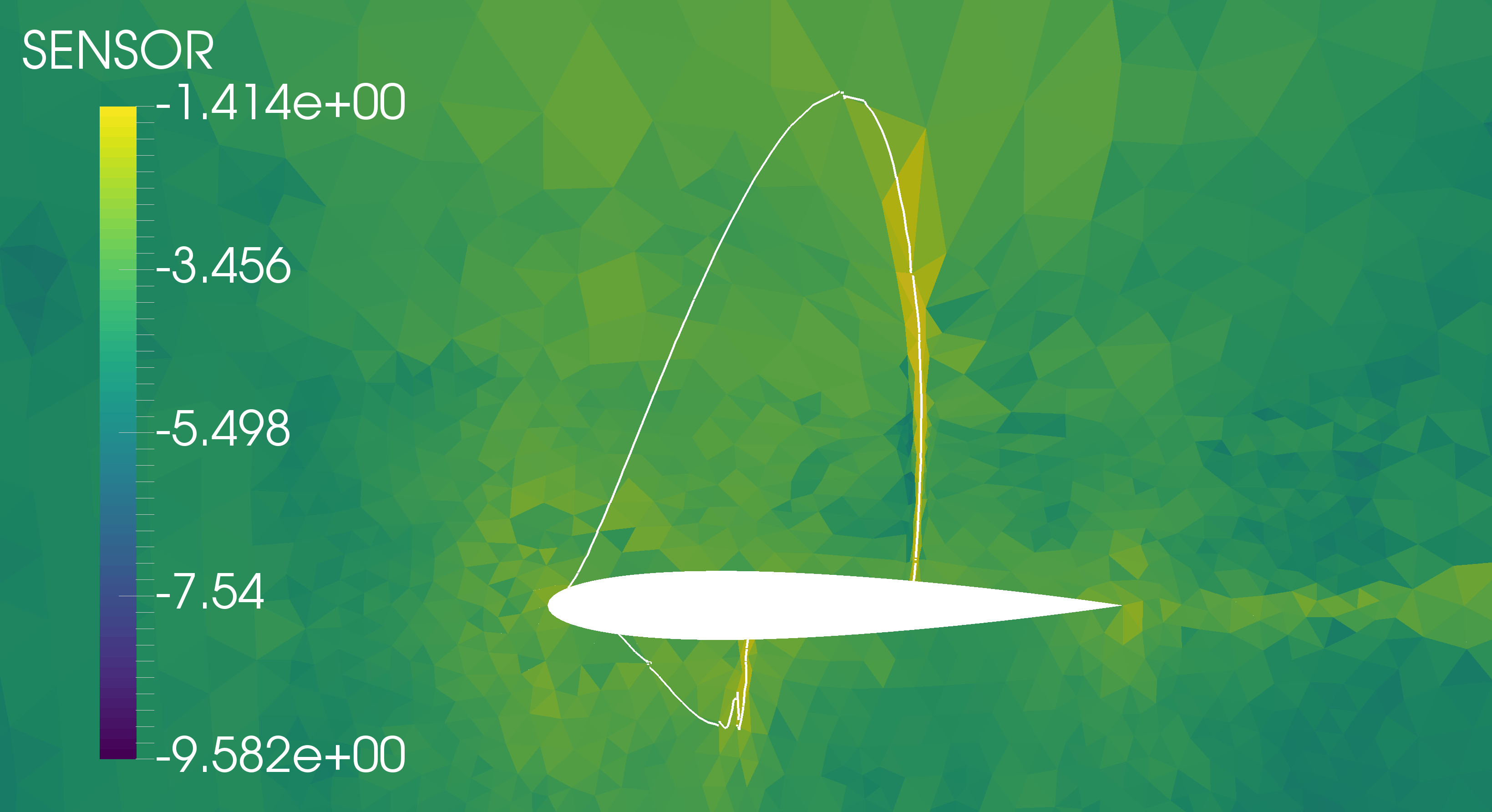}
        \caption{\adaptation{p} with lowest order restriction.}\label{fig:naca-p-modes-4}
      \end{center}
    \end{subfigure}

    \caption{Comparison of the local number of modes (\( = p+1 \); left) and post-adaptation sensor fields (right) for the three test scenarios of the NACA 0012 test case. A white line denotes the \( M=1 \) line in all figures.}\label{fig:naca-p-modes}
  \end{center}
\end{figure}

After better resolving the shocks, we can now apply local \adaptation{p} for the smooth field.
For this test case, we compare three scenarios.
In the first scenario, we apply local \adaptation{p} without any restriction (see Fig.~\ref{fig:naca-p-modes-2}) while, in the other two scenarios, we restrict the local polynomial order inside the shock areas.
In the second scenario, we preserve the local polynomial order of the uniform \( p=4 \) order simulation of Sect.~\ref{sec:naca-r} (see Fig.~\ref{fig:naca-p-modes-3}).
In the third and last scenario, we decrease the local polynomial order inside the shock areas to the lowest user-allowed order (see Fig.~\ref{fig:naca-p-modes-4}).

For these tests, we start from the field obtained at \( p=4 \) in Sect.~\ref{sec:naca-r} and use values of \( p_{\min} = 2 \) and \( p_{\max} = 6 \).
The sensor is based on the density field \( \rho \) and the solver default values of lower and upper sensor tolerances are respectively \( 10^{-8} \) and \( 10^{-6} \).
Fig.~\ref{fig:naca-p-modes} shows the results.
The figures on the left show a final map of the local number of modes (\( = p+1 \)) after a steady solution is reached and, by extension, when the local polynomial order remains constant throughout \adaptation{p} steps.
The number of DOF for each simulation is shown in Table~\ref{tab:naca-dof}.
All scenarios produce fewer DOF than the simulation at uniform \textit{p}, thanks to local \textit{p}-coarsening in low-error regions.
By design, the unrestricted \adaptation{p} scenario increases the local polynomial order of elements in the shock
region to the maximum user-allowed value.
This leads to a higher global number of DOF than the other two scenarios.
Then follows the second scenario while the last scenario has the smallest number of DOF.\@
Each of these DOF counts also translates into similar increases or decreases in computing times.

We compare these solutions to a reference solution computed on a very fine mesh.
To evaluate the performance of each mesh and \textit{p}-adaptation scenario, we look at the Mach number distribution on the surface of the aerofoil.
We use the \(L^2\)-norm of the error, defined as
\[
  \lVert e \rVert^2_{L^2(S)} = \int_S {\left( M-M_{ref} \right)}^2 \, dS
\]
where \(M\) is the Mach number of the test solution, \(M_{ref}\) is the Mach number of the reference solution and \(S\) is the chord.
Results are reported in Table~\ref{tab:naca-dof}.
Note that central processing unit (CPU) times per time step are reported as run on a 16-core machine, once convergence is reached.
We first note that \textit{r}-adaptation alone provides an important boost in terms of accuracy.
Scenarios \#1 and \#2 both suffer a loss of accuracy due to the coarsening of the solution in large parts of the domain.
This slight increase in the error, however, allows us to cut the number of DOF in half.
Scenario \#3, on the other hand, performs very poorly, with the error going even higher than on the original mesh.
Decreasing the polynomial order inside the shock --- a rather small region --- allows us to save a few more DOF but at too great a cost.

\begin{table}
  \caption{Number of DOF, error and CPU time per time step at convergence for the NACA 0012 profile.}\label{tab:naca-dof}
  \centering
  \begin{tabular}{c c c c c c}
    \toprule
    \multirow{2}{*}{Simulation} & \multirow{2}{*}{Number of DOF} & \multicolumn{3}{c}{\(\lVert e \rVert^2_{L^2(S)} \left({10^{-4}}\right)\)} & \multirow{2}{*}{CPU time (ms)}               \\
                                &                                & Pressure                                                                  & Suction                        & Total &     \\
    \midrule
    Initial mesh                & 65\,550                        & 0.547                                                                     & 5.28                           & 5.83  & 65  \\
    \textit{r}-adaptated mesh   & 65\,550                        & 0.751                                                                     & 1.36                           & 2.11  & 118 \\
    Scenario \#1                & 29\,201                        & 0.875                                                                     & 2.54                           & 3.41  & 48  \\
    Scenario \#2                & 29\,117                        & 0.919                                                                     & 1.82                           & 2.74  & 56  \\
    Scenario \#3                & 27\,736                        & 1.045                                                                     & 6.61                           & 7.65  & 45  \\
    \bottomrule
  \end{tabular}
\end{table}

\begin{figure}[htbp!]
  \begin{center}

    \begin{subfigure}[]{\textwidth}
      \begin{center}
        \includegraphics[width=0.49\textwidth]{figs/naca_mach_1_r_p_uni}
        \includegraphics[width=0.49\textwidth]{figs/naca_visc_1_r_p_uni}
        \caption{Uniform \(p\).}\label{fig:naca-p-mach-1}
      \end{center}
    \end{subfigure}

    \begin{subfigure}[]{\textwidth}
      \begin{center}
        \includegraphics[width=0.49\textwidth]{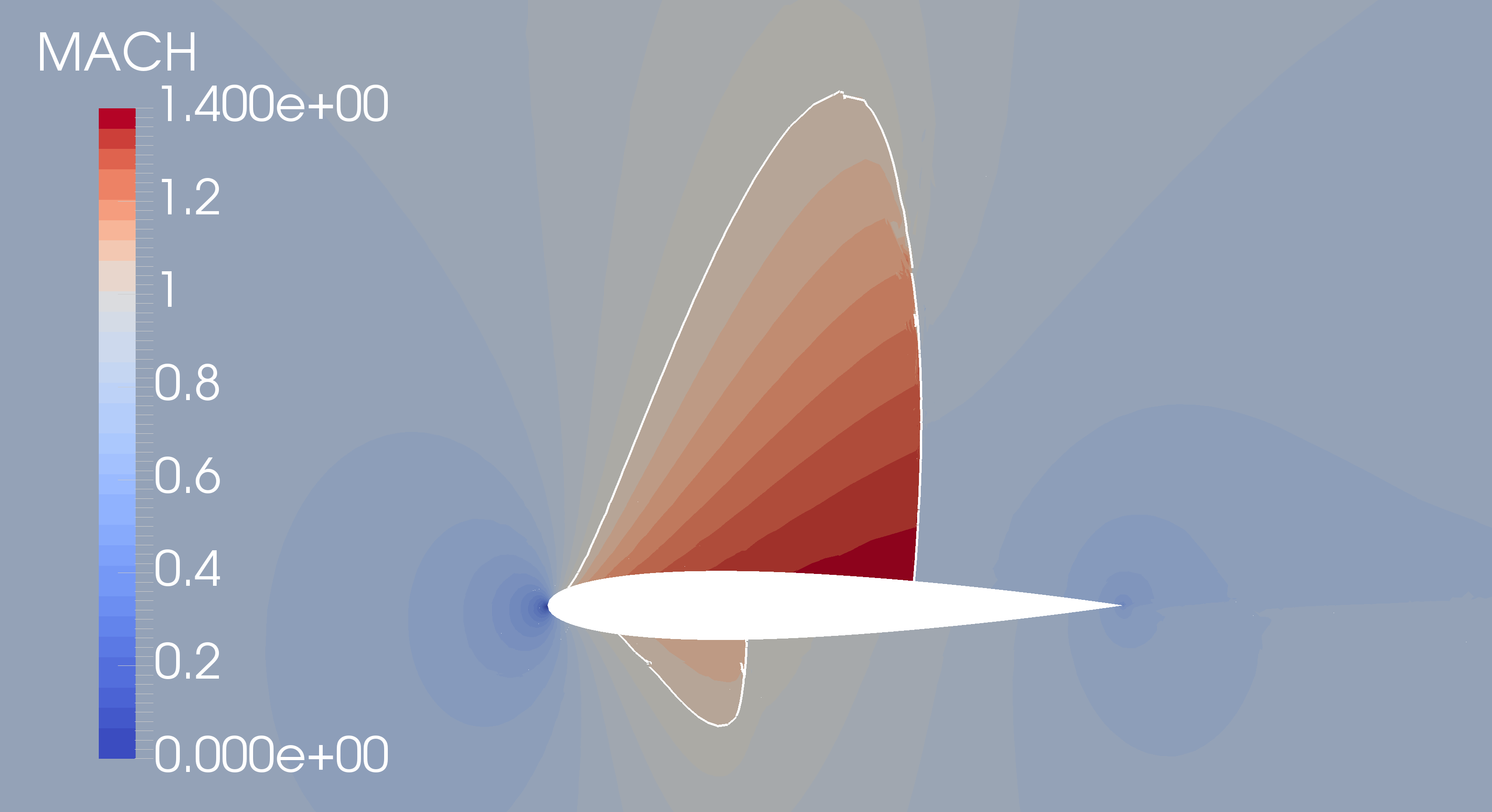}
        \includegraphics[width=0.49\textwidth]{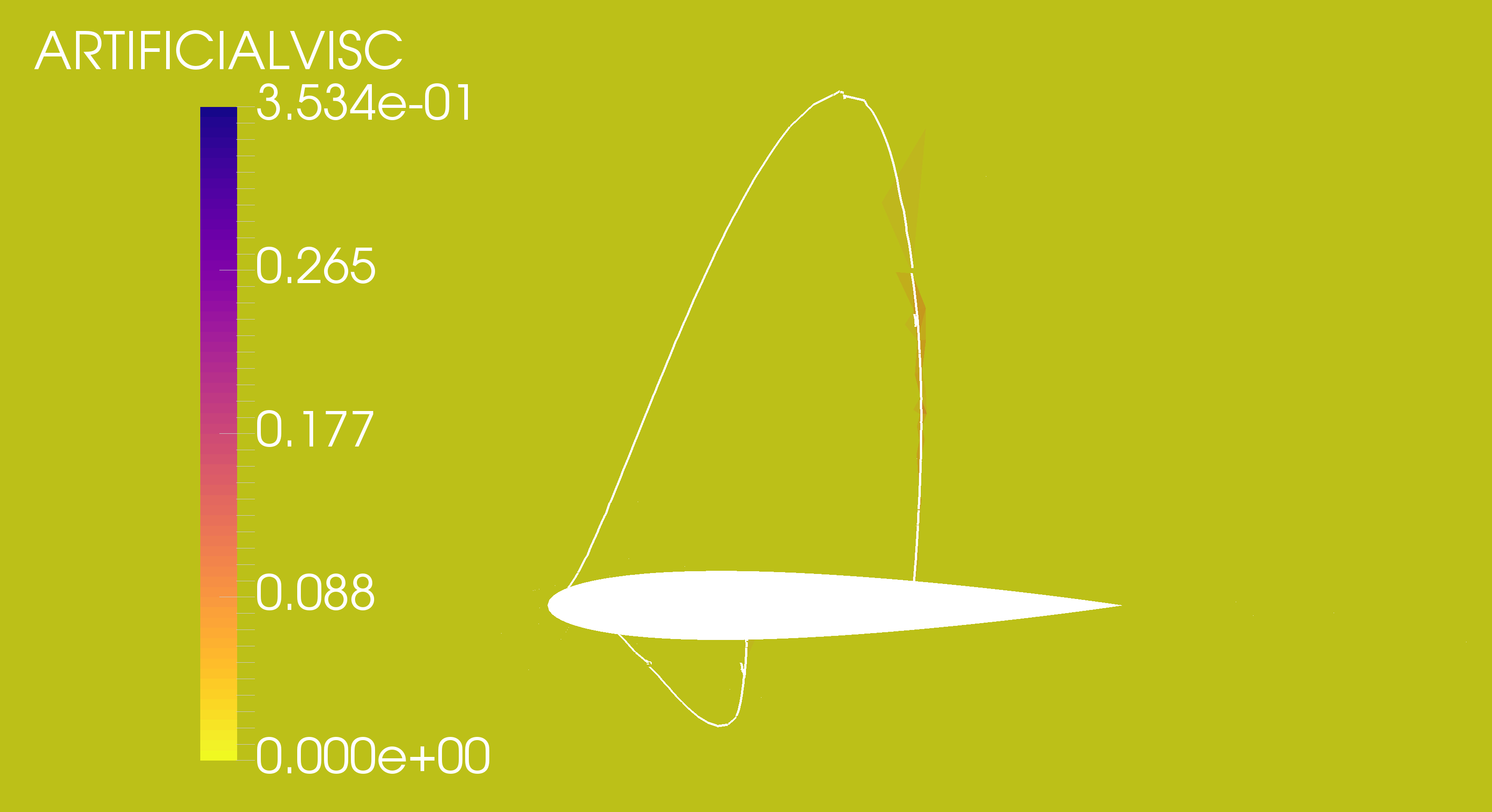}
        \caption{Full \adaptation{p}.}\label{fig:naca-p-mach-2}
      \end{center}
    \end{subfigure}

    \begin{subfigure}[]{\textwidth}
      \begin{center}
        \includegraphics[width=0.49\textwidth]{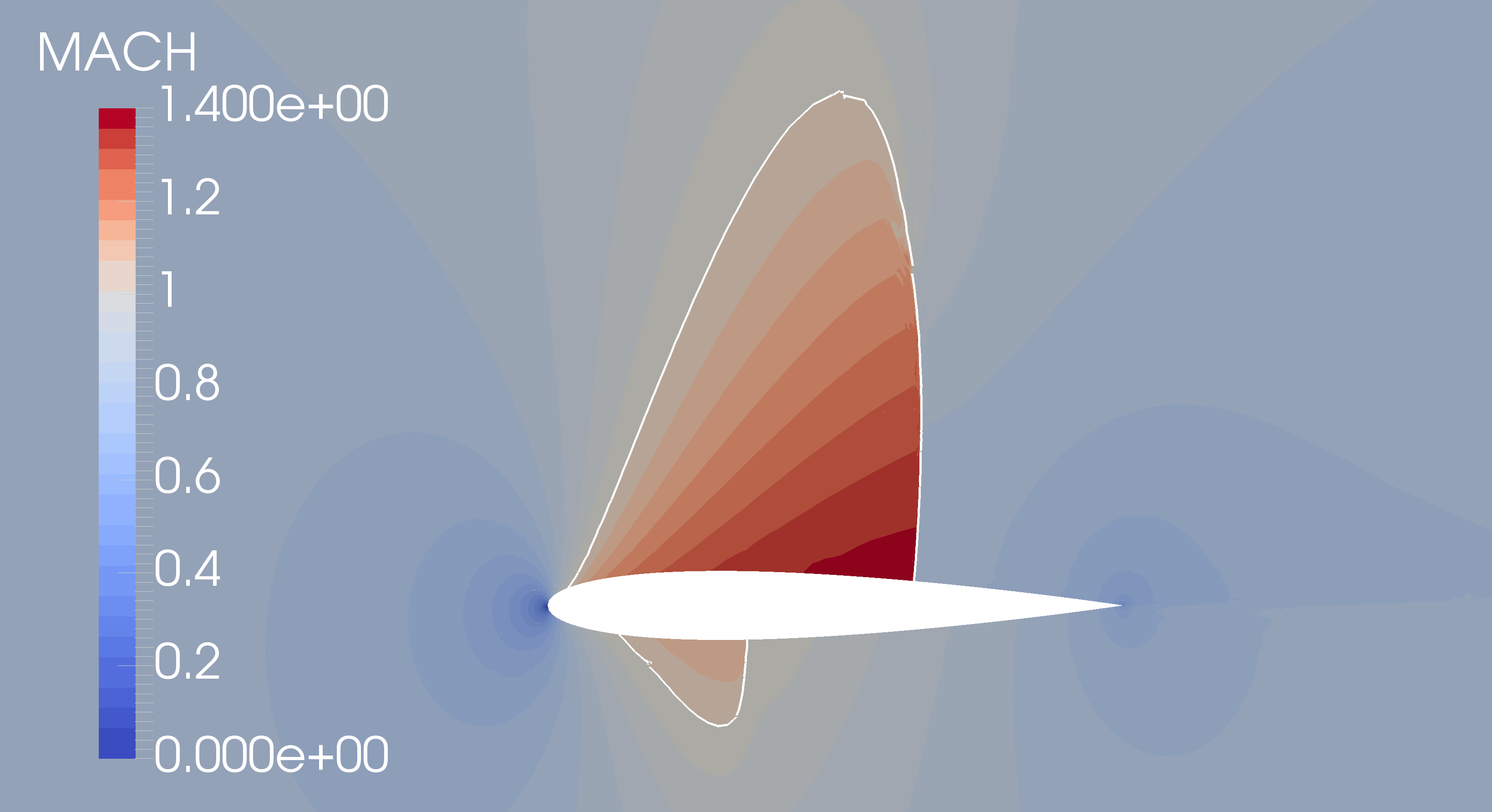}
        \includegraphics[width=0.49\textwidth]{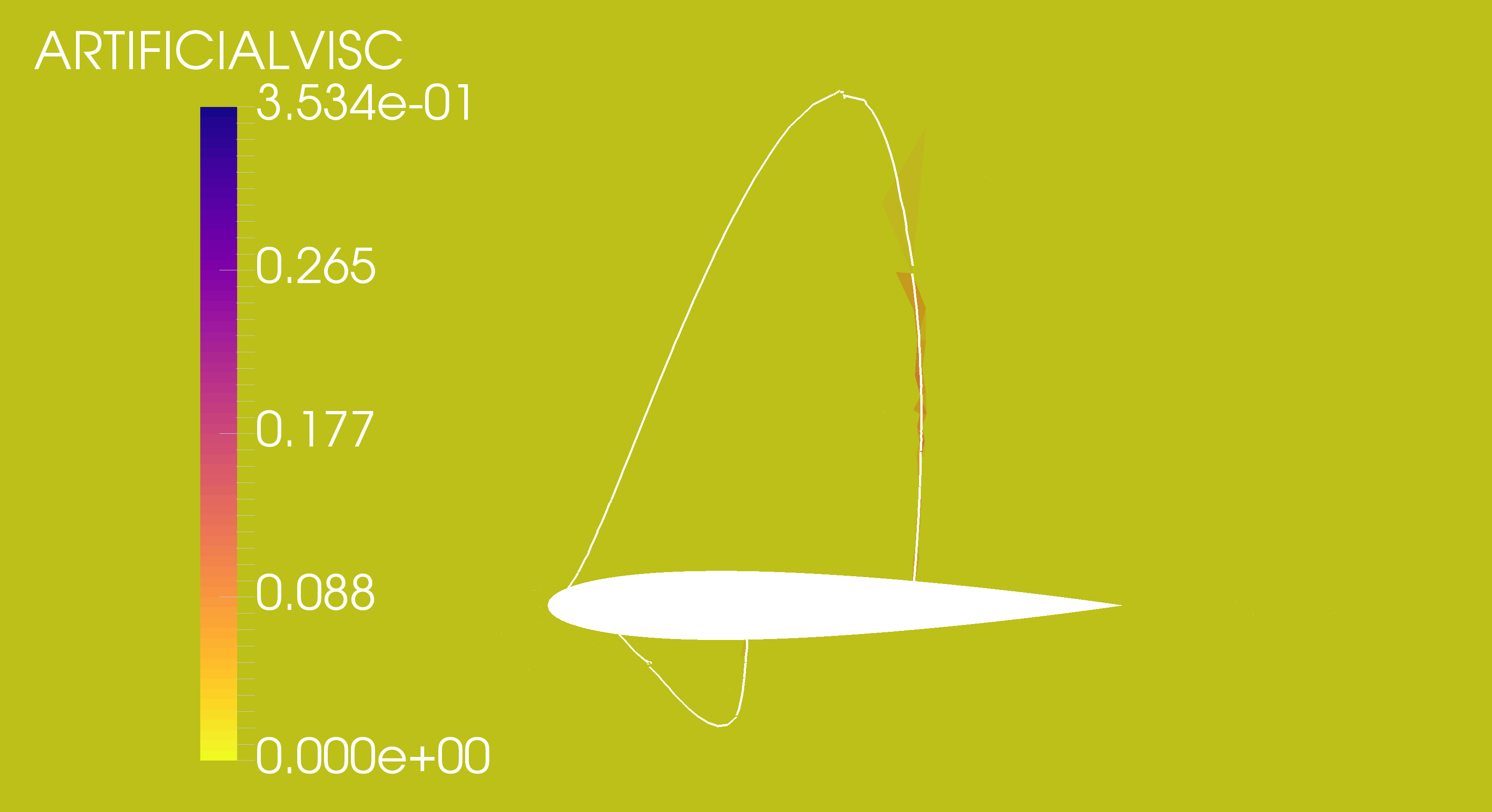}
        \caption{\adaptation{p} with original order restriction.}\label{fig:naca-p-mach-3}
      \end{center}
    \end{subfigure}

    \begin{subfigure}[]{\textwidth}
      \begin{center}
        \includegraphics[width=0.49\textwidth]{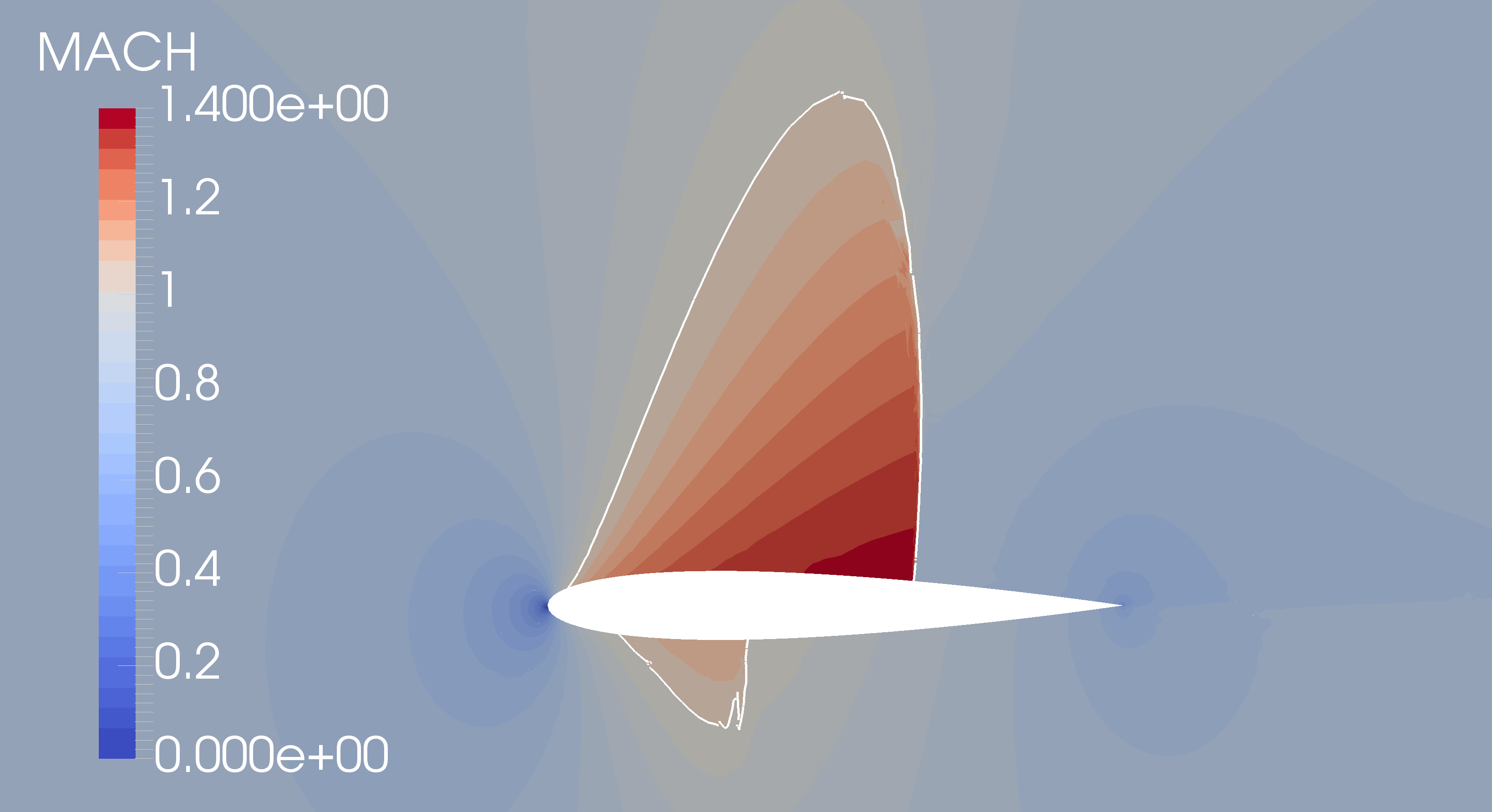}
        \includegraphics[width=0.49\textwidth]{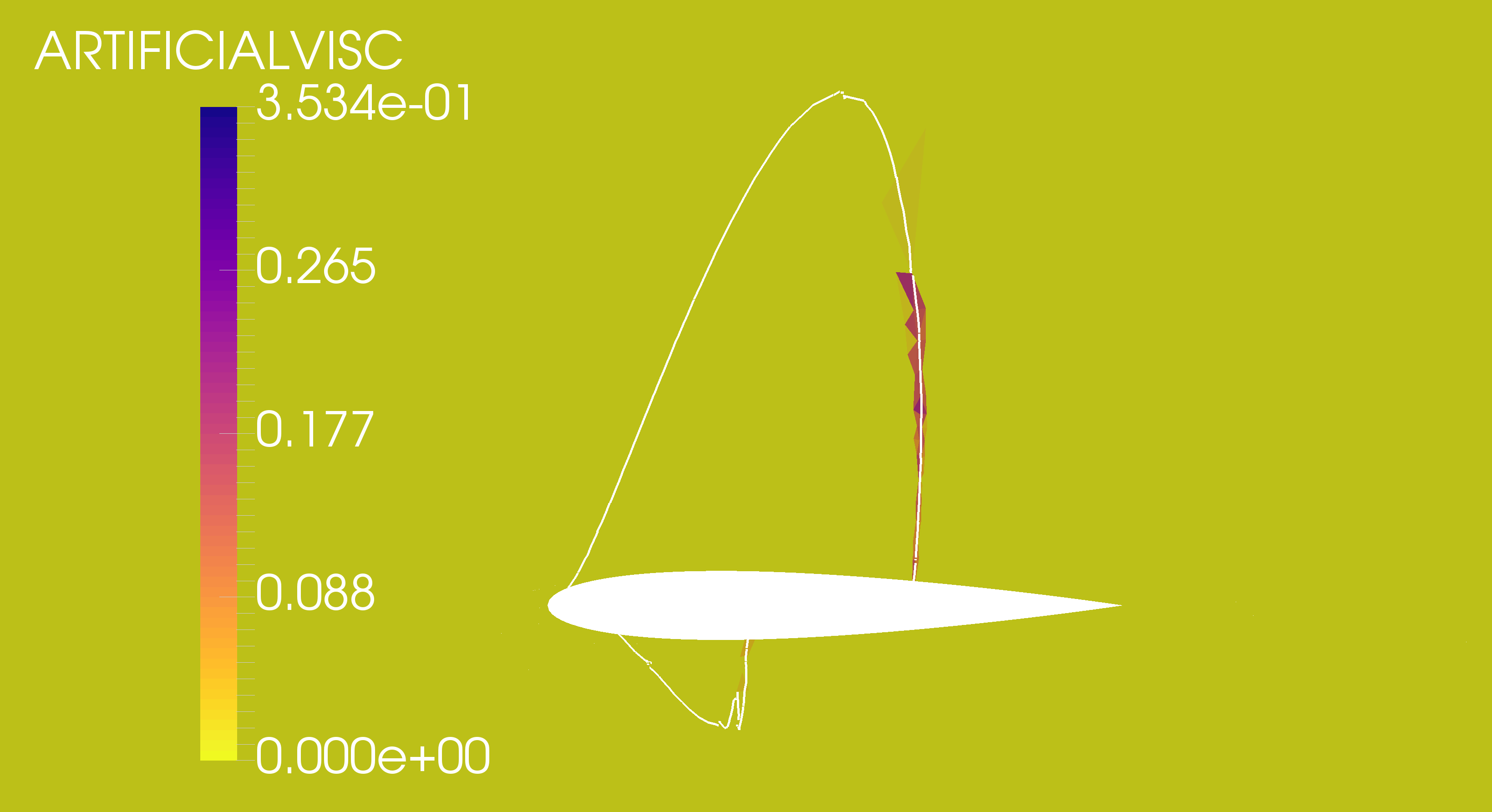}
        \caption{\adaptation{p} with lowest order restriction.}\label{fig:naca-p-mach-4}
      \end{center}
    \end{subfigure}

    \caption{Comparison of the Mach number (left) and artificial viscosity (right) fields for the uniform \(p\) simulation and the three test scenarios of the NACA 0012 test case. A white line denotes the \( M=1 \) line in all figures.}\label{fig:naca-p-mach}
  \end{center}
\end{figure}

Fig.~\ref{fig:naca-p-mach} shows a comparison of the Mach number (left) and artificial viscosity (right) fields for the
uniform \(p\) simulation and the three test scenarios.
We observe little difference between scenarios \#1 and \#2.
Scenario \#3 on the other hand exhibits under-resolution of the shock, seen through its thicker profile.
This is consistent with the local element size and the lack of DOF in the thickness of the shock at lower order.
As a result, the last lower-order scenario exhibits some oscillations in the wake, due to the generated entropy in the shock area.
We can also observe that lower-order scenarios induce more artificial viscosity.
This phenomenon is consistent with the previous assessment of the lack of resolution of the shock.
The discontinuity sensor detects a certain lack of resolution and therefore more artificial viscosity is added to the system.

Overall all scenarios expectedly exhibit similar distributions of local polynomial order in the smooth field regions in Fig.~\ref{fig:naca-p-modes}.
When analysing the distribution of local polynomial orders, we observe higher orders in the area above the strong shock and below the weak shock, in all scenarios.
These areas were not detected in Sect.~\ref{sec:naca-r} as part of the shock due to the then under-resolved and therefore too short shocks.
Now that the shocks are better resolved, they reach further out and require additional resolution, in the form of higher polynomial order in this case.
We also observe, in the lower-order scenarios, that parasite higher-order zones are created.
This is especially obvious around the weak shock in the third scenario.
This is due, as we noted above, to the thicker shock profile and therefore the need to add resolution around the shock.
Fig.~\ref{fig:naca-p-modes-4}~(left) is consistent with this explanation as we observe a larger area of high sensor values, extending beyond the shock areas determined in Sect.~\ref{sec:naca-r}.

\subsection{Supersonic intake}\label{sec:intake}

This section illustrates the new approach on a test case with a more complicated shock pattern.
The test case is that of a supersonic intake at \( M_\infty = 3.0 \) first studied experimentally~\cite{Anderson1970} and later numerically~\cite{Jain2006}.
The intake consists of two straight ramps inclined with respect to the incoming free-stream flow at angles of \( 7^\circ \) and \( 14^\circ \) respectively.
The first ramp creates an oblique shock which impinges on the horizontal cowl and in turn leads to a complex pattern of reflecting oblique shocks throughout the diffuser of the intake.
The difficulty here is the presence of multiple shocks with different orientations in the very narrow regions of the diffuser.

We discretise the domain uniformly in the stream-wise direction.
We set an element of size \(0.01L\) (\(L\) being the length of the intake) inside the intake and let it coarsen outside the intake up to an element of size \(0.05L\) in the far-field.
The mesh is curvilinear of order \(p=4\) and it is optimised in the throat.
Fig.~\ref{fig:intake-r1-mesh}~(left) shows what the mesh looks like inside the intake and in its immediate surrounding.

We run the solver at uniform order \(p=3\) on the initial mesh to obtain a base solution.
We impose wall BC on the surfaces of the intake, at the intake outlet BC we set a low enough pressure until a fully supersonic field is obtained (\( P_b = 0.9 P_{\inf} \)) and far-field BC at the external boundaries of the domain.
We use Roe's approximate Riemann solver~\cite{Toro2009}.
For the artificial viscosity, we tuned the solver parameters to \(s_\kappa = 0.0\), \( \kappa = 0.0 \) and \(\mu_0 = 0.1\).
Fig.~\ref{fig:intake-r1-sensor}~(left) shows that large values of the sensor are obtained in all shocks and that moderate values are obtained everywhere after the first upstream shock.
However, artificial viscosity only really triggers in the vicinity of the shocks as Fig.~\ref{fig:intake-r1-visc}~(left) shows, proving adequate tuning of the artificial viscosity parameters.
Just like for the NACA 0012 test case, the shocks demonstrate a thick profile, as can be seen in Fig.~\ref{fig:intake-r1-mach}~(left), due to the relatively coarse local mesh as well as some oscillations near the leading edge of the cowl.

\subsubsection{\emph{rr}-adaptation}\label{sec:intake-r}

\begin{figure}[htbp!]
  \begin{center}

    \begin{subfigure}[]{\textwidth}
      \begin{center}
        \includegraphics[width=0.49\textwidth]{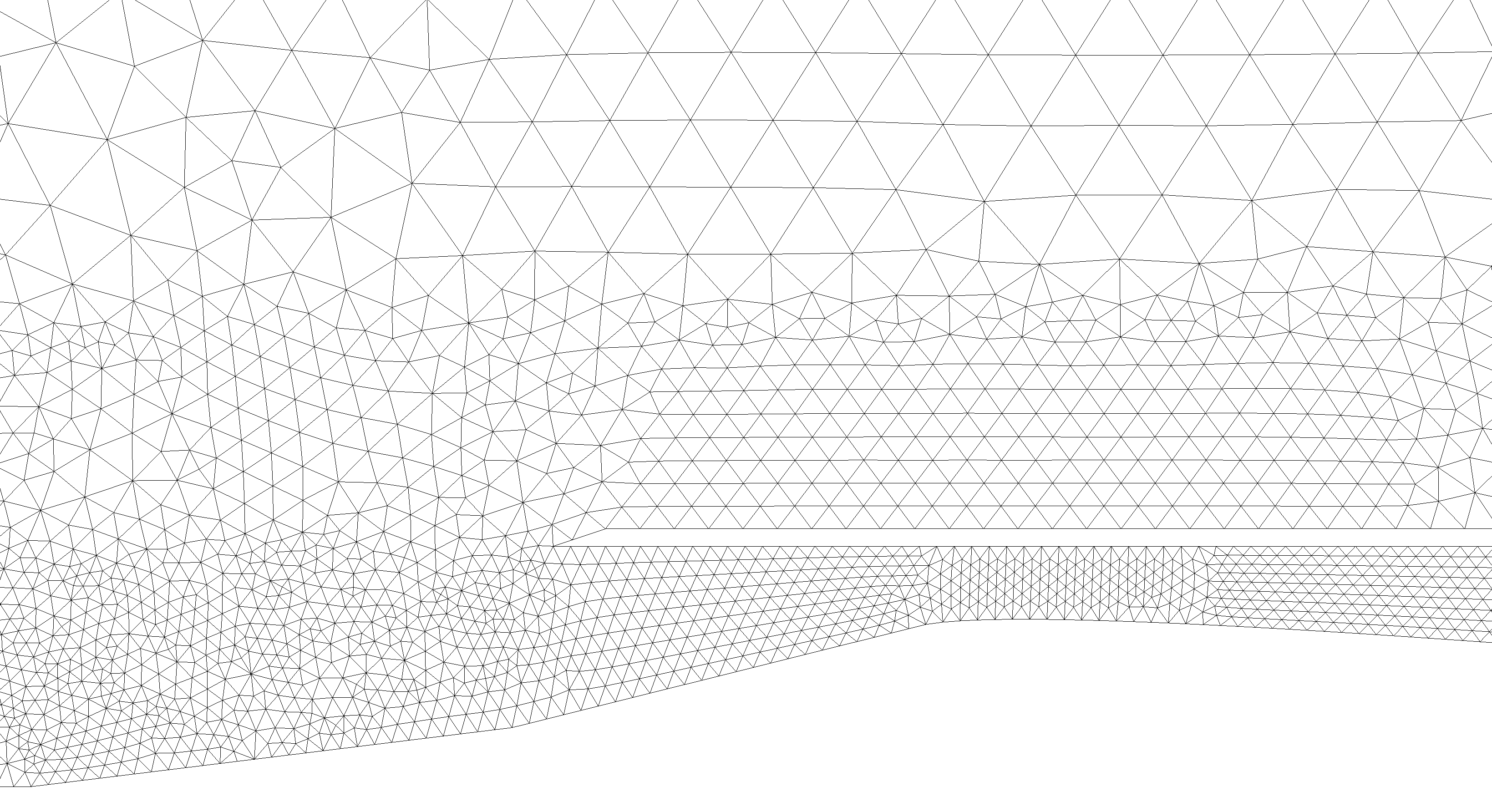}
        \includegraphics[width=0.49\textwidth]{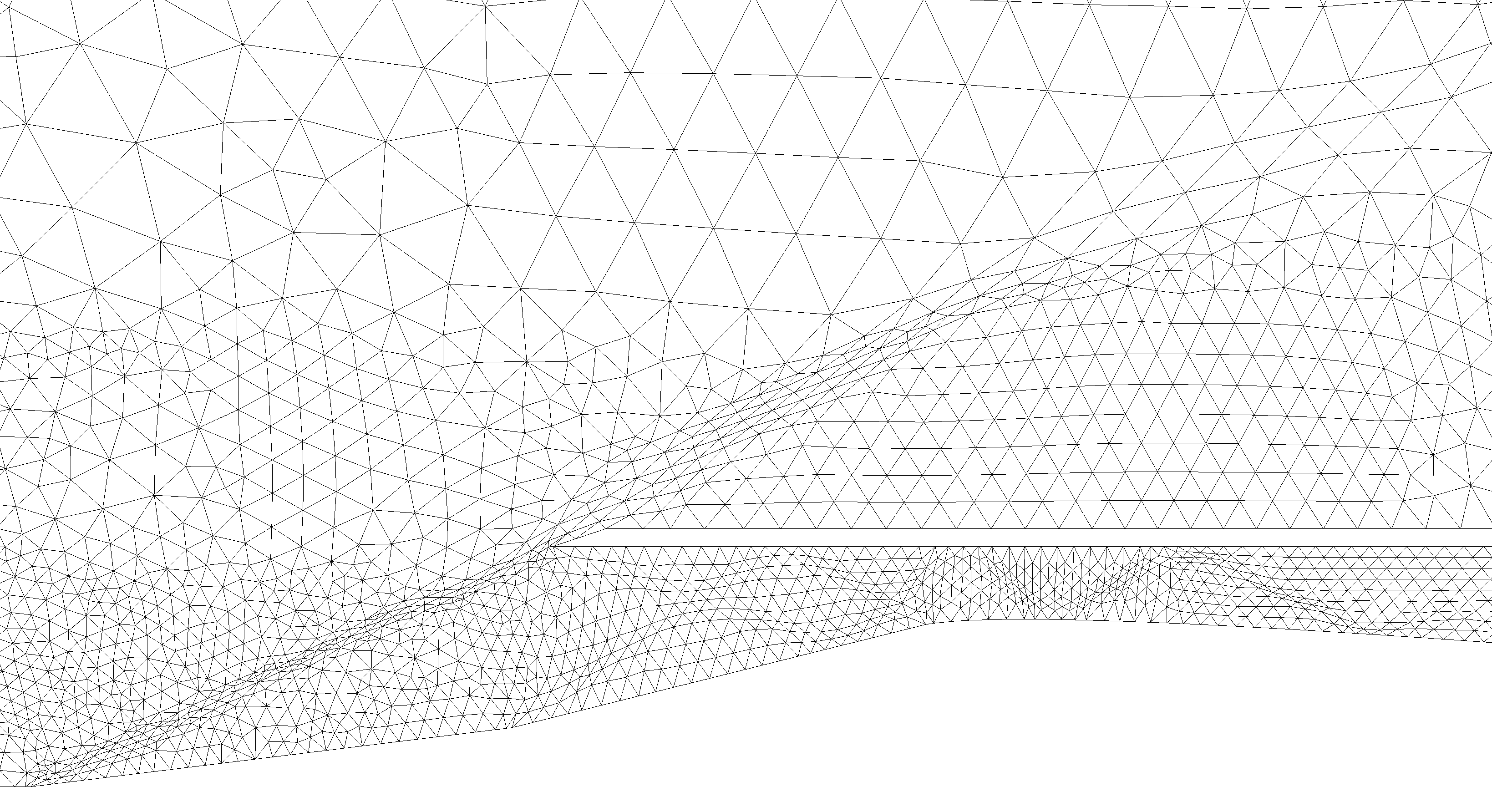}
        \caption{Mesh.}\label{fig:intake-r1-mesh}
      \end{center}
    \end{subfigure}

    \begin{subfigure}[]{\textwidth}
      \begin{center}
        \includegraphics[width=0.49\textwidth]{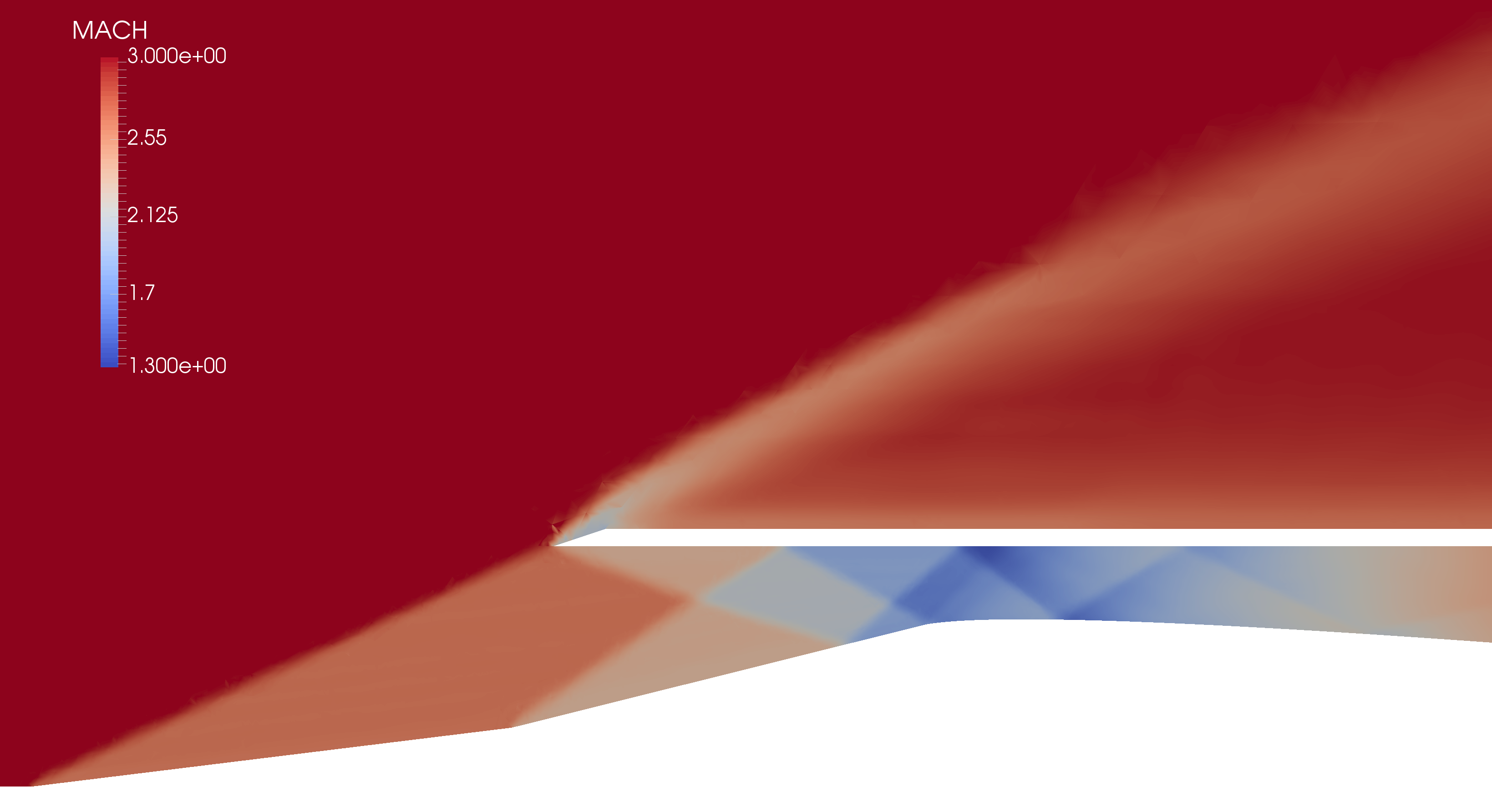}
        \includegraphics[width=0.49\textwidth]{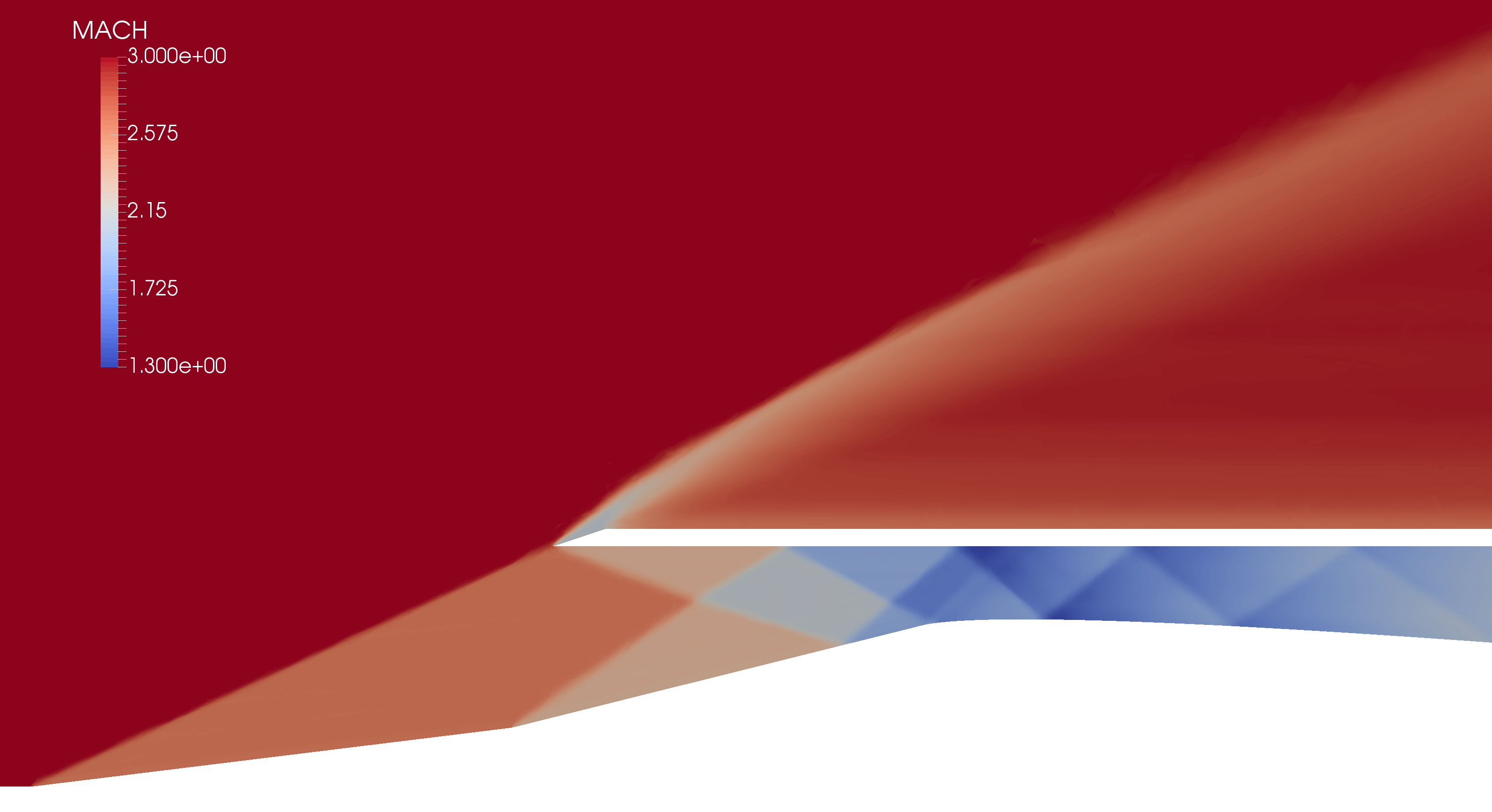}
        \caption{Mach number field.}\label{fig:intake-r1-mach}
      \end{center}
    \end{subfigure}

    \begin{subfigure}[]{\textwidth}
      \begin{center}
        \includegraphics[width=0.49\textwidth]{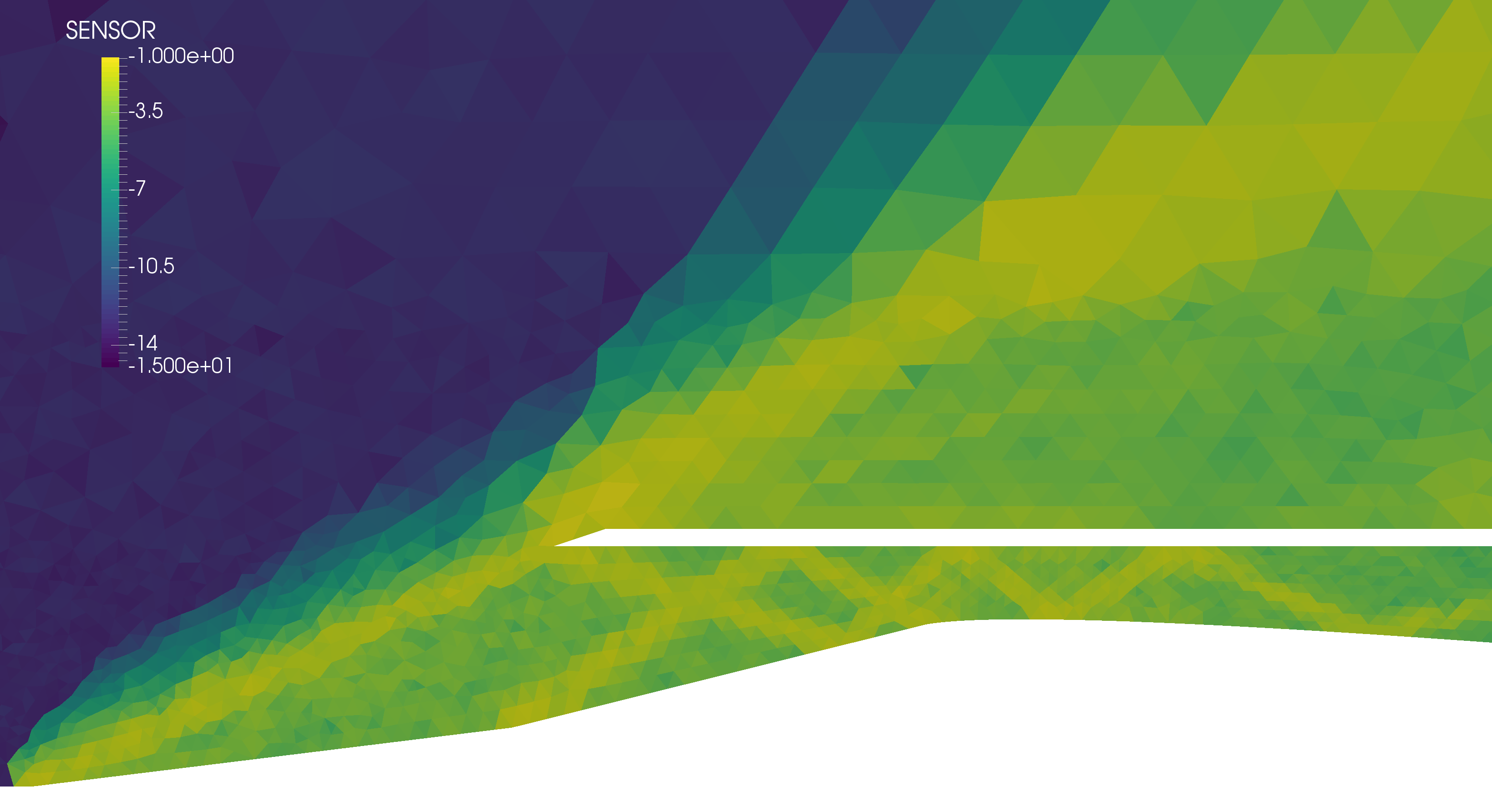}
        \includegraphics[width=0.49\textwidth]{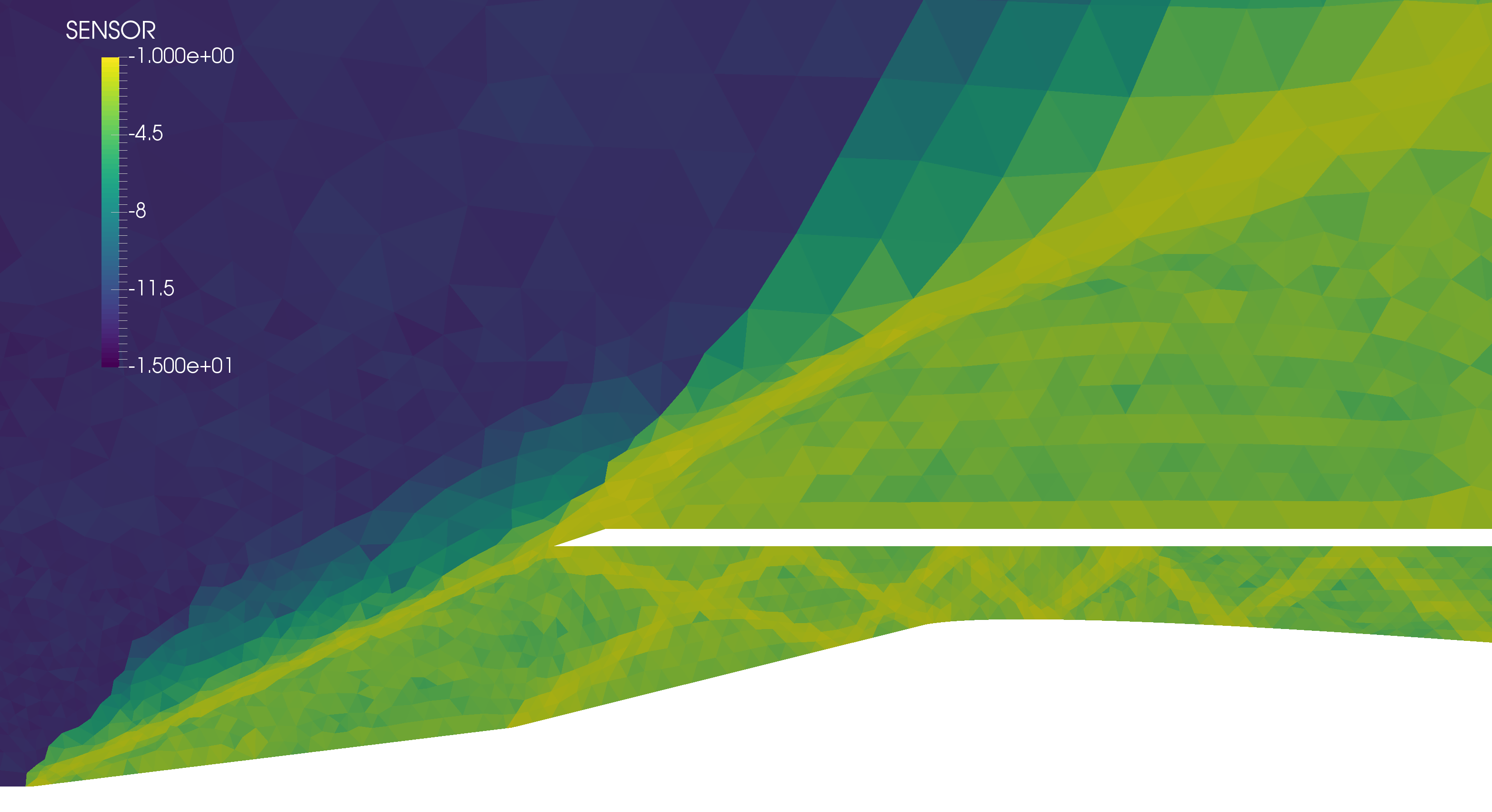}
        \caption{Sensor field.}\label{fig:intake-r1-sensor}
      \end{center}
    \end{subfigure}

    \begin{subfigure}[]{\textwidth}
      \begin{center}
        \includegraphics[width=0.49\textwidth]{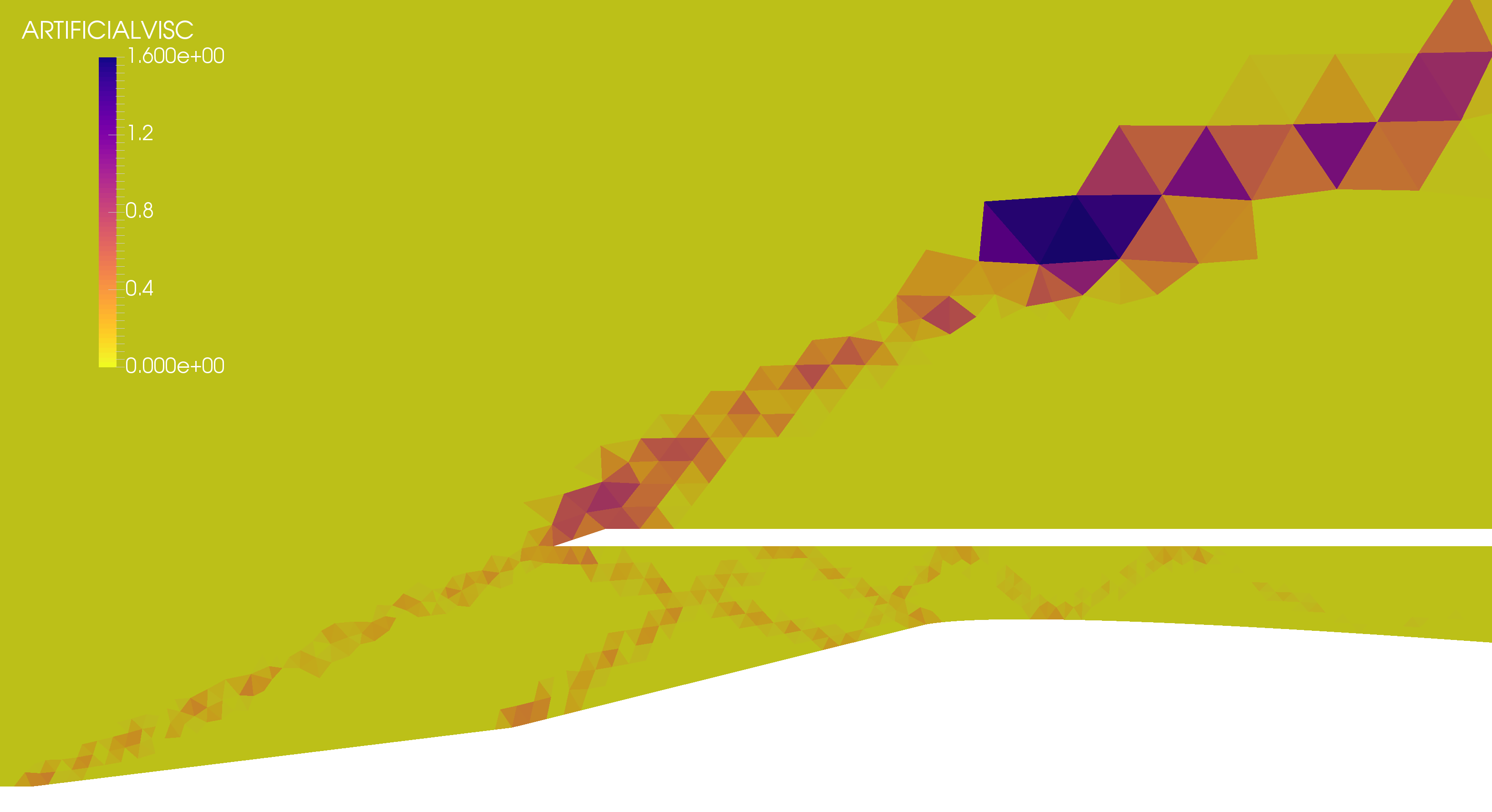}
        \includegraphics[width=0.49\textwidth]{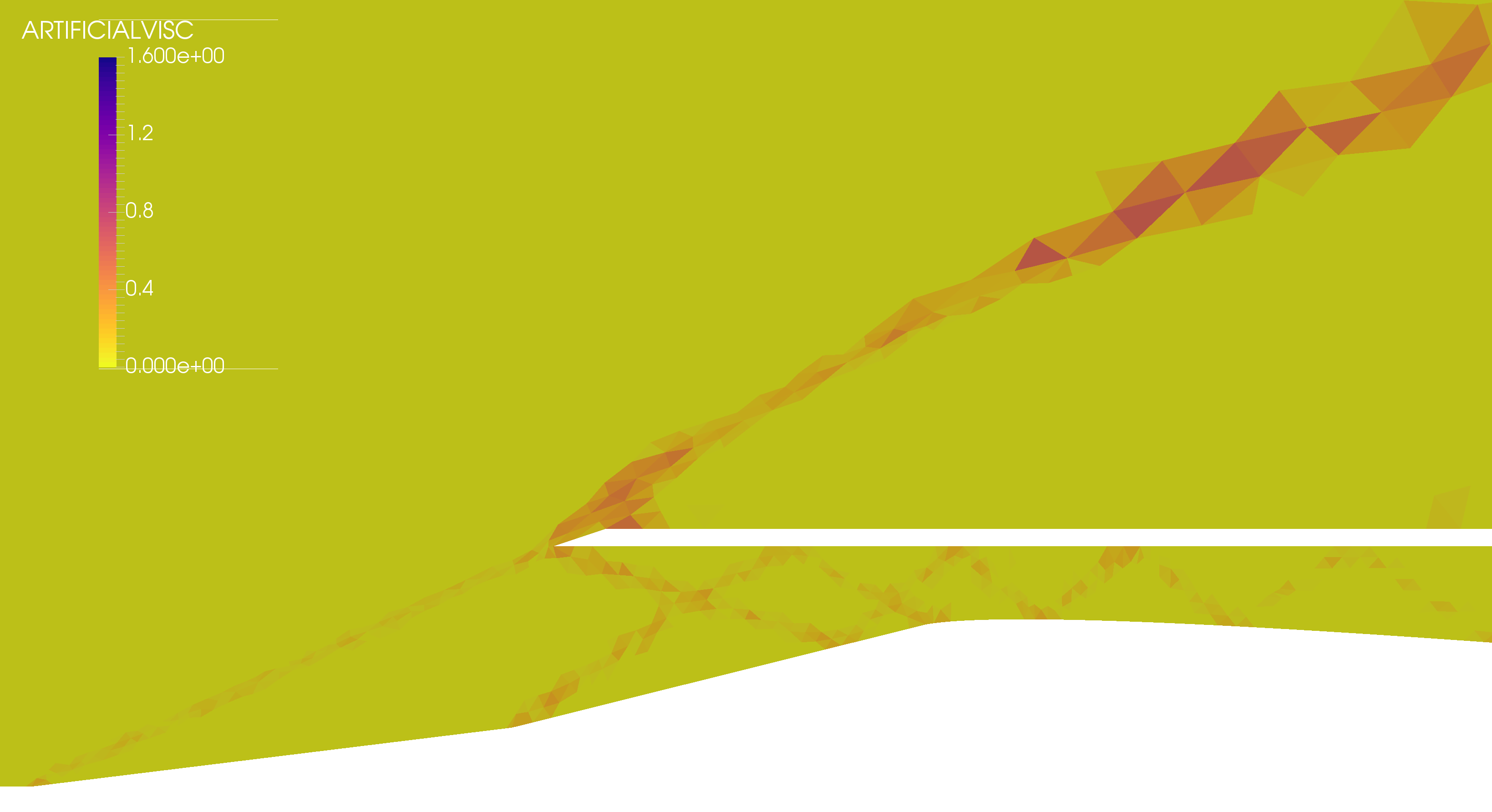}
        \caption{Artificial viscosity field.}\label{fig:intake-r1-visc}
      \end{center}
    \end{subfigure}

    \caption{Comparison of the mesh and fields for the intake before (left) and after (right) the first round of \adaptation{r}.}\label{fig:intake-r1-adapt}
  \end{center}
\end{figure}

Once more we follow the workflow laid out in Sect.~\ref{sec:workflow} except that we decide to run two rounds of \adaptation{r}.
Each round uses a less aggressive shrinking factor \(r_{\text{scale}}=0.5\).
Before each simulation, we again carry out mesh optimisation to improve the quality of the high-order mesh.
The new mesh after one round of \adaptation{r} is depicted in Fig.~\ref{fig:intake-r1-mesh}~(right).
We observe refinement in all areas of interest and note that refinement is stronger in the area of the first upstream shock.
Indeed, elements in the first shock are able to pull DOF from the freestream areas whereas elements inside the intake are interacting with each other.
Refinement is nonetheless obtained in all shock areas and anisotropy naturally appears such that elements are shrunk in mostly the shock normal direction.
The \adaptation{r} strategy works by pulling nodes together.
Even though the optimiser is not aware of the shock structures,
nodes are naturally moved normally rather than tangentially to the underlying shock, because the shrinking areas are long and narrow.

We now run the solver on the new adapted mesh using the same solver parameters.
A stable flow solution is obtained and shown in Fig.~\ref{fig:intake-r1-mach}~(right).
All shocks now appear sharper and the oscillations observed near the leading edge of the cowl have disappeared.
Figs.~\ref{fig:intake-r1-sensor}~\&~\ref{fig:intake-r1-visc}~(right) also show that discontinuity, as per the sensor, occurs in a narrower region.

\begin{figure}[htbp!]
  \begin{center}

    \begin{subfigure}[]{\textwidth}
      \begin{center}
        \includegraphics[width=0.49\textwidth]{figs/intake_mesh_1_r1_p_uni}
        \includegraphics[width=0.49\textwidth]{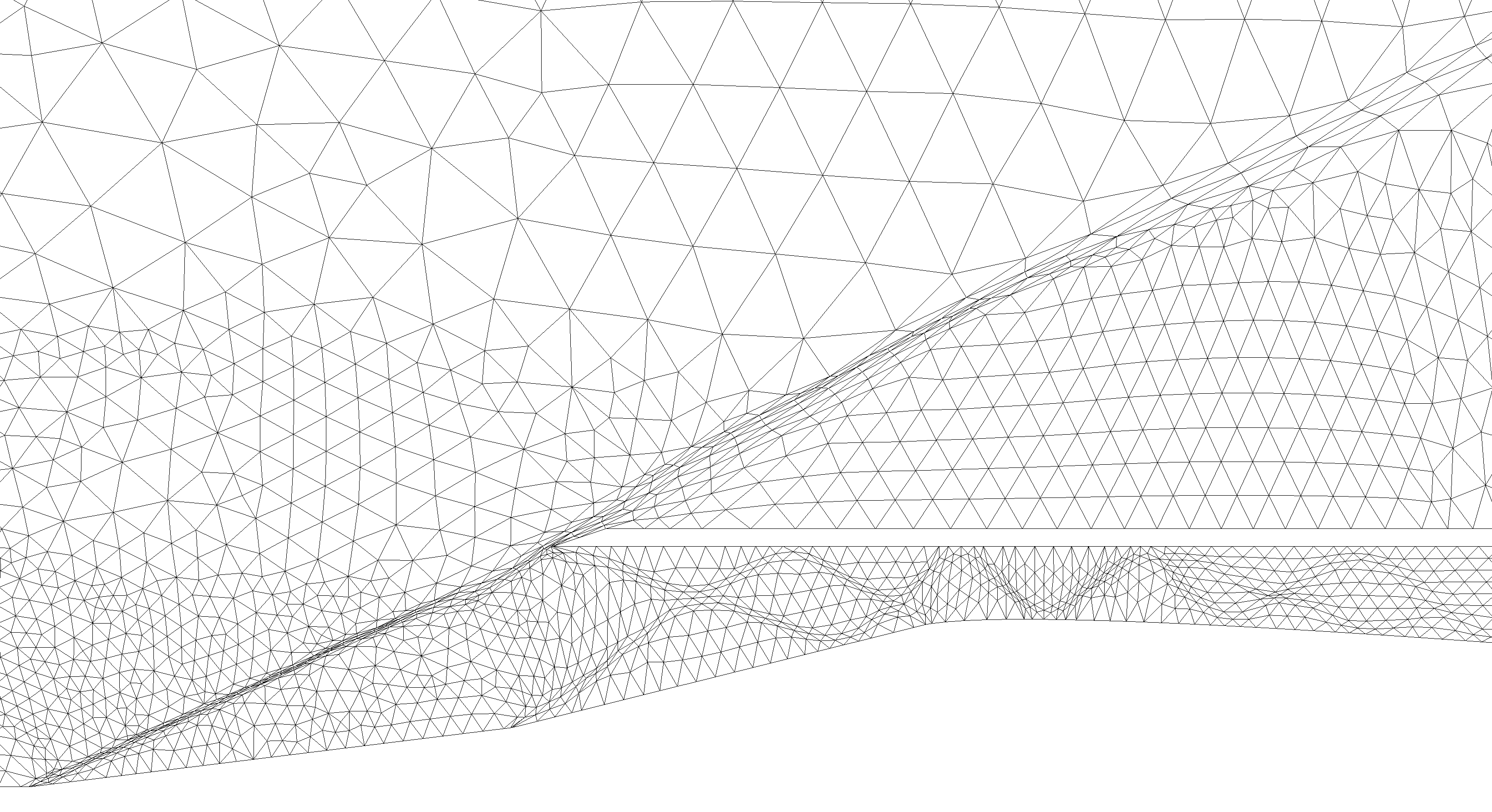}
        \caption{Mesh.}\label{fig:intake-r2-mesh}
      \end{center}
    \end{subfigure}

    \begin{subfigure}[]{\textwidth}
      \begin{center}
        \includegraphics[width=0.49\textwidth]{figs/intake_mach_1_r1_p_uni}
        \includegraphics[width=0.49\textwidth]{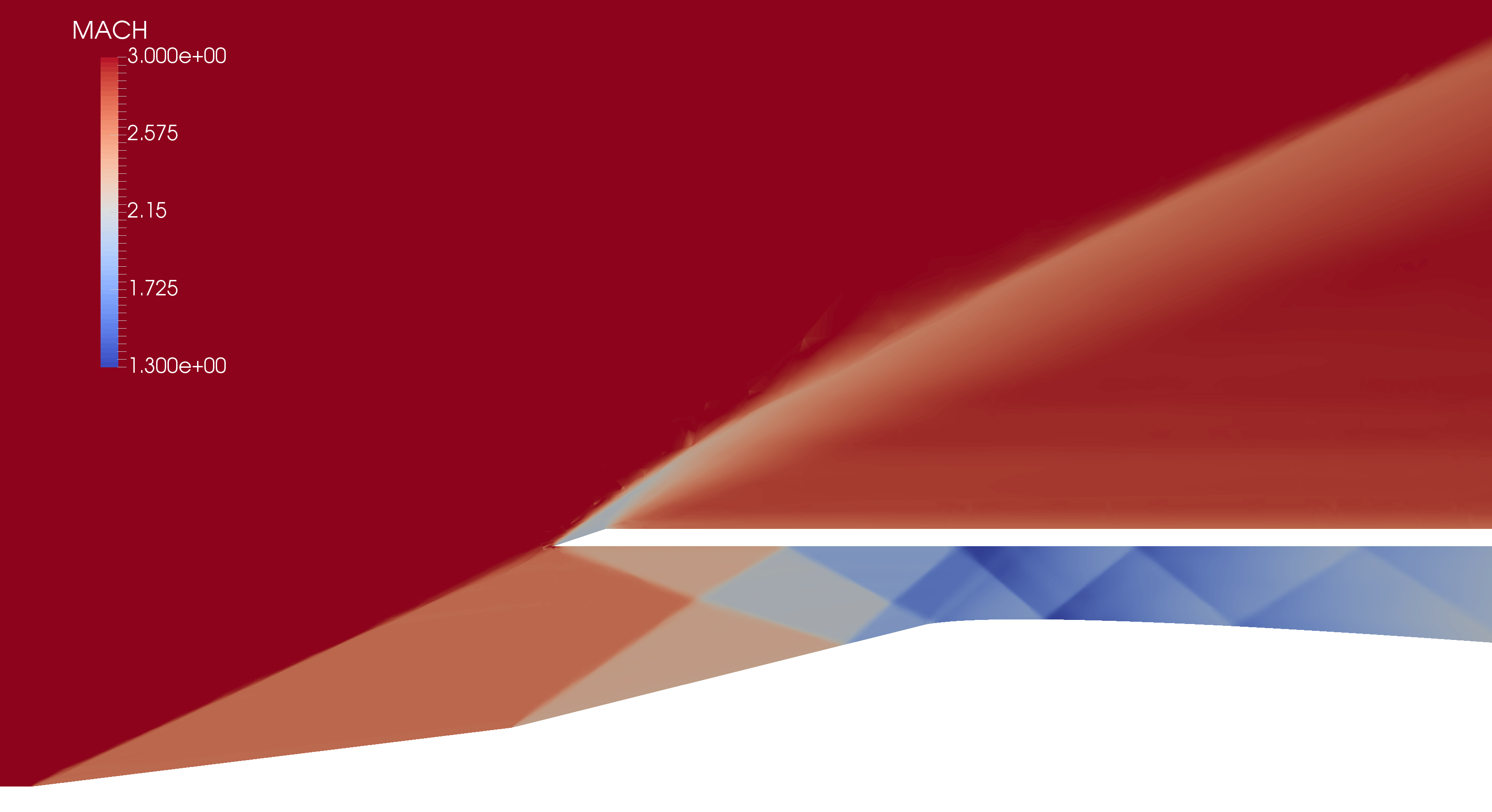}
        \caption{Mach number field.}\label{fig:intake-r2-mach}
      \end{center}
    \end{subfigure}

    \begin{subfigure}[]{\textwidth}
      \begin{center}
        \includegraphics[width=0.49\textwidth]{figs/intake_sensor_1_r1_p_uni}
        \includegraphics[width=0.49\textwidth]{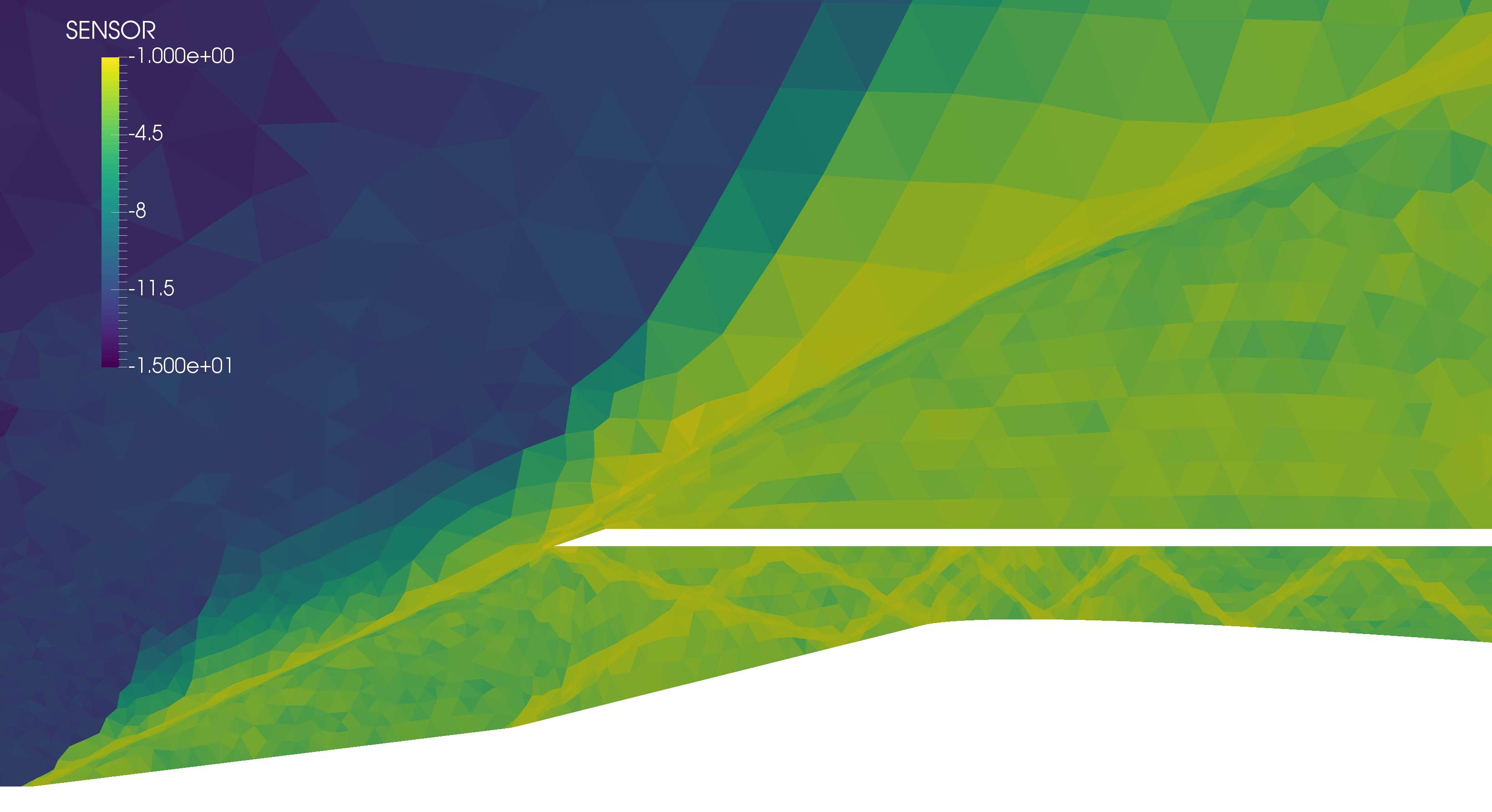}
        \caption{Sensor field.}\label{fig:intake-r2-sensor}
      \end{center}
    \end{subfigure}

    \begin{subfigure}[]{\textwidth}
      \begin{center}
        \includegraphics[width=0.49\textwidth]{figs/intake_visc_1_r1_p_uni}
        \includegraphics[width=0.49\textwidth]{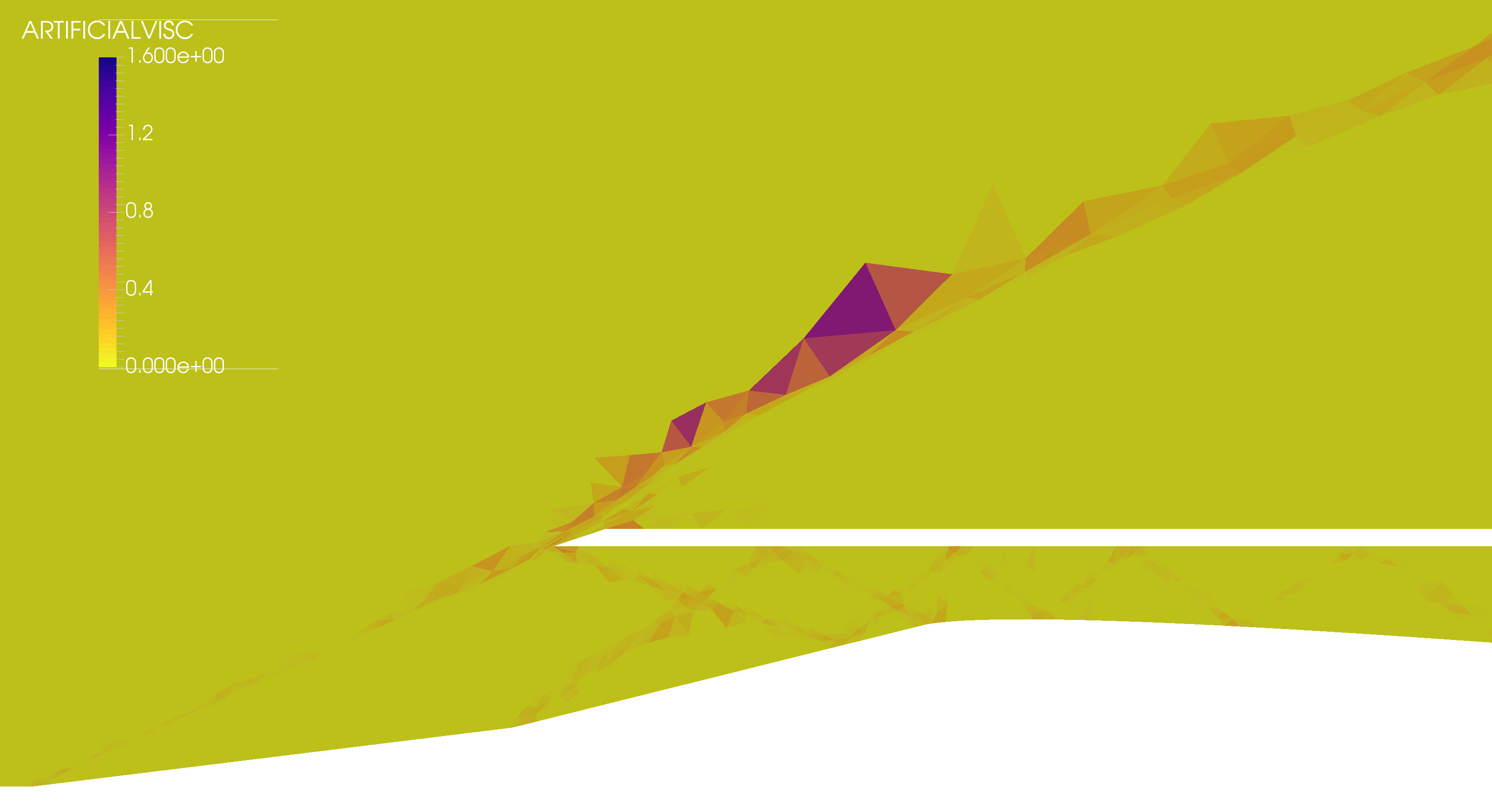}
        \caption{Artificial viscosity field.}\label{fig:intake-r2-visc}
      \end{center}
    \end{subfigure}

    \caption{Comparison of the mesh and fields for the intake before (left) and after (right) the second round of \adaptation{r}.}\label{fig:intake-r2-adapt}
  \end{center}
\end{figure}

We then apply a second round of \adaptation{r} in the exact similar fashion: we isolate shock areas and use them as input for the optimiser.
Fig.~\ref{fig:intake-r2-mesh}~(right) depicts the final adapted mesh which shows further refinement of the shock regions.
We also notice that the oblique shocks inside the intake past the throat have moved upstream due to the refinement of
the oblique shocks located upstream of the throat.
While the \emph{r}-adapted mesh could not capture these downstream shocks, the \emph{rr}-adapted mesh can.
By using a two-step approach, we are also able to pull more mesh nodes together than when using a one-step approach.

\begin{figure}[htbp!]
  \begin{center}
    \begin{tikzpicture}[spy using outlines={rectangle,magnification=2,connect spies}]
      \begin{axis}[
          xmin=-0.1, xmax=1, ymin=0, ymax=3.2,
          width=0.99\textwidth,
          xlabel=\(x/L\),
          ylabel=\(M\),
          legend pos=north east
        ]

        \addplot[color=red,no marks]
        table[x=x,y=Mach,col sep=comma] {plots/intake_init.csv};
        \addlegendentry{Initial mesh}

        \addplot[color=blue,no marks]
        table[x=x,y=Mach,col sep=comma] {plots/intake_r1.csv};
        \addlegendentry{\emph{r}-adapted mesh}

        \addplot[color=black,no marks]
        table[x=x,y=Mach,col sep=comma] {plots/intake_r2.csv};
        \addlegendentry{\emph{rr}-adapted mesh}

        \addplot[color=olive,no marks]
        table[x=x,y=Mach,col sep=comma] {plots/intake_p.csv};
        \addlegendentry{\emph{rrp}-adapted mesh}

        \coordinate (spypoint1) at (axis cs:0,2.87);
        \coordinate (spyviewer1) at (axis cs:0.1,0.65);
        \spy[width=0.25\textwidth,height=0.25\textwidth] on (spypoint1) in node [fill=white] at (spyviewer1);

        \coordinate (spypoint2) at (axis cs:0.28,2.5);
        \coordinate (spyviewer2) at (axis cs:0.45,0.65);
        \spy[width=0.25\textwidth,height=0.25\textwidth] on (spypoint2) in node [fill=white] at (spyviewer2);

        \coordinate (spypoint3) at (axis cs:0.54,1.63);
        \coordinate (spyviewer3) at (axis cs:0.8,0.65);
        \spy[width=0.25\textwidth,height=0.25\textwidth] on (spypoint3) in node [fill=white] at (spyviewer3);

      \end{axis}
    \end{tikzpicture}
    \caption{Plot of the Mach number \(M\) on the lower surface throughout the \adaptation{rrp} process.}\label{fig:intake-plot}
  \end{center}
\end{figure}
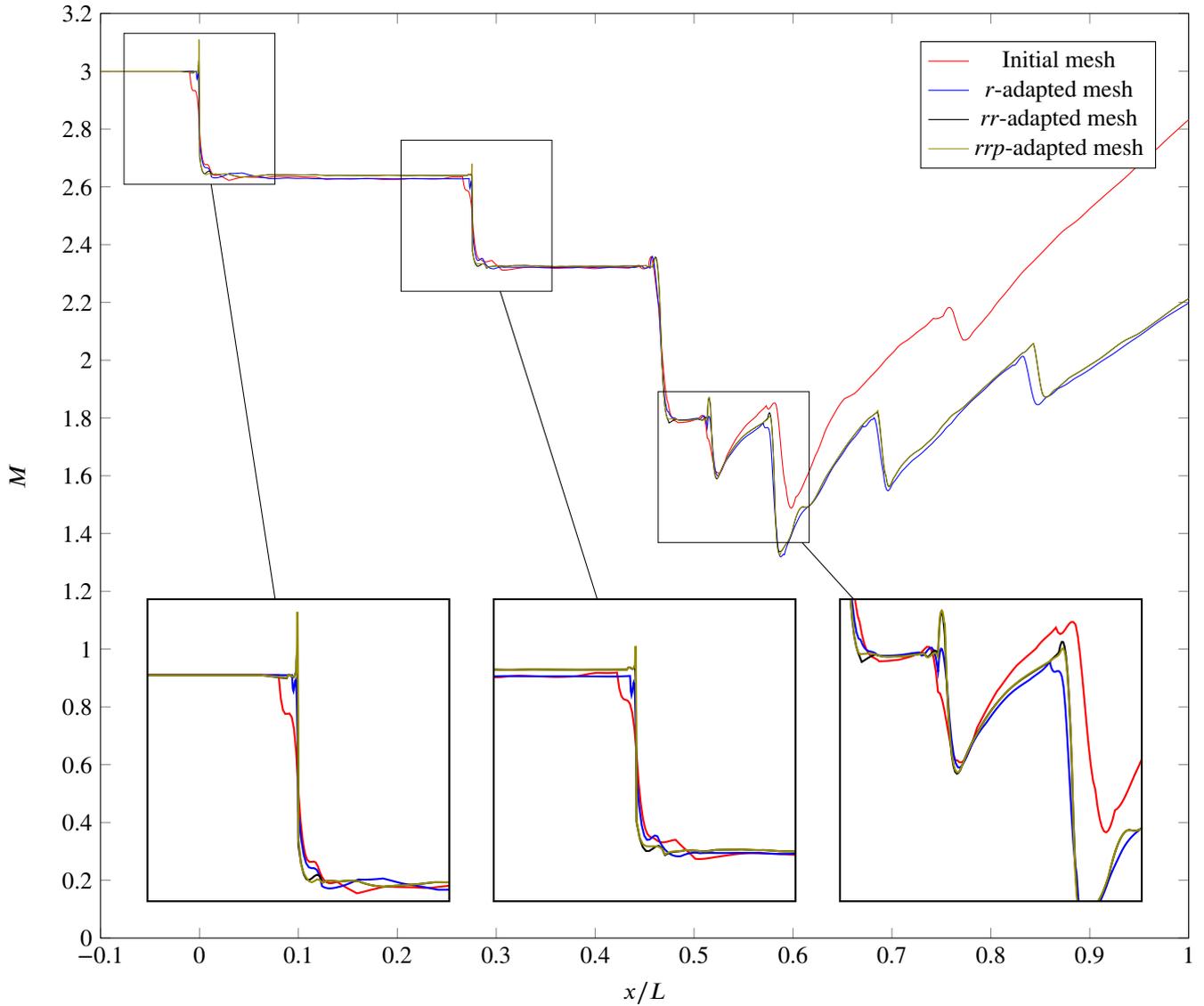

This becomes even more obvious when looking at the plot of the Mach number in Fig.~\ref{fig:intake-plot}.
The initial mesh largely overestimates the values of the Mach number past the throat (\( x/L \approx 0.57\)).
This is mostly solved by \adaptation{r} although further improvement is obtained through \adaptation{rr}-adaptation.
This is due to the good resolution of upstream shocks through the clustering of DOF.\@

\subsubsection{\emph{p}-adaptation}\label{sec:intake-p}

We now apply \adaptation{p} to the \emph{rr}-adapted mesh.
For this test case, we focus on unrestricted \adaptation{p} where the local polynomial order inside elements
is left free to change, even in shock areas.
We start from the field obtained at \(p=4\) in Sect.~\ref{sec:intake-r} and use values of \( p_{\min} = 2 \) and \( p_{\max} = 6 \).
We again use a sensor based on the density field \(\rho \) and solver default values for the thresholds.

\begin{figure}[htbp!]
  \begin{center}

    \begin{subfigure}[]{\textwidth}
      \begin{center}
        \includegraphics[width=0.49\textwidth]{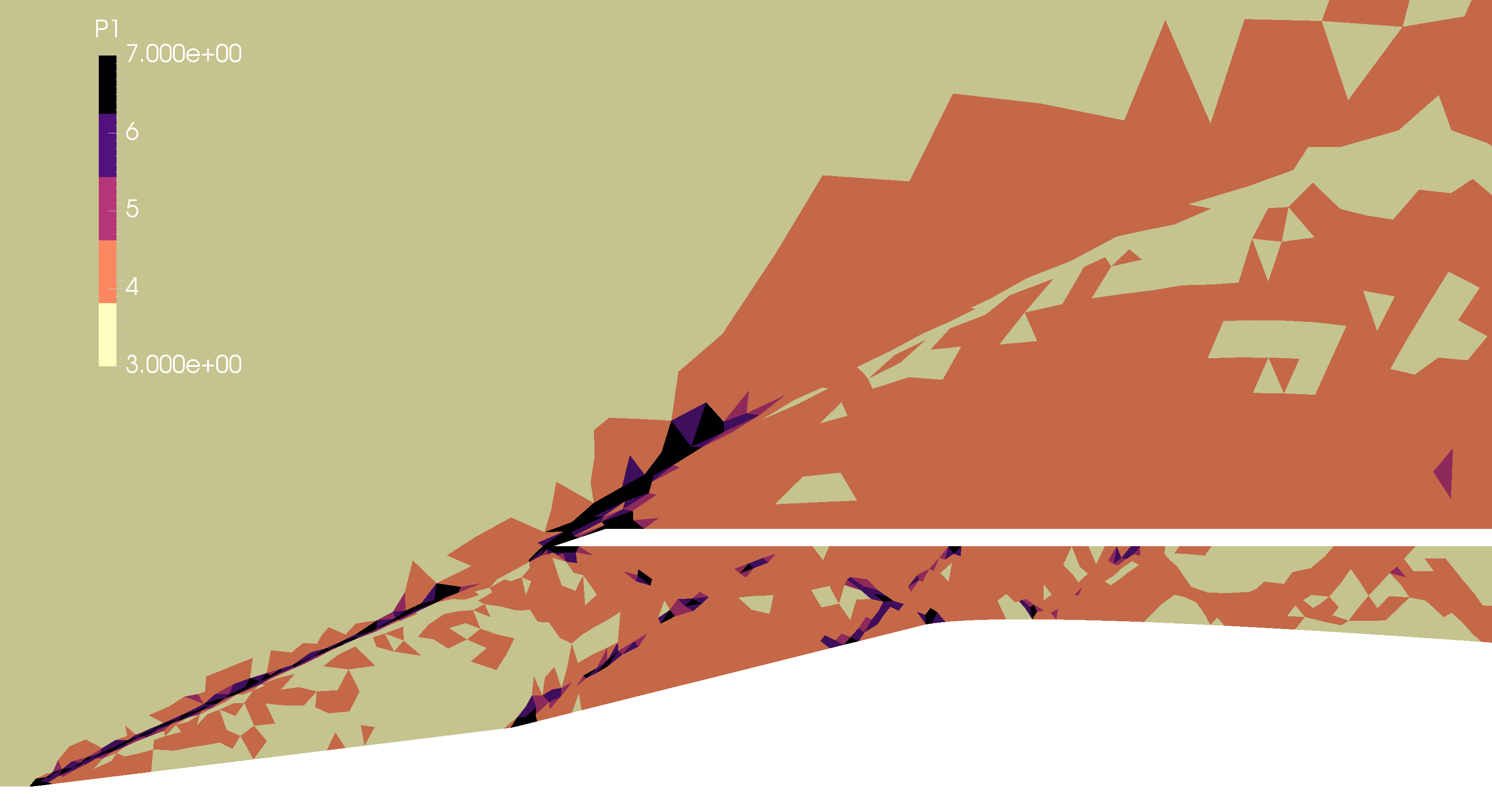}
        \includegraphics[width=0.49\textwidth]{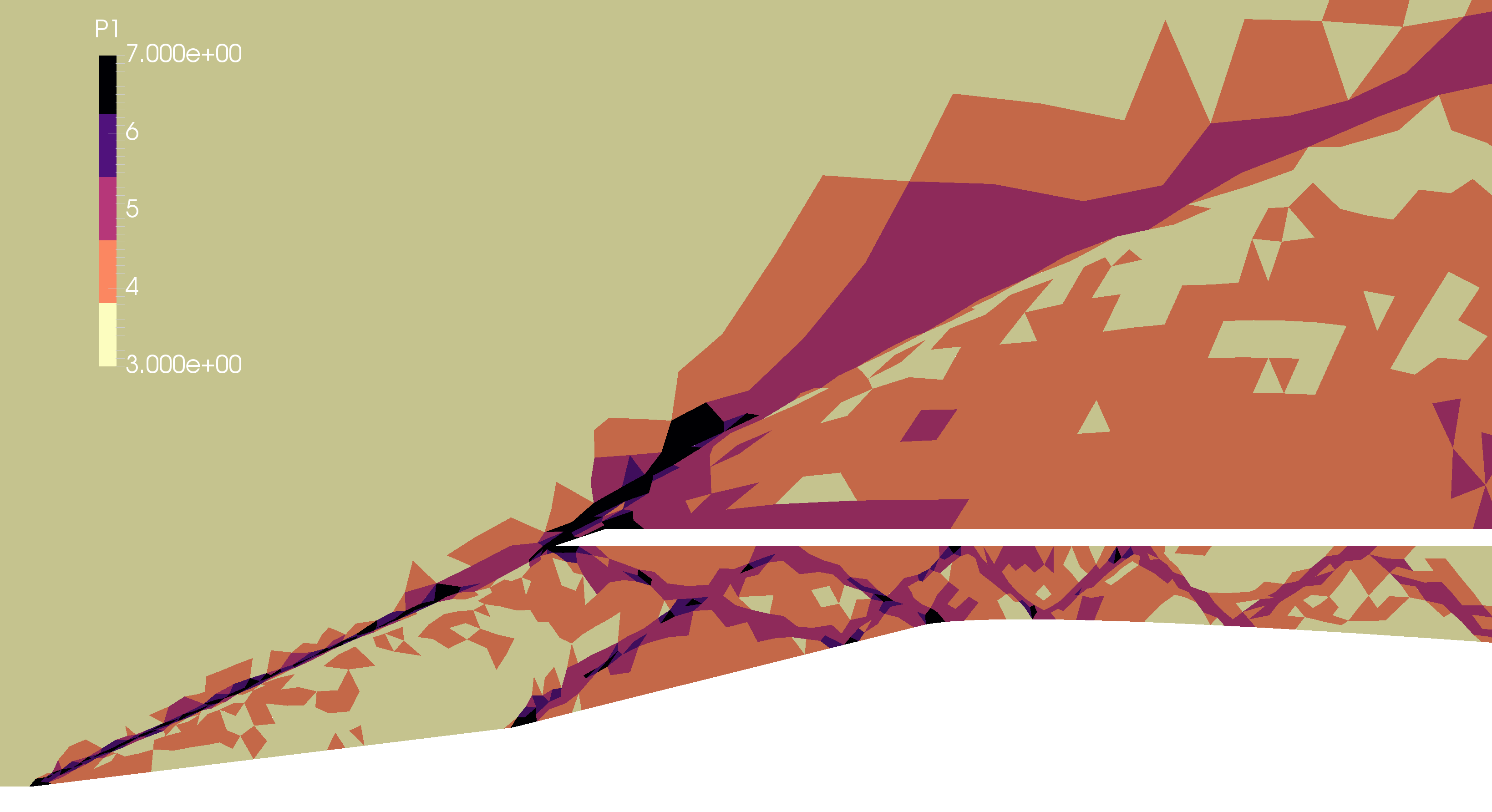}
        \caption{Number of local modes (\(=p+1\)).}\label{fig:intake-p-modes}
      \end{center}
    \end{subfigure}

    \begin{subfigure}[]{\textwidth}
      \begin{center}
        \includegraphics[width=0.49\textwidth]{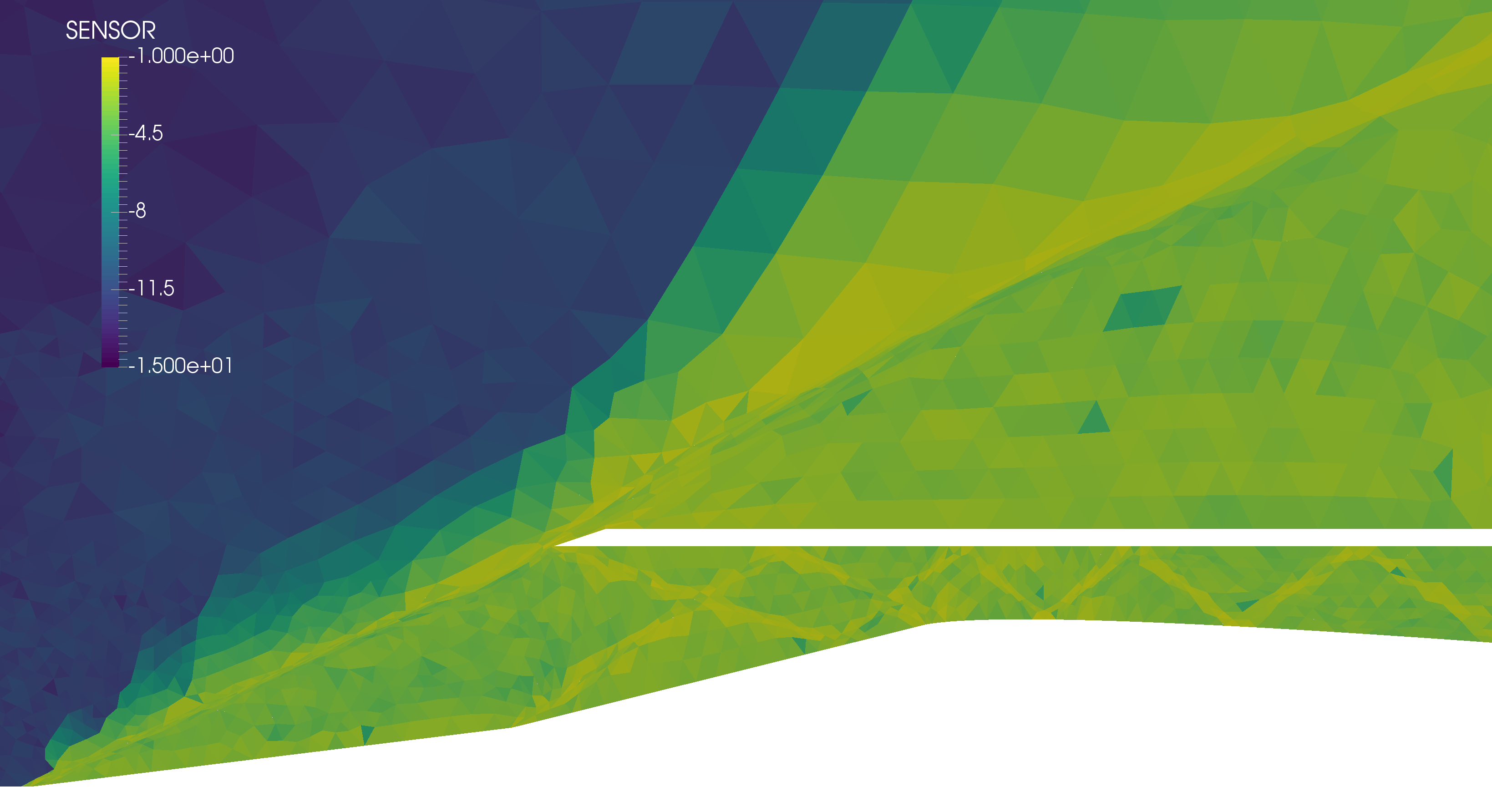}
        \includegraphics[width=0.49\textwidth]{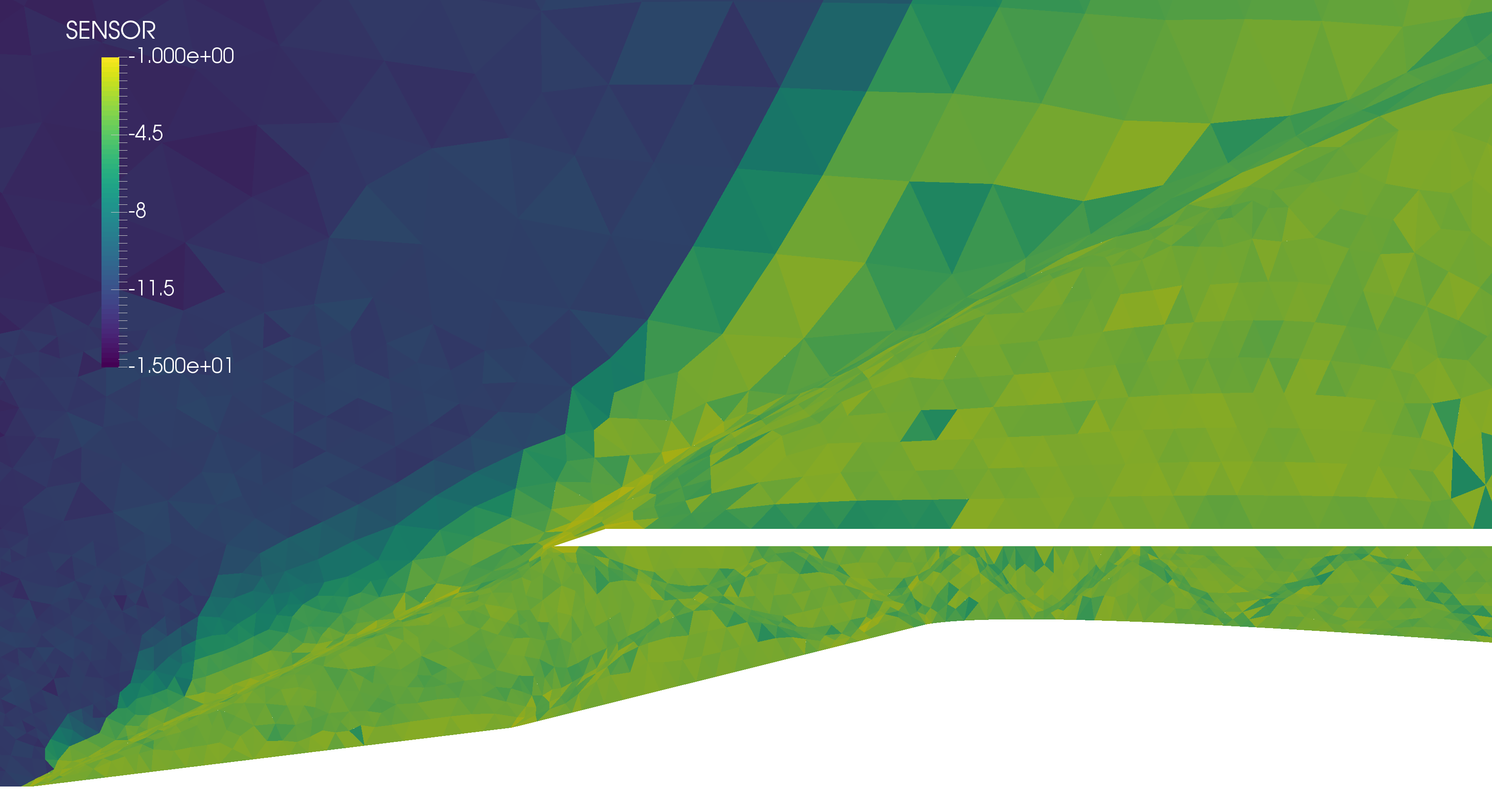}
        \caption{Sensor field.}\label{fig:intake-p-sensor}
      \end{center}
    \end{subfigure}

    \begin{subfigure}[]{\textwidth}
      \begin{center}
        \includegraphics[width=0.49\textwidth]{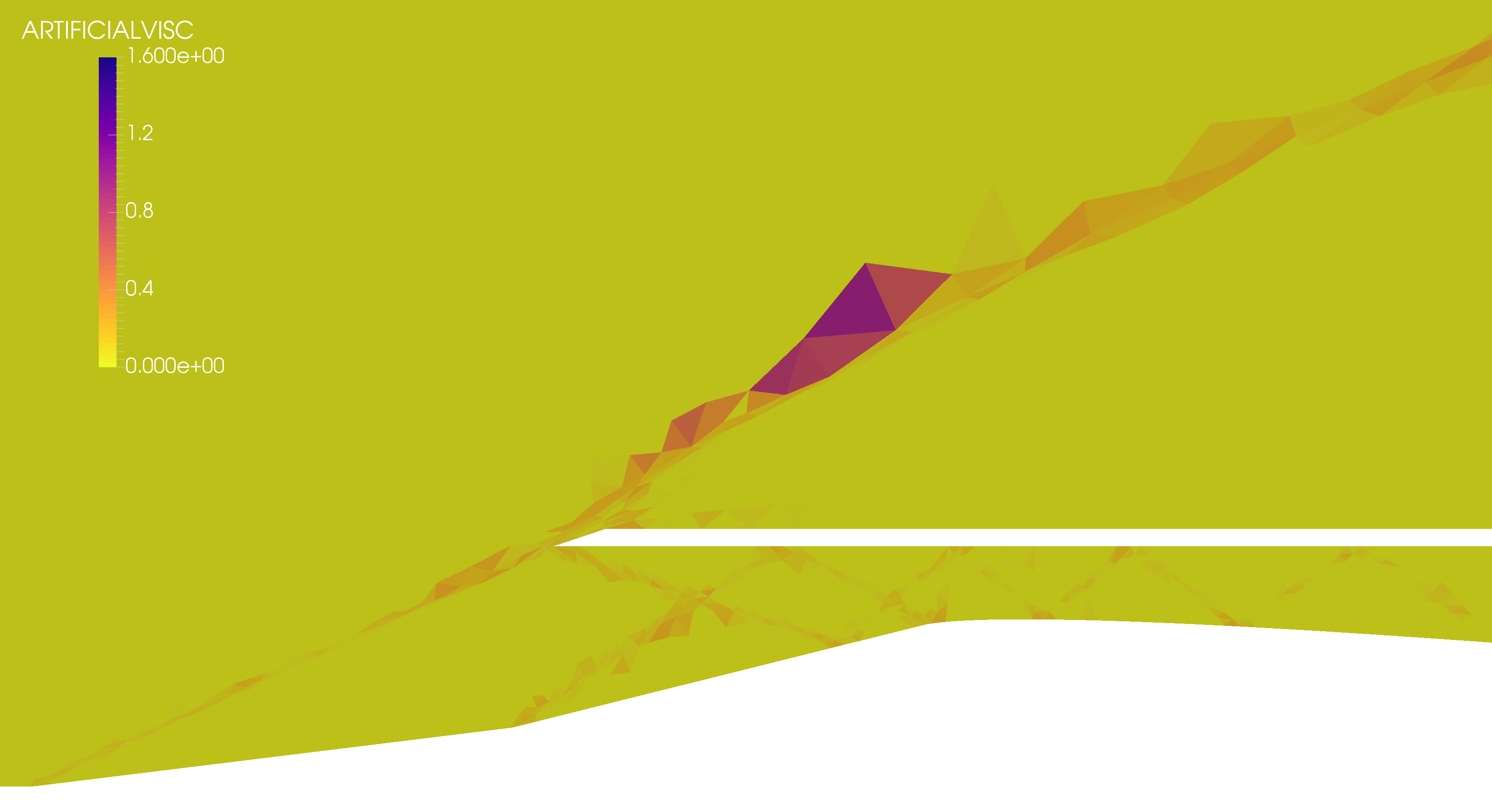}
        \includegraphics[width=0.49\textwidth]{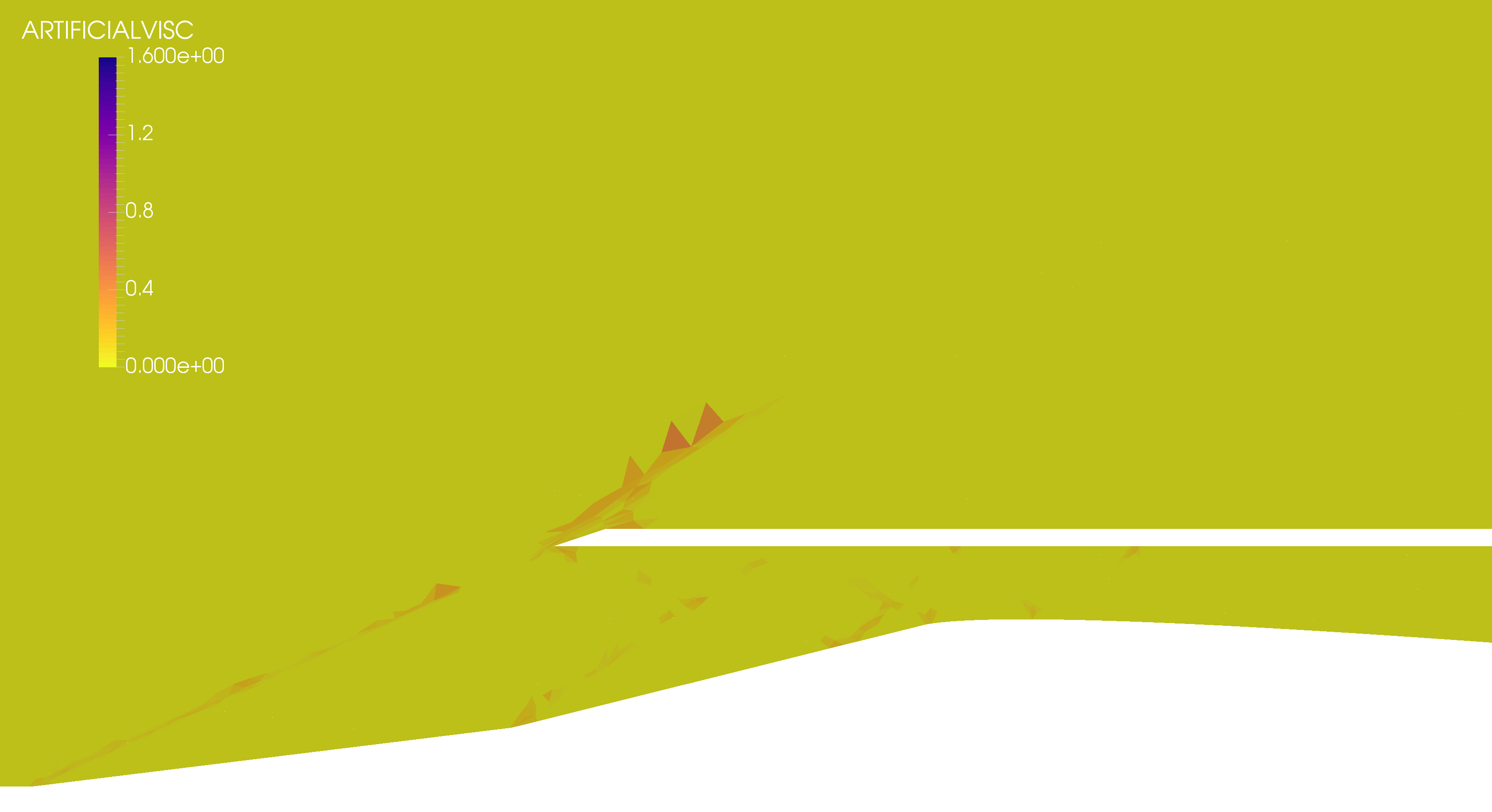}
        \caption{Artificial viscosity field.}\label{fig:intake-p-visc}
      \end{center}
    \end{subfigure}

    \begin{subfigure}[]{\textwidth}
      \begin{center}
        \includegraphics[width=0.49\textwidth]{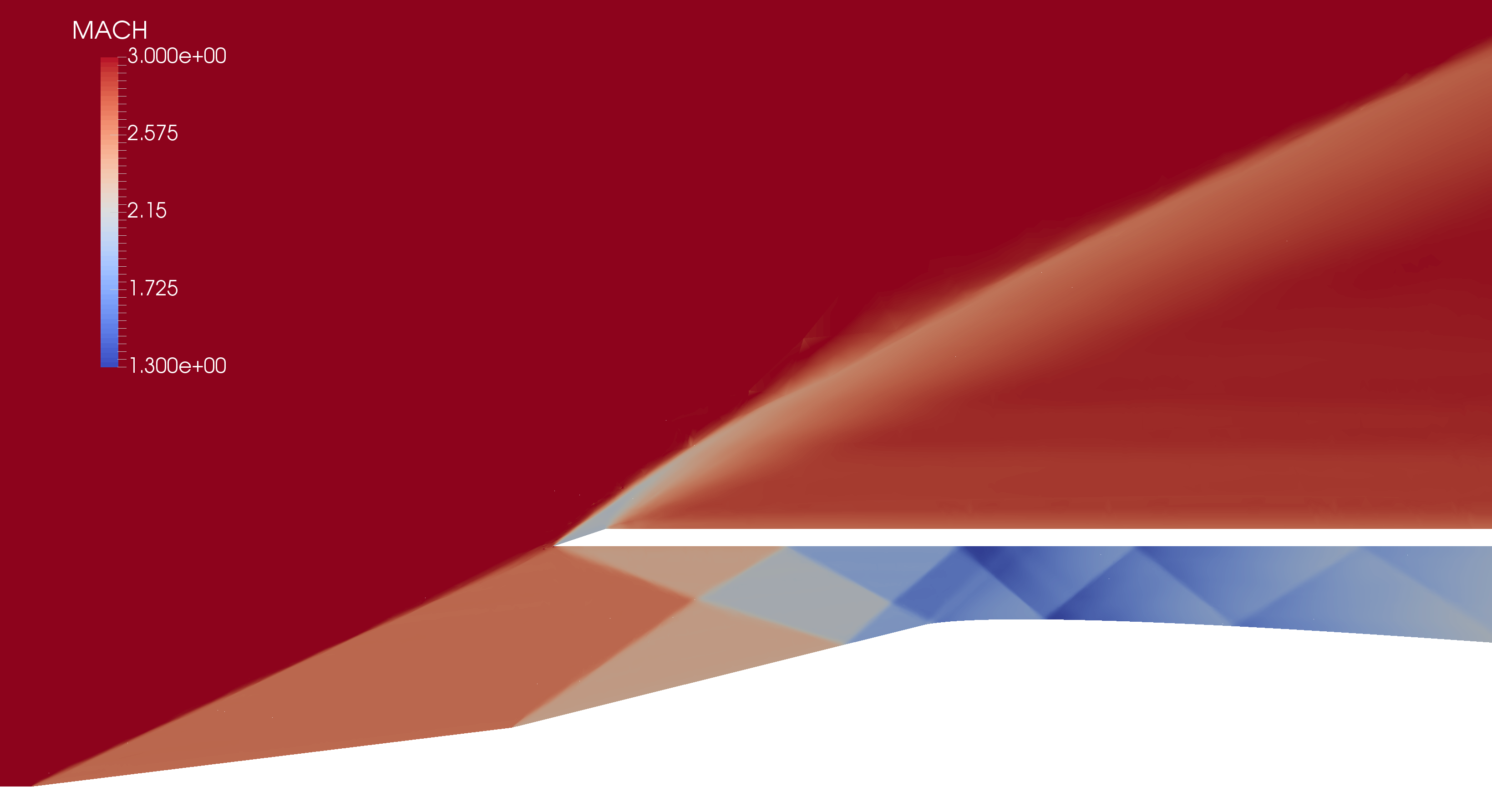}
        \includegraphics[width=0.49\textwidth]{figs/intake_mach_2_r_p_adapt_full_a}
        \caption{Mach number field.}\label{fig:intake-p-mach}
      \end{center}
    \end{subfigure}

    \caption{Comparison of the fields for the intake in its coarser (left) and finer (right) resolved states during unrestricted \adaptation{p}.}\label{fig:intake-p-adapt}
  \end{center}
\end{figure}

First, we observe that no steady state is achieved. Upon inspection, we notice that the system jumps back and forth between two
states at each \adaptation{p} cycle. The two states correspond roughly to coarser and finer resolved fields.
In the coarser resolved state, sensor values in shock areas are high.
At the end of the \adaptation{p} cycle, these large sensor values trigger an increase in local polynomial order of a number of elements.
Simulation goes on and the finer resolved state is obtained where sensor values are low.
This in turn triggers a decrease in local polynomial order of the same elements, returning the system to the former coarser resolved field.
This is shown in Fig.~\ref{fig:intake-p-adapt} with the coarser resolved state on the left and the finer one on the right.
We explain this behaviour by a naive \adaptation{p} approach using simple sensor thresholds.
The problem is highly non-linear and non-local and error from refining/coarsening regions propagates along characteristics.
The non-adjoint nature of the refinement strategy is bound to produce this sort of behaviour.

Nevertheless we observe that additional resolution in the form of higher local polynomials is found in sensible areas: in the shocks, inside the intake (especially in the throat) and right above the coil.
The only very high polynomial orders are obtained in the shocks whereas smooth regions reach order \(p=3\) at most.
Note that CPU times per time step are reported as run on a 16-core machine, once convergence is reached.
The number of DOF for each simulation and state is shown in Table~\ref{tab:intake-dof}.
Referring back to Fig.\ref{fig:intake-plot}, we can see that little difference in the solution appears from \adaptation{rr} to \adaptation{rrp} despite the decrease in number of DOF.\@

\begin{table}
  \caption{Number of DOF and CPU time per time step at convergence for the intake.}\label{tab:intake-dof}
  \centering
  \begin{tabular}{c c c c}
    \toprule
    Simulation                         & State   & Number of DOF & CPU time (ms) \\
    \midrule
    Uniform \emph{p}                   &         & 40\,210       & 27            \\
    Unrestricted \textit{p}-adaptation & Finer   & 39\,527       & 74            \\
    Unrestricted \textit{p}-adaptation & Coarser & 36\,696       & 69            \\
    \bottomrule
  \end{tabular}
\end{table}

\section{Conclusions}\label{sec:concl}

In this paper, we have presented a novel strategy for adaptive simulations, based on a combination of both \(r\)- and \adaptation{p}.
The proof-of-concept work applied here takes advantage of both strategies in different manners, as appropriate for the simulation of compressible flows containing shocks.
We achieve mesh movement required for \adaptation{r} through the use of a variational optimisation strategy, using the combination of a local discontinuity sensor and a target element size in order to effectively cluster DOF in the presence of shocks and more sharply simulate their features.
At the same time, we apply a \adaptation{p} technique in the rest of the domain in order to benefit from the spectral rate of convergence of high-order discretisations for smooth solutions.
The simulation is effectively stabilised through the use of an artificial diffusion term, again using the local discontinuity sensor.

The proposed strategy exhibits a number of benefits from a computational perspective, as seen in the results presented in the previous section, where the canoncial NACA 0012 test case and a more challenging supersonic intake have been examined.
The main benefit of this dual-adaptive technique is that we are able to significantly reduce the number of DOF required to resolve a given simulation, when compared to a uniformly refined grid or using solely \adaptation{r}.
Table~\ref{tab:naca-dof} shows that, for the various \(p\)-refinement strategies considered, the error when compared to a very fine solution remains roughly the same, whilst the simulation requires only 50\% of the DOF of the original simulation.
This has important consequences from the perspective of computational efficiency, since a significant reduction in the number of DOF will lead to a reduction in runtimes.
Likewise, the cost of operations per DOF is reduced as the polynomial order decreases, which offers the opportunity to further reduce computational cost.
The \adaptation{rp} technique therefore permits an effective balance to be achieved between the attained error and simulation expense.

From the context of more general conclusions of our results, we demonstrate that care must be taken when selecting a \adaptation{p} strategy.
In particular, the NACA 0012 simulations demonstrate that \emph{p}-coarsening can have important negative effects on the solution for minimal computational gains.
Additionally, the supersonic intake exhibits a complex shock pattern.
Because of the complexity and strength of the reflecting shocks, we show that multiple \adaptation{r} steps are not only possible but desirable.
Despite the lack of nodes to redistribute inside the intake, sufficient mesh deformation is achieved to better capture the different shocks.

Although the overall strategy has been shown to be effective, it is important to emphasise that some of the benefits we highlight in this work can be attributed to our particular implementation of the \adaptation{r} technique.
In particular, the use of the variational framework yields several advantages.
Firstly, the use of a target element size allows the grid to deform in an anisotropic manner within restricted regions of the domain.
Even when the deformation is substantial, this still permits a valid grid to be obtained, as shown in Fig.~\ref{fig:naca-r-adapt}.
Secondly, the ability to conform to complex CAD surfaces and curves whilst permitting nodes to slide across them is clearly important in the context of this work, where shocks arise at or near solid surfaces.
This functionality can be difficult to achieve in other mesh deformation techniques, particularly those that require the solution of a partial differential equations (PDE) system of an appropriate solid body model.

Finally, we note that there are a number of clear directions for potential future work in this area.
An extension of this method to transient flows, especially with moving shocks, would constitute an interesting application of this \adaptation{rp} strategy.
The variational moving mesh framework would be able to track shocks throughout the simulation without the need to generate a new mesh.
With preserved mesh connectivities, the system of equations would not need to be re-built at each adaptation step.
This is especially desirable on large meshes and large simulations based on high-performance computing (HPC) where input-output (I/O) and inter-node communication can incur significant expense.

The results presented have also unearthed some of the limitations of the approach.
In flow simulations with very complex shock patterns, if the original mesh does not contain enough points, increasing the polynomial order on its own will not provide enough DOF to capture these complex shock patterns with sufficient accuracy.
Therefore we posit that incorporating \adaptation{h} will be required and that a combination of the three approaches, namely \adaptation{hrp}, will be required for optimal results.

\section*{Acknowledgements}

\ack{
  This project has received funding from the European Union's Horizon 2020 research and innovation programme under the Marie Sk\l{}odowska-Curie grant agreement No 675008.
  DM, SJS and JP acknowledge support from the PRISM project under EPSRC grant EP/R029423/1.
}

\bibliography{refs}

\begin{thebibliography}{10}
\providecommand \doibase [0]{http://dx.doi.org/}%

\bibitem{degrazia-2016}
{de Grazia} D, Moxey D, Sherwin SJ, Kravtsova MA, Ruban AI. DNS of a
  compressible boundary layer flow past an isolated three-dimensional hump in a
  high-speed subsonic regime. {\it Physical Review Fluids} 2018\string;
  3\string: 024101.
\newblock \href {\doibase 10.1103/PhysRevFluids.3.024101} {doi:
  10.1103/PhysRevFluids.3.024101}

\bibitem{Burbeau2005}
Burbeau A, Sagaut P. {A dynamic p-adaptive discontinuous Galerkin method for
  viscous flow with shocks}. {\it Computers {\&} Fluids} 2005\string;
  34(4-5)\string: 401--417.
\newblock \href {\doibase 10.1016/j.compfluid.2003.04.002} {doi:
  10.1016/j.compfluid.2003.04.002}

\bibitem{Li2010}
Li Y, Premasuthan S, Jameson A. Comparison of adaptive h and p refinements for
  spectral difference methods. In: 40th Fluid Dynamics Conference and Exhibit.
  American Institute of Aeronautics and Astronautics; 2010; Reston, Virigina

\bibitem{Ekelschot2016}
Ekelschot D, Moxey D, Sherwin S, Peir{\'{o}} J. {A p-adaptation method for
  compressible flow problems using a goal-based error indicator}. {\it
  Computers {\&} Structures} 2016\string; 181\string: 55--69.
\newblock \href {\doibase 10.1016/j.compstruc.2016.03.004} {doi:
  10.1016/j.compstruc.2016.03.004}

\bibitem{Marcon2017}
Marcon J, Turner M, Moxey D, Sherwin S, Peir{\'{o}} J. A variational approach
  to high-order r-adaptation. In: 26th International Meshing Roundtable.
  Elsevier; 2017.

\bibitem{Mitran2001}
Mitran S. {A comparison of adaptive mesh refinement approaches for large eddy
  simulation}. tech. rep., Washington University, Seattle; {Fort Belvoir, VA}:
   2001.

\bibitem{Oden1995}
Oden J, Wu W, Legat V. {An hp-adaptive strategy for finite element
  approximations of the Navier-Stokes equations}. {\it International Journal
  for Numerical Methods in Fluids} 1995\string; 20\string: 831--851.
\newblock \href {\doibase 10.1002/fld.1650200810} {doi: 10.1002/fld.1650200810}

\bibitem{Persson2006}
Persson P, Peraire J. Sub-cell shock capturing for discontinuous Galerkin
  method. In: 44th AIAA Aerospace Sciences Meeting and Exhibit. American
  Institute of Aeronautics and Astronautics; 2006; Reno, Nevada, US.
\newblock AIAA Paper 2006-112

\bibitem{Fidkowski2011}
Fidkowski K, Darmofal D. Review of output-based error estimation and mesh
  adaptation in computational fluid dynamics. {\it AIAA Journal} 2011\string;
  49\string: 673--694.
\newblock \href {\doibase 10.2514/1.J050073} {doi: 10.2514/1.J050073}

\bibitem{Yano2012}
Yano M. {\it An optimization framework for adaptive higher-order
  discretizations of partial differential equations on anisotropic simplex
  meshes}. PhD thesis. Massachusetts Institute of Technology, {Cambridge, MA};
  2012.

\bibitem{Naddei2019}
Naddei F, {de la Llave Plata} M, Couaillier V, Coquel F. {A comparison of
  refinement indicators for p-adaptive simulations of steady and unsteady flows
  using discontinuous Galerkin methods}. {\it Journal of Computational Physics}
  2019\string; 376\string: 508--533.
\newblock \href {\doibase 10.1016/j.jcp.2018.09.045} {doi:
  10.1016/j.jcp.2018.09.045}

\bibitem{offermans2020adaptive}
Offermans N, Peplinski A, Marin O, Schlatter P. Adaptive mesh refinement for
  steady flows in Nek5000. {\it Computers \& Fluids} 2020\string; 197\string:
  104352.

\bibitem{jacobs2018error}
Jacobs CT, Zauner M, De~Tullio N, Jammy SP, Lusher DJ, Sandham ND. An error
  indicator for finite difference methods using spectral techniques with
  application to aerofoil simulation. {\it Computers \& Fluids} 2018\string;
  168\string: 67--72.

\bibitem{Moxey2017}
Moxey D, Cantwell C, Mengaldo G, et al. Towards p-adaptive spectral/hp element
  methods for modelling industrial flows. In: Spectral and High Order Methods
  for Partial Differential Equations ICOSAHOM 2016. Springer, Cham;
  2017\string: 63--79

\bibitem{Bassi1996}
Bassi F, Rebay S. A high-order accurate discontinuous finite element method for
  the numerical solution of the compressible {N}avier-{S}tokes equations. {\it
  Journal of Computational Physics} 1996\string; 131\string: 267-279.
\newblock \href {\doibase 10.1006/jcph.1996.5572} {doi: 10.1006/jcph.1996.5572}

\bibitem{Cantwell2015}
Cantwell CD, Moxey D, Comerford A, et al. Nektar++: {A}n open-source
  spectral/hp element framework. {\it Computer Physics Communications}
  2015\string; 192\string: 205--219.
\newblock \href {\doibase 10.1016/j.cpc.2015.02.008} {doi:
  10.1016/j.cpc.2015.02.008}

\bibitem{moxey2020}
Moxey D, Cantwell CD, Bao Y, et al. Nektar++: {E}nhancing the capability and
  application of high-fidelity spectral/hp element methods. {\it Computer
  Physics Communications} 2020\string; 249\string: 107110.

\bibitem{CockburnShu1997}
Cockburn B, Shu C. The local discontinuous Galerkin method for time-dependent
  convection-diffusion systems. {\it SIAM Journal on Numerical Analysis}
  1998\string; 35(6)\string: 2440-2463.
\newblock \href {\doibase 10.1137/S0036142997316712} {doi:
  10.1137/S0036142997316712}

\bibitem{barter2010shock}
Barter GE, Darmofal DL. Shock capturing with PDE-based artificial viscosity for
  {DGFEM}: {P}art I. Formulation. {\it Journal of Computational Physics}
  2010\string; 229(5)\string: 1810--1827.
\newblock \href {\doibase 10.1016/j.jcp.2009.11.010} {doi:
  10.1016/j.jcp.2009.11.010}

\bibitem{Turner2018}
Turner M, Peir{\'{o}} J, Moxey D. {Curvilinear mesh generation using a
  variational framework}. {\it Computer-Aided Design} 2017\string; 103\string:
  73--91.
\newblock \href {\doibase 10.1016/j.cad.2017.10.004} {doi:
  10.1016/j.cad.2017.10.004}

\bibitem{Kloeckner2011}
Kl\"ockner A, Warburton T, Hesthaven JS. Viscous shock capturing in a
  time-explicit discontinuous {G}alerkin method. {\it Mathematical Modelling of
  Natural Phenomena} 2011\string; 6(3)\string: 57--83.

\bibitem{Vassberg2010}
Vassberg J, Jameson A. In pursuit of grid convergence for two-dimensional
  {E}uler solutions. {\it Journal of Aircraft} 2010\string; 47(4)\string:
  1152--1166.
\newblock \href {\doibase 10.2514/1.46737} {doi: 10.2514/1.46737}

\bibitem{Turner2016a}
Turner M, Moxey D, Sherwin S, Peir{\'{o}} J. {Automatic generation of 3D
  unstructured high-order curvilinear meshes}. In: VII European Congress on
  Computational Methods in Applied Sciences and Engineering. ECCOMAS; 2016;
  Crete Island, Greece

\bibitem{Toro2009}
Toro EF. {\it {Riemann Solvers and Numerical Methods for Fluid Dynamics}}.
\newblock Berlin, Heidelberg: Springer .
\newblock 2009

\bibitem{OpenCascadeSAS2018}
{Open Cascade SAS} . {Open Cascade}. Software;  2019.

\bibitem{Anderson1970}
Anderson W, Wong N, Field M. Experimental investigation of a large-scale,
  two-dimensional, mixed-compression inlet system -- {P}erformance at design
  Conditions, {M} = 3.0. tech. rep., NASA Ames Research Center; Mountain View,
  CA:   1970.

\bibitem{Jain2006}
Jain M, Mittal S. {Euler flow in a supersonic mixed-compression inlet}. {\it
  International Journal for Numerical Methods in Fluids} 2006\string;
  50(12)\string: 1405--1423.
\newblock \href {\doibase 10.1002/fld.1109} {doi: 10.1002/fld.1109}

\end{thebibliography}

\end{document}